\documentclass[
aps,prd,twocolumn,preprintnumbers,
nofootinbib,floatfix,
amsmath,amssymb,
longbibliography
]{revtex4-2}

\usepackage[utf8]{inputenc}
\usepackage{siunitx}
\usepackage{amsfonts}
\usepackage{braket}
\usepackage{mathtools}
\usepackage{cancel}
\usepackage{slashed}
\usepackage{pifont}
\usepackage{soul}
\usepackage{comment}

\usepackage{yhmath}

\usepackage{booktabs,array}
\usepackage{hhline}
\usepackage{dcolumn}
    \newcolumntype{d}[1]{D{.}{.}{#1}}
\usepackage{graphicx}
\usepackage[caption=false]{subfig}
\usepackage{multirow}

\usepackage{dcolumn}
\usepackage{mathtools}
\usepackage{upgreek}
\newcolumntype{T}{D{.}{.}{10}}
\newcolumntype{E}{D{.}{.}{11}}
\newcolumntype{F}{D{.}{.}{5}}

\usepackage[normalem]{ulem}

\usepackage[dvipsnames]{xcolor}
\usepackage{hyperref}
\hypersetup{
    colorlinks=true,
    linkcolor=blue,
    citecolor=blue,
    filecolor=blue,
    urlcolor=blue
}

\usepackage{bm}

\newcommand{\vev}[1]{\ensuremath{\langle #1 \rangle}}
\renewcommand{\vec}[1]{{\mathbf{#1}}}

\newcommand{\MS}{\overline{\mathrm{MS}}}

\makeatletter 
\renewcommand\onecolumngrid{
\do@columngrid{one}{\@ne}
\def\set@footnotewidth{\onecolumngrid}
\def\footnoterule{\kern-6pt\hrule width 1.5in\kern6pt}
}
\renewcommand\twocolumngrid{
        \def\footnoterule{
        \dimen@\skip\footins\divide\dimen@\thr@@
        \kern-\dimen@\hrule width.5in\kern\dimen@}
        \do@columngrid{mlt}{\tw@}
}
\makeatother

\begin{document}
\title{Gravitational form factors of the pion from lattice QCD}

\newcommand{\getMITAffiliation}{\affiliation{Center for Theoretical Physics, Massachusetts Institute of Technology, Cambridge, MA 02139, U.S.A.}}

\author{Daniel C. Hackett}
\author{Patrick R. Oare}
\author{Dimitra A. Pefkou}
\author{Phiala E. Shanahan}

\getMITAffiliation

\begin{abstract}

The two gravitational form factors of the pion, $A^{\pi}(t)$ and $D^{\pi}(t)$, are computed as functions of the momentum transfer squared $t$ in the kinematic region $0\leq -t< 2~\text{GeV}^2$ on a lattice QCD ensemble with quark masses corresponding to a close-to-physical pion mass $m_{\pi}\approx 170~\text{MeV}$ and $N_f=2+1$ quark flavors. The flavor decomposition of these form factors into gluon, up/down light-quark, and strange-quark contributions is presented in the $\overline{\text{MS}}$ scheme at energy scale $\mu=2~\text{GeV}$, with renormalization factors computed nonperturbatively via the RI-MOM scheme. Using monopole and $z$-expansion fits to the gravitational form factors, we obtain estimates for the pion momentum fraction and $D$-term that are consistent with the momentum fraction sum rule and the next-to-leading order chiral perturbation theory prediction for $D^{\pi}(0)$.

\end{abstract}

\preprint{MIT-CTP/5585}

\maketitle

\section{INTRODUCTION}
\label{sec:intro}

Quantum chromodynamics (QCD)~\cite{Fritzsch:1972jv,Fritzsch:1973pi,Wilczek,Politzer} provides a rigorous description of hadrons as composite particles made up of quarks and gluons interacting via the strong force. The complexity of QCD, however, is such that constraints on quark and gluon contributions to many aspects of hadron structure are difficult to obtain, and the landscape of experiment and theory efforts on this front continues to evolve. For example, recent years have seen considerable progress in understanding how the physical properties of hadrons, like their mass and internal forces, are generated by their fundamental constituents~\cite{Burkert:2023wzr,Polyakov:2018zvc}.

These aspects of hadron structure can be addressed through determinations of gravitational form factors (GFFs)~\cite{Burkert:2023wzr,Pagels:1966zza}. These quantities contain nonperturbative information about the coupling of a hadron state to the energy-momentum tensor (EMT) of QCD, the symmetric part of which can be decomposed  as~\cite{PhysRev.128.2832,RosenfeldFrench,Belitsky:2005qn,Freese:2021jqs} ${\hat{T}^{\mu\nu} = \hat{T}_{g}^{\mu\nu}+\hat{T}_{q}^{\mu\nu}}$, where
\begin{align} \label{eq:belifante}
\begin{split}\;
\hat{T}_{g}^{\mu\nu} &= 2\; \mathrm{Tr}\left[- F^{\mu\alpha}F^{\nu}_{\;\alpha} + \frac{1}{4}
g^{\mu\nu}F^{\alpha\beta}F_{\alpha\beta}\right] \;,\\
\hat{T}_{q}^{\mu\nu} &=\sum_{f}\left[ i\bar{\psi}_{f}D^
{\{\mu}\gamma^{\nu\}} \psi_{f}\right] \;.
\end{split}
\end{align}
Here $F^{\mu\nu}$ is the gluon
field strength tensor, $\psi_f$ is a quark field of flavor $f$, $D^{\mu} = \partial^{\mu}+igA^{\mu}$, $A^{\mu}$ are the gluon fields, $g$ is the strong coupling, and $\gamma^{\mu}$ are the Dirac matrices. The repeated indices are contracted with the 
Minkowski space-time metric $g^{\mu\nu}$, the trace is over color space, and $a^{\{\mu}b^{\nu\}}=(a^{\mu}b^{\nu}+a^{\nu}b^{\mu})/2$. The GFFs of a hadron are defined from the matrix elements of the EMT in  the hadron state, and can be decomposed into quark and gluon contributions. They encode the distribution of energy, spin, pressure, and shear forces\footnote{The physical significance of this interpretation as mechanical forces is debated~\cite{Ji:2021mfb}.} within hadrons~\cite{Polyakov:2002yz}.
The total EMT of QCD is conserved, i.e.~$\partial_{\mu} \hat{T}^{\mu\nu} = 0$, but the individual quark and gluon terms are not, and therefore the resulting quark and gluon GFFs mix under renormalization and depend on the renormalization scheme and scale.

For a pion state specifically, the matrix element of the symmetric EMT can be decomposed in terms of two GFFs, $A^{\pi}(t)$ and $D^{\pi}(t)$, as
\begin{equation} \label{eq:pionME}
\begin{aligned}
\bra{\pi(p')} \hat{T}^{\mu\nu} \ket{\pi(p)} 
&= 2P^{\mu}P^{\nu}A^{\pi}(t) \\
& ~~ +\frac{1}{2}
(\Delta^{\mu}\Delta^{\nu}-g^{\mu\nu}\Delta^2)D^{\pi}(t) \;,
\end{aligned}
\end{equation}
where $\ket{\pi(p)}$ is a pion state carrying four-momentum $p$, $P=(p+p')/2$, $\Delta=p'-p$, and $t=\Delta^2$.
In the forward limit, the contributions to the $A^{\pi}$ GFF define the momentum fraction carried by the quark and gluon constituents, and therefore $A^{\pi}(0)=A_g^{\pi}(0)+A_q^{\pi}(0) = 1$, while $D^{\pi}(0)=D_g^{\pi}(0)+D_q^{\pi}(0)$ is the so-called $D$-term, which is related to the internal forces of hadrons~\cite{Polyakov:2002yz}, and is predicted to be $-1$ for the pion up to chiral-symmetry breaking effects~\cite{Hudson:2017xug,Polyakov:1999gs,Donoghue:1991qv}. 

The importance of the GFFs in characterizing hadron structure has driven a targeted experimental program in recent years, with the first extractions of proton quark~\cite{Burkert:2018bqq} and gluon~\cite{Duran:2022xag} GFFs achieved from deeply virtual Compton scattering and $J/\psi$ photoproduction measurements respectively. Progress towards the determination of the pion GFFs has been more limited, with the first phenomenological constraints of the pion quark GFFs attained using data from the Belle experiment at KEKB \cite{Belle:2015oin,Savinov:2013hda,Kumano:2017lhr}.
Further constraints on various hadron GFFs can be expected from current and future facilities, including the JLab 12 GeV program~\cite{JeffersonLabHallA:2022pnx,doi:10.1146/annurev-nucl-101917-021129,CLAS:2022syx} and the Electron-Ion Collider (EIC)~\cite{AbdulKhalek:2021gbh}. 

The theoretical determination of GFFs from first principles is possible through the computation of EMT matrix elements using lattice QCD~\cite{Wilson:1974sk}, a numerical framework that defines the QCD path integral on a discrete Euclidean space-time lattice, allowing the calculation of nonperturbative hadronic properties. The pion quark GFFs have previously been calculated using lattice methods in Refs.~\cite{Brommel:2007zz,Brommel:2005jC}. Lattice calculations have also provided predictions for the gluon GFFs of the pion~\cite{Shanahan:2018pib,Pefkou:2021fni}, for the pion quark momentum fraction~\cite{Loffler:2021afv,Yang:2014xsa}, and for the complete flavor decomposition of the pion momentum fraction~\cite{ExtendedTwistedMass:2021rdx}. Predictions of the quark GFFs of the pion have also been obtained from chiral perturbation theory~\cite{Donoghue:1991qv,Hudson:2017xug}, chiral quark models~\cite{Broniowski:2008hx,Freese:2019bhb,Son:2014sna}, the large-$N_c$ approach~\cite{Masjuan:2012sk}, the extended holographic light-front QCD framework~\cite{deTeramond:2021lxc}, a relativistic composite-particle theory~\cite{Krutov:2020ewr,Krutov:2022zgg}, and algebraic GPD ansatz~\cite{Raya:2021zrz}.

In this work, we present the first lattice QCD calculation of the full flavor decomposition of the pion GFFs $A^{\pi}(t)$ and $D^{\pi}(t)$ in the kinematic region $0\leq -t < 2~\text{GeV}^2$ on a single ensemble with $N_f=2+1$ quark flavors, and quark masses corresponding to a close-to-physical pion mass of $m_{\pi}\approx 170~\text{MeV}$. The extraction of the bare matrix elements is presented in Sec.~\ref{sec:barematel}, while the nonperturbative renormalization is discussed in Sec.~\ref{sec:renorm}. Our final results for the renormalized GFFs are given in Sec.~\ref{sec:renormGFFs}, along with a discussion of the forward limits of the GFFs.

\begin{table}
\begin{center}
\begin{tabular}{ccccccccccccccc}
\toprule
 & $L/a$   & $T/a$ & $\beta$ & $a m_l$ & $a m_s$ & $a$ [fm] & $m_{\pi}$ [MeV]  \\ \midrule
A & $48$ & $96$ & $6.3$ & $-0.2416$ &
$-0.2050$ & $0.091(1)$ & $169(1)$ &  \\ \midrule
B & $12$ & $24$ & $6.1$ & $-0.2800$ &
$-0.2450$ & $0.1167(16)$ & 
 $450(5)$ \\
\bottomrule
\end{tabular}
\end{center}
\caption{\label{tab:ensemble}Specifics of the lattice ensembles used in this work. Ensemble A, generated by the JLab/LANL/MIT/WM groups~\cite{ensembles}, is used for the calculation of the bare matrix elements presented in Sec.~\ref{sec:barematel}. The calculation of the renormalization coefficients, presented in Sec.~\ref{sec:renorm}, is performed on ensemble B.}
\end{table}

\section{EXTRACTION OF BARE MATRIX ELEMENTS}
\label{sec:barematel}

In this section, we discuss the extraction of the bare matrix elements of the EMT, which are combined with the renormalization coefficients presented in Sec.~\ref{sec:renorm} to produce the renormalized GFFs presented in Sec.~\ref{sec:renormGFFs}. The bare matrix elements are calculated on a single (2+1)-flavor lattice QCD ensemble~\cite{ensembles} of volume $L^3 \times T = 48^3 \times 96$, with light quark mass tuned to produce pion mass $m_{\pi} \approx 170~\text{MeV}$, and lattice spacing ${a = 0.091(1)~\text{fm}}$~\cite{Park:2021ypf,BMW:2012hcm}. The ensemble was generated using the L\"uscher-Weisz gauge action~\cite{Luscher:1984xn} and clover-improved Wilson quarks~\cite{Sheikholeslami:1985ij} with the clover coefficient set to the tree-level tadpole-improved value and constructed using stout-smeared links~\cite{Morningstar:2003gk}. The specifics of this ensemble, referred to as ensemble A, are summarized in Table~\ref{tab:ensemble}. The configurations were taken to be independent, and the number used was different for the calculations of the bare quark and gluon contributions, and will be specified in the corresponding subsections below. 

\subsection{Two-point functions}
\label{subsec:twopointfn}

\begin{figure}
    \centering
    \includegraphics[width=0.48\textwidth]{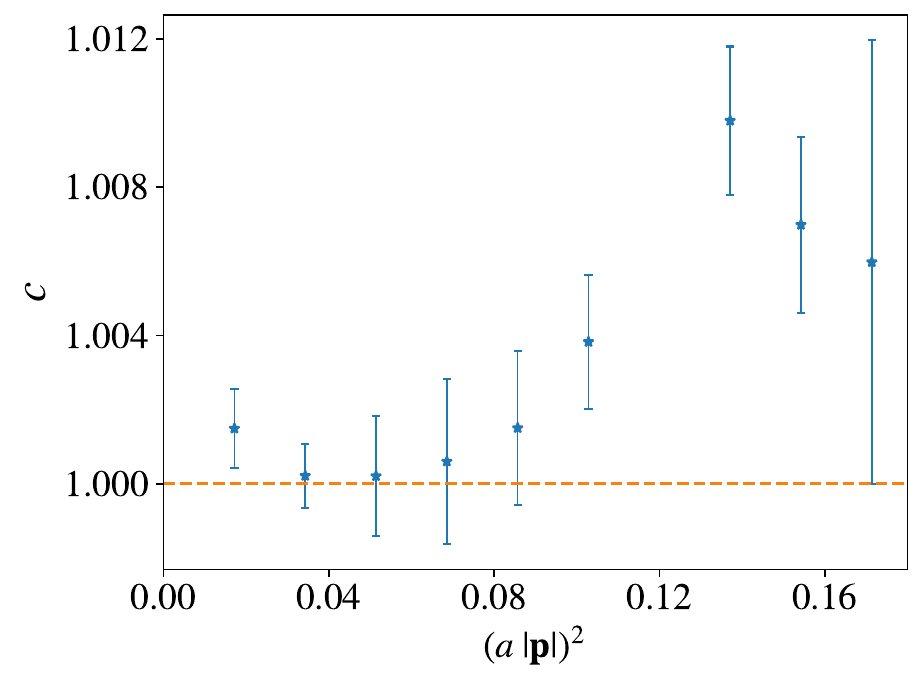}
    \caption{
        The values of $c$ obtained from the dispersion relation $E_{\vec{p}}^{\pi} = \sqrt{m_{\pi}^2+|c\vec{p}|^2}$ using the ground-state energies extracted from fits to pion two-point correlation functions, as described in the text. }
    \label{fig:dispersion}
\end{figure}

We construct momentum-projected two-point correlation functions as \begin{equation} \label{eq:twopoint_pi}
\begin{split}
C^{\text{2pt}}(\vec{p},t_s; \vec{x}_0,t_0) &=
\sum_{\vec{x}} e^{-i\vec{p} \cdot (\vec{x} - \vec{x}_0) }
 \vev{ \chi(\vec{x},t_s\!+\!t_0)  \chi^{\dagger}(\vec{x}_0,t_0) } ,
\end{split}
\end{equation}
where the pion interpolating operator is chosen to be
\begin{equation}
    \chi(x) = \overline{\psi}_u(x) \gamma_5 \psi_d(x) \;.
\end{equation}
In this expression, $\psi_u$ and $\psi_d$ are smeared quark fields obtained by applying gauge-invariant Gaussian smearing to radius $4.5a$, constructed using links stout-smeared~\cite{Morningstar:2003gk} in the spatial directions only.
Correlation functions are computed for 1024 source positions $(\vec{x}_0, t_0)$ on each of a total of $2511$ configurations, separated by at least four trajectories. The 1024 sources are arranged in two $4^3 \times 8$ grids offset by $(6,6,6,6)$ lattice units, with an overall random offset for each configuration. 
Correlation functions are constructed for all three-momenta $\vec{p} = 2\pi\vec{n}/L$ with $|\vec{n}|^2 \leq 10$, i.e., 147 distinct momenta. 

To extract the pion spectrum, we average the two-point correlation functions on each configuration over all sources, shifting each such that $(\vec{x}_0,t_0)=(\vec{0},0)$, and form $500$ bootstrap~\cite{efron1986bootstrap,efron1994introduction} ensembles from the $2511$ measurements, sampling with replacement and sample size for each bootstrap equal to the total number of configurations. We extract the energies by considering the spectral decomposition
\begin{equation}
\label{eq:twoptspectral}
C^{\text{2pt}}(\vec{p},t_s) =
 \sum_n \frac{e^{-E^{n}_\vec{p} t_s}+e^{E^{n}_\vec{p} (t_s-T)}}{2 E^{n}_\vec{p}} |Z^n_{\vec{p}}|^2 \;,
\end{equation}
where $t_s$ denotes the sink time and $E_{\vec{p}}^{n}$ denotes the energy of the $n{\text{th}}$ state with the same quantum numbers as the interpolating operator, $\ket{n(\vec{p})}$. Overlap factors are defined as
\begin{equation} \label{eq:interp}
\bra{0} \chi(\vec{x},0) \ket{n(\vec{p})} = Z^n_{\vec{p}}e^{i\vec{p}\cdot\vec{x}} \;.
\end{equation}
The lowest state in the spectrum ($n=0$) is expected to be the pion $\ket{\pi(\vec{p})}$ with energy $E^{\pi}_{\vec{p}}$. To extract $E_{\vec{p}}^{\pi}$, we perform correlated multiexponential fits incorporating up to three states to the two-point functions averaged over all sink momenta with the same magnitude $|\vec{p}|$. The effective values of $c$ obtained by taking $m_\pi = E^\pi_\vec{0}$ and inverting the dispersion relation $E^{\pi}_{\vec{p}}=\sqrt{m_{\pi}^2+|c\vec{p}|^2}$ deviate by $1\%$ at most from unity, as shown in Fig.~\ref{fig:dispersion}. Both the uncertainties on the determination of the energies $E^{\pi}_{\vec{p}}$ and their deviation from the dispersion relation values are negligible in comparison to the statistical uncertainties of the matrix elements that define the GFFs. Thus, the dispersion relation with the central value of $am_{\pi}=0.0779$ and $c=1$ is used to set the energies $E_{\vec{p}}^{\pi}$ used in the remainder of the analysis. 

\subsection{Operators}

In discrete Euclidean space-time, symmetric quark and gluon EMT operators can be expressed as
\begin{equation}
\begin{split}
\hat{T}^E_{f,\mu\nu}(x) &= \bar{\psi}_f(x)\overleftrightarrow{D}^E_{\{\mu}\gamma^E_{\nu\}}\psi_f(x) \;, \\
\hat{T}^E_{g,\mu\nu}(x) &= 2\,\text{Tr}\left[F^E_{\mu\rho}(x)F^E_{\nu\rho}(x)-\frac{1}{4}\delta_{\mu\nu}F^E_{\alpha\beta}(x)F^E_{\alpha\beta}(x)\right] 
\end{split}
\end{equation}
where $f$ denotes the quark flavor, the repeated Greek indices are summed over Euclidean components, and the quantities with an $E$ superscript are the Euclidean versions of those defined in Sec.~\ref{sec:intro}. The symmetrized lattice covariant derivative $\overleftrightarrow{D}^E
_{\mu} = (\overrightarrow{D}^E_{\mu}-
\overleftarrow{D}^E_{\mu})/2$ and the gluon field strength tensor can be expressed up to discretization effects using
\begin{align}
\begin{split}
\label{eq:covariantderiv}
\overrightarrow{D}^E_{\mu}\psi(x) &= \frac{1}{2}(U_{\mu}(x)\psi(x+\hat{\mu}) - U_{\mu}^{\dagger}(x-\hat{\mu})\psi(x-\hat{\mu})) \;,\\
\bar{\psi}(x)\overleftarrow{D}^E_{\mu} &= \frac{1}{2}(\bar{\psi}(x+\hat{\mu})U^{\dagger}_{\mu}(x) -\bar{\psi}(x-\hat{\mu}) U_{\mu}(x-\hat{\mu})) \;,\\
F^E_{\mu\nu}(x) &= \frac{i}{8 g_0}(Q_{\mu\nu}(x) - Q^{\dagger}_{\mu\nu}(x)) \;,
\end{split}
\end{align}
where $g_0$ is the bare lattice coupling\footnote{For the tadpole-improved L\"uscher-Weisz gauge action used here, $1/g_0^2=\beta(1-2/5u_0^2)/2N_c$, where $u_0$ is the tadpole parameter, and $N_c$ the number of colors.}, $U_{\mu}(x)$ are the lattice gluon link fields, and
\begin{equation}
\begin{split}
Q_{\mu\nu}(x) = &U_{\mu}(x)U_{\nu}(x+\hat{\mu})U^{\dagger}_{\mu}(x+\hat{\nu})
U^{\dagger}_{\nu}(x) \\
+ &U_{\nu}(x)U_{\mu}^{\dagger}(x-\hat{\mu}-\hat{\nu})U_{\nu}^{\dagger}(x-\hat{\mu})
U_{\mu}(x-\hat{\mu}) \\
+ &U^{\dagger}_{\mu}(x-\hat{\mu})U^{\dagger}_{\nu}(x-\hat{\mu}-\hat{\nu})
U_{\mu}(x-\hat{\mu}-\hat{\nu})U_{\nu}(x-\hat{\nu}) \\
+ &U^{\dagger}_{\nu}(x-\hat{\nu})U_{\mu}(x-\hat{\nu})U_{\nu}(x-\hat{\nu}+\hat{\mu})
U^{\dagger}_{\mu}(x) 
\end{split}
\end{equation}
is the clover term.
The Euclidean EMT components are related to the Minkowski ones by
\begin{equation} \hat{T}_{00} = \hat{T}^E_{44}, \quad \hat{T}_{0j} = -i\hat{T}^E_{4j},\quad \hat{T}_{jk} = -\hat{T}^E_{jk} \;,
\end{equation}
for $j,k\in\{1,2,3\}$. To determine the flavor decomposition of the pion GFFs into light and strange quark components, we consider the isosinglet $\hat{T}_{q,\mu\nu}$ and nonsinglet $\hat{T}_{v,\mu\nu}$ EMT operators, defined as
\begin{equation}\label{eq:isovectorT}
\begin{split}
\hat{T}_{q,\mu\nu} &=\hat{T}_{u,\mu\nu}+\hat{T}_{d,\mu\nu}+ \hat{T}_{s,\mu\nu} \;,\\
\hat{T}_{v,\mu\nu} &=\hat{T}_{u,\mu\nu}+\hat{T}_{d,\mu\nu}-2 \hat{T}_{s,\mu\nu} \;.
\end{split}
\end{equation} 
The quark isosinglet EMT mixes with the gluon one under renormalization, while the quark nonsinglet EMT does not. Lorentz symmetry is broken on a discrete hypercubic lattice, and traceless linear combinations of diagonal and off-diagonal components of the EMT transform under different irreducible representations (irreps) of the hypercubic group. Two different irreps protected from mixing with lower-dimensional operators are available, $\tau_1^{(3)}$ and $\tau_3^{(6)}$~\cite{Mandula:1983ut,Gockeler:1996mu}. A basis of operators for $\tau_1^{(3)}$ is
 \begin{equation} \label{eq:irrep3basis}
\begin{split}
\hat{T}_{\tau_{1,1}^{(3)}} &=  \frac{1}{2}(\hat{T}_{11}+\hat{T}_{22}-\hat{T}_{33}+\hat{T}_{00}),\\ 
\hat{T}_{\tau_{1,2}^{(3)}} = \frac{1}{\sqrt{2}}&(\hat{T}_{11}-\hat{T}_{22}),\; \hat{T}_{\tau_{1,3}^{(3)}} = \frac{1}{\sqrt{2}}(\hat{T}_{33}+\hat{T}_{00})\;,
\end{split}
\end{equation}
and a basis for $\tau_3^{(6)}$ is
\begin{equation}
\begin{split} \label{eq:irrep6basis}
\hat{T}_{\tau_{3,1}^{(6)}} = \frac{1}{\sqrt{2}}(\hat{T}_{12}+\hat{T}_{21}),&\;\hat{T}_{\tau_{3,2}^{(6)}} =\frac{1}{\sqrt{2}}(\hat{T}_{13}+\hat{T}_{31}) \;,\\
\hat{T}_{\tau_{3,3}^{(6)}} =\frac{-i}{\sqrt{2}}(\hat{T}_{10}+\hat{T}_{01}),&\;\hat{T}_{\tau_{3,4}^{(6)}} =\frac{1}{\sqrt{2}}(\hat{T}_{23}+\hat{T}_{32}) \;,\\
\hat{T}_{\tau_{3,5}^{(6)}} =\frac{-i}{\sqrt{2}}(\hat{T}_{20}+\hat{T}_{02}),&\;\hat{T}_{\tau_{3,6}^{(6)}} =\frac{-i}{\sqrt{2}}(\hat{T}_{30}+\hat{T}_{03}) \;,
\end{split}
\end{equation}
both in Minkowski space.
These bare operators must be renormalized, accounting for mixing between the quark and gluon operators within the same irrep. In the rest of this section, we discuss the computation of matrix elements of the bare lattice operators, while the renormalization and mixing is discussed in Sec.~\ref{sec:renorm}.

\subsection{Three-point functions}
\label{subsec:threepointfunctions}

The three-point correlation functions needed in order to isolate the bare matrix elements of the operators of Eqs.~\eqref{eq:irrep3basis} and~\eqref{eq:irrep6basis} are
\begin{equation} \label{eq:pion3pt}
\begin{split}
C^{3\text{pt}}_{\mathcal{R}\ell}(\vec{p}',t_s; \vec{\Delta}, \tau; \vec{x_0},t_0) = \sum_{\vec{x},\vec{y}}  e^{-i\vec{p}'\cdot(\vec{x}-\vec{x}_0)}e^{i\vec{\Delta}\cdot(\vec{y}-\vec{x}_0)}& \\
\times \langle \chi(\vec{x},t_s+t_0)\hat{T}_{\mathcal{R}\ell}(\vec{y},\tau+t_0)\chi^{\dagger}(\vec{x}_0,t_0) \rangle& \;,
\end{split}
\end{equation}
where $\mathcal{R}\in\{\tau_1^{(3)},\tau_3^{(6)}\}$ and $\ell$ runs over the corresponding irrep operator basis. Using translational invariance to set $(\vec{x}_0,t_0)=(\vec{0},0)$, the spectral representation of the three-point function in the limit where $(t_s -\tau) \ll T$ and  $t_s \ll T$ can be expanded as
\begin{equation}\label{eq:expanded3pt}
\begin{split}
C^{3\text{pt}}_{\mathcal{R}\ell}(\vec{p}',t_s; \vec{\Delta}, \tau) =
\sum_{n,n'} Z_\vec{p}^{n*} Z_{\vec{p}'}^{n'}\frac{e^{-E^{n'}_{\vec{p}'}t_s}e^{-(E^{n}_\vec{p}-E^{n'}_{\vec{p}'})\tau}}{4E^{n'}_{\vec{p'}}E^{n}_{\vec{p}}} &\\
\times\bra{n'(\vec{p}')}\hat{T}_{\mathcal{R}\ell}(\vec{\Delta})\ket{n(\vec{p})} &\;,
\end{split}
\end{equation}
where $\vec{p} = \vec{p'}-\vec{\Delta}$.
The three-point function contains light-quark connected, light-quark disconnected, strange (disconnected), and gluon terms,
\begin{equation} \label{eq:3ptdecomp}
C^{3\text{pt}}_{\mathcal{R}\ell}= 2 C^{3\text{pt},\text{conn}}_{l\mathcal{R}\ell}+2 C^{3\text{pt},\text{disco}}_{l\mathcal{R}\ell}+C^{3\text{pt}}_{s\mathcal{R}\ell} + C^{3\text{pt}}_{g\mathcal{R}\ell} \;.
\end{equation}
The spectral decomposition, Eq.~\eqref{eq:expanded3pt}, holds for each individual piece,\footnote{$C^{3\text{pt},\text{conn}}_{l\mathcal{R}\ell}+C^{3\text{pt},\text{disco}}_{l\mathcal{R}\ell}$ always admits a spectral decomposition, and $C^{3\text{pt},\text{disco}}_{l\mathcal{R}\ell}$ does because it is identical to a fully disconnected three-point function in a partially quenched theory with an additional light valuence quark, so their difference $C^{3\text{pt},\text{conn}}_{l\mathcal{R}\ell}$ does as well.} allowing the corresponding matrix elements $\bra{\pi(\vec{p}')}\hat{T}_{i\mathcal{R}\ell}(\vec{\Delta})\ket{\pi(\vec{p})}$ to be considered separately, where $i\in\{l^{\text{conn}},l^{\text{disco}},s,g\}$. The factor of 2 multiplying the $l^{\text{conn}}$ and $l^{\text{disco}}$ terms is due to the identical contributions of the up and down quarks in the case of the pion.

We apply the summation method \cite{Maiani:1987by,Dong:1997xr,Capitani:2012gj} to extract the bare matrix elements from the three-point functions. We first form the ratios
\begin{equation}\label{eq:ratio}
\begin{split}
R_{i\mathcal{R}\ell}(\vec{p}',t_s; \vec{\Delta},\tau)
= \frac{
    C^{\text{3pt}}_{i\mathcal{R}\ell}(\vec{p}',t_s; \vec{\Delta},\tau)
}{
    C^{\text{2pt}}(\vec{p}',t_s)
}& \\
\times\sqrt{\frac{
    C^{\text{2pt}}(\vec{p},t_s-\tau) ~
    C^{\text{2pt}}(\vec{p}',t_s) ~
    C^{\text{2pt}}(\vec{p}',\tau)
}{
    C^{\text{2pt}}(\vec{p}',t_s-\tau) ~
    C^{\text{2pt}}(\vec{p},t_s) ~
    C^{\text{2pt}}(\vec{p},\tau)
}}&
\end{split}
\end{equation}
in which the overlap factors and the time dependence of the ground state terms in Eqs.~\eqref{eq:expanded3pt} and~\eqref{eq:twoptspectral} cancel, up to finite-$T$ effects. When the source, operator, and sink are well-separated in Euclidean time, the ratio can be expanded as
\begin{equation}
\begin{split}
\label{eq:ratiospectral}
\lim_{t_s,(t_s-\tau)\rightarrow\infty}R_{i\mathcal{R}\ell}(\vec{p}',t_s; \vec{\Delta},\tau) = \frac{\bra{\pi(\vec{p}')}\hat{T}_{i\mathcal{R}\ell}\ket{\pi(\vec{p})}}{2\sqrt{E^{\pi}_{\vec{p}}E^{\pi}_{\vec{p'}}}} &\\
+ \mathcal{O}\left(e^{-\Delta E \tau}\right) + \mathcal{O}\left(e^{-\Delta E' (t_s-\tau)}\right) &\;,
\end{split}
\end{equation} 
where $\Delta E$ and $\Delta E'$ are energy differences which depend on $\vec{\Delta}$ and $\vec{p}'$.
We then sum the averaged ratios over operator insertion time for $\tau_{\text{cut}}\leq\tau\leq t_s-\tau_{\text{cut}}$ to form
\begin{equation} \label{eq:summation}
\begin{split}
\Sigma_{i\mathcal{R}\ell}(\vec{p}',t_s;\vec{\Delta},\tau_{\text{cut}}) &= \sum_{\tau=\tau_{\text{cut}}}^{t_s-\tau_{\text{cut}}} R_{i\mathcal{R}\ell}(\vec{p}',t_s;\vec{\Delta},\tau) \\
 &=\frac{t_s-2\tau_{\text{cut}}+1}{2\sqrt{E^{\pi}_{\vec{p}}E^{\pi}_{\vec{p'}}}}\bra{\pi(\vec{p}')}\hat{T}_{i\mathcal{R}\ell}\ket{\pi(\vec{p})} \\
&\;\;\;+ \Lambda(\vec{\Delta},\vec{p}') + \mathcal{O}( e^{-t_s \delta}) \;,
\end{split}
\end{equation}
where $\Lambda$ is a $t_s$-independent function of momenta and $\delta$ is an energy difference which depends on $\vec{\Delta}$ and $\vec{p}'$.
To improve the signal, we average within each flavor and irrep all ratios for choices $(\ell,\vec{p}',\vec{\Delta})$ that correspond to identical linear combinations of the GFFs up to an overall sign, which we call ``$c$-bins''. The coefficients of the GFFs in these linear combinations are defined in Eq.~\eqref{eq:pionME}, rescaled to account for the factor of $2\sqrt{E_{\vec{p}}^{\pi}E_{\vec{p}'}^{\pi}}$ in the denominator of Eq.~\eqref{eq:ratiospectral}. 
The resulting summed averaged ratios are denoted as $\bar{\Sigma}_{i\mathcal{R}c}$. We further partition these into $25$ discrete groups, denoted as ``$t$-bins'', based on the proximity of their $t$ values, using $k$-means clustering \cite{kmeans1d}. We form $\bar{\Sigma}_{i\mathcal{R}ct}$ for all contributions to the bare EMT three-point function of Eq.~\eqref{eq:3ptdecomp}, and fit them to obtain the corresponding bare matrix elements, $\text{ME}_{i\mathcal{R}ct}$. The bare GFFs for each irrep $\mathcal{R}$ and $t$-bin are constrained by the system of linear equations
\begin{equation} \label{eq:KAKDME}
\vec{K}^A_{\mathcal{R}t}A^{\pi,B}_{i\mathcal{R}t}+\vec{K}^D_{\mathcal{R}t}D^{\pi,B}_{i\mathcal{R}t} = \text{\bf{ME}}_{i\mathcal{R}t} \;,
\end{equation}
where $(K^A_{\mathcal{R}ct}, K^D_{\mathcal{R}ct})$ are the (unique) coefficients of the $c$-bin corresponding to matrix element $\text{ME}_{i\mathcal{R}ct}$, and bold symbols are vectors in the space of $c$-bins. We discuss the extraction for each contribution $i \in\{l^{\text{conn}},l^{\text{disco}},s,g\}$ individually in the rest of this section. 

\subsection{Connected quark contribution}
\label{subsec:connbare}

The connected light-quark contribution $(i=l^{\text{conn}})$ to the three-point function of Eq.~\eqref{eq:3ptdecomp} can be constructed as
\begin{equation} \label{eq:connected3pt}
\begin{split}
C^{3\text{pt},\text{conn}}_{l\mathcal{R}\ell}(\vec{p}',t_s; \vec{\Delta}, \tau; \vec{x_0},t_0) = \sum_{\vec{x},\vec{y}}  e^{-i\vec{p}'\cdot(\vec{x}-\vec{x}_0)}e^{i\vec{\Delta}\cdot(\vec{y}-\vec{x}_0)}& \\
\times\text{Tr}\bigg[S_l(\vec{x_0},t_0;\vec{y},\tau+t_0) \left(\gamma\overleftrightarrow{D}\right)_{\mathcal{R}\ell}(\vec{y},\tau+t_0)&  \\  S_l(\vec{y},\tau+t_0; \vec{x},t_s+t_0)\gamma_5 S_l(\vec{x},t_s+t_0;\vec{x_0},t_0) \gamma_5&\bigg] \;,
\end{split}
\end{equation}
where $\left(\gamma\overleftrightarrow{D}\right)_{\mathcal{R}\ell}$ represents the linear combination of $\gamma_{\mu}\overleftrightarrow{D}_{\nu}$ components corresponding to irrep $\mathcal{R}$ and basis element $\ell$, defined in Eqs.~\eqref{eq:irrep3basis} and~\eqref{eq:irrep6basis}, and $S_l(a;b)$ are light-quark propagators from lattice coordinate $a$ to $b$. The symmetric discretized covariant derivative in Eq.~\eqref{eq:connected3pt} acts on the $(\vec{y},\tau+t_0)$ argument of the quark propagators directly on its left and right, and shifts them as defined in Eq.~\eqref{eq:covariantderiv}, which is left implicit in the equation for simplicity. Quark smearing at the source and sink is also left implicit. We compute these on 1381 configurations, separated by at least ten trajectories, using the sequential source method, inverting through the sink. On each configuration, we compute correlation functions for $7$ distinct temporal source-sink separations $t_s$. The number of source positions computed varies with $t_s$ as tabulated in Table~\ref{tab:conn-counts}. For every source position and sink time, we compute correlation functions with three different sink momenta, $(L/2\pi) \vec{p}' \in \{(1,0,-1), (-2,-1,0),(-1,-1,-1)\}$. We use the full set of 9 operators of Eqs.~\eqref{eq:irrep3basis} and~\eqref{eq:irrep6basis} and all operator insertion momenta with $|\vec{\Delta}|^2 < 25 (2\pi/L)^2$, and form $500$  bootstrap ensembles from the 1381 source-averaged correlation functions.

\begin{table}
    \begin{ruledtabular}
    \begin{tabular}{c|rrrrrrr}
    $\; t_s \;$ & 6 &  8 & 10 & 12 & 14 & 16 & 18 \\ 
    $\; N_s  \;$ & 6 & 16 & 16 & 16 & 32 & 32 & 32
    \end{tabular}
    \end{ruledtabular}
    \caption{Number of sources $N_s$ for which the connected quark contribution to the three-point function, Eq.~\eqref{eq:connected3pt}, is computed for each sink time $t_s$.}
    \label{tab:conn-counts}
\end{table}

In order to extract the bare ground-state matrix elements, we first form the summed ratios of Eqs.~\eqref{eq:ratio} and \eqref{eq:summation}, using source-averaged three-point correlators and two-point correlators computed on the same set of $1381$ configurations. Ratios are binned as described in Sec.~\ref{subsec:threepointfunctions}, yielding $\bar{\Sigma}^{\text{conn}}_{l\mathcal{R}ct}$ with $708$ $c$-bins for $\tau_1^{(3)}$ and $671$ for $\tau_3^{(6)}$. These are fit to the functional form of Eq.~\eqref{eq:summation}, including the exponential term (cf.~Ref.~\cite{Djukanovic:2021cgp}). To enforce a gap over the ground state, the log of $\Delta E$ is fit with a prior $\log \Delta E= \log[0.2\pm0.5]$, where $0.2 \approx 3 E^\pi - E^\pi$ is chosen because the first excited state in the spectrum is expected to be a three-pion state. The other parameters are fit with wide, noninformative priors. Model averaging~\cite{Jay:2020jkz,Rinaldi:2018osy,Beane:2015yha} with Akaike information criterion~\cite{Akaike:1998zah} (AIC) weights over different data cuts is employed to obtain the final estimated matrix element for each $c$-bin.
The fit ranges averaged over are $\tau_\text{cut} \in\{2,3,4\}$ and $t_{s,\text{min}}\leq t_s\leq t_{s,\text{max}}$, where $t_{s,\text{max}}\in\{16,18\}$, and $t_{s,\text{min}} \in\{6,8,10\}$. Redundant fits where $t_{s,\text{min}} < 2 \tau_\text{cut}$ are not double-counted. Fits with less than five distinct $t_s$ values are not included in the pool of values for averaging. More details are provided in Appendix~\ref{app:bare-conn}.

\subsection{Disconnected quark contribution}
\label{subsec:discobare}

\begin{figure*}
\centering
\subfloat[\centering  ]
{{\includegraphics[width=0.48\textwidth]{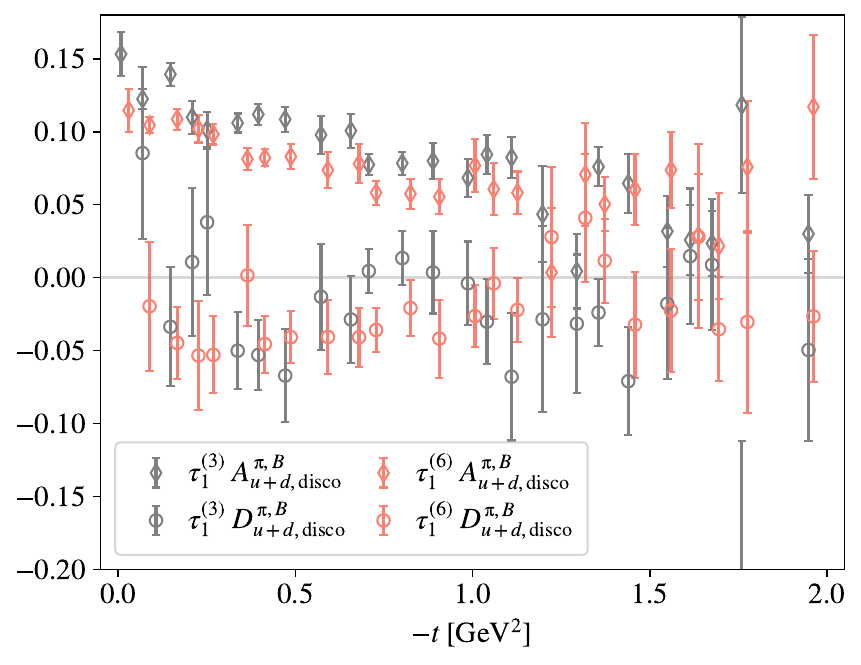}}}
\!
\subfloat[\centering ]
{{\includegraphics[width=0.48\textwidth]{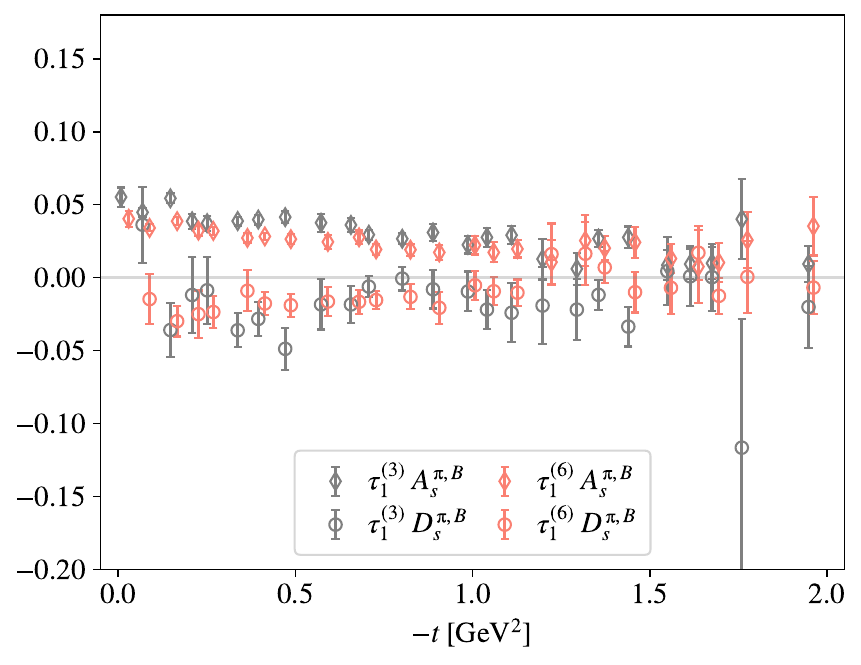}}}
\caption{Light (a) and strange (b) quark contributions to the bare disconnected GFFs, shown for both irreps, $\tau_1^{(3)}$ and $\tau_3^{(6)}$.}
\label{fig:bareGFFdisco}
\end{figure*}

The disconnected quark three-point function receives a contribution from the light quark and  strange quark terms of the EMT. These can be written as:
\begin{equation}
\begin{split}
C^{3\text{pt},\text{disco}}_{s/l\mathcal{R}\ell}(\vec{p}',t_s; \vec{\Delta}, \tau; \vec{x_0},t_0) =\small{-} \sum_{\vec{x},\vec{y}}  e^{-i\vec{p}'\cdot(\vec{x}-\vec{x}_0)}e^{i\vec{\Delta}\cdot(\vec{y}-\vec{x}_0)} &\\
\times\text{Tr}[S_l(\vec{x_0},t_0;\vec{x},t_s+t_0)\gamma_5 S_l(\vec{x},t_s+t_0; \vec{x_0},t_0)\gamma_5] &\\
\times\text{Tr}[S_{s/l}(\vec{y},\tau+t_0;\vec{y},\tau+t_0)(\gamma\overleftrightarrow{D})_{\mathcal{R}\ell}(\vec{y},\tau+t_0)] &\;,
\end{split}
\end{equation}
where $S_s(a;b)$ is the strange quark propagator. Like in Eq.~\eqref{eq:connected3pt}, the shifting of the quark propagator on the third line due to the action of the covariant derivative and the quark smearing at the source and sink are left implicit. We measure both light and strange contributions on the same 1381 configurations used for the measurement of the connected contribution discussed in Sec.~\ref{subsec:connbare}. We stochastically estimate the trace in the third line of the expression above using one independent sample of $Z_4$ noise~\cite{doi:10.1080/03610919008812866} per configuration. We dilute in spacetime using hierarchical probing~\cite{Stathopoulos:2013aci,Gambhir:2017the} with a basis of $512$ Hadamard vectors, and compute the spin-color trace exactly. The three-point function for each configuration is then computed by convolving the trace estimate with the grid of 1024 two-point correlation functions described in Sec.~\ref{subsec:twopointfn} to obtain a source-averaged estimate. The result is then averaged over forward/backward signals and vacuum subtracted. We use all operators in Eqs.~\eqref{eq:irrep3basis} and~\eqref{eq:irrep6basis}, all momenta with $|\vec{\Delta}|^2\leq25(2\pi/L)^2$ and $|\vec{p}'|^2\leq 10(2\pi/L)^2$, and all sink times $t_s$ and operator insertion times $\tau$.

To extract the matrix elements, we fit summed averaged ratios $\bar{\Sigma}^{\text{disco}}_{s/l\mathcal{R}ct}$ using the functional form of Eq.~\eqref{eq:summation} with linear terms only, as the data has insufficient precision to model excited states. We perform fits to summed ratios constructed with all  $\tau_{\text{cut}}\geq 2$, for all ranges of $4$ or more consecutive sink times in the range $[t_{s,\text{min}},t_{s,\text{max}}]$. The minimum sink time is set to $t_{s,\text{min}}=11$ for $\tau_3^{(6)}$ and $t_{s,\text{min}}=7$ for $\tau_1^{(3)}$, and the maximum sink time to $t_{s,\text{max}}=24$. These choices are motivated in Appendix~\ref{app:bare-disco}. Model averaging with AIC weights~\cite{Jay:2020jkz} then yields the bare matrix elements.

Quark matrix elements that are well-defined under renormalization can be constructed by forming the nonsinglet and isosinglet combinations defined in Eq.~\eqref{eq:isovectorT}, combining the appropriate connected and disconnected components. Unlike the connected three-point function dataset, the disconnected data includes all values of $t_s$, which, with the momenta $\vec{\Delta}$ and $\vec{p}'$ defined above, yield a total of $2179$ $c$-bins for irrep $\tau_1^{(3)}$, and $1185$ for irrep $\tau_3^{(6)}$. Combining the connected and disconnected data at the level of $c$-bins would thus involve discarding a considerable fraction of the disconnected data for which no corresponding connected results were computed. Instead, the bare disconnected matrix elements are first fit using Eq.~\eqref{eq:KAKDME} to yield the bare GFFs\footnote{$A^{\pi,\text{disco},B}_{u+d,\mathcal{R}t} = 2A^{\pi,\text{disco},B}_{l\mathcal{R}t}$, and similarly for all quantities with a $u+d$ subscript.} $A^{\pi,\text{disco},B}_{u+d,\mathcal{R}t}$, $D^{\pi,\text{disco},B}_{u+d,\mathcal{R}t}$, $A^{\pi,B}_{s\mathcal{R}t}$, and $D^{\pi,B}_{s\mathcal{R}t}$, with results shown in Fig.~\ref{fig:bareGFFdisco}. These are then used to reconstruct improved estimates of the disconnected matrix elements $\text{ME}^{\text{disco}}_{u+d,\mathcal{R}ct}$ and $\text{ME}_{s\mathcal{R}ct}$ matching the subset of $c$-bins that are available for the connected quarks. The resulting matrix elements are combined at the bootstrap level with the connected ones, to yield the nonsinglet and isosinglet operator matrix elements, which are renormalized as discussed in Sec.~\ref{sec:renormGFFs}.

\subsection{Gluon contribution}
\label{subsec:gluonbare}

The gluon contribution to the three-point function of Eq.~\eqref{eq:3ptdecomp} can be computed as
\begin{equation}
\begin{split}
C^{3\text{pt}}_{g\mathcal{R}\ell}(\vec{p}',t_s; \vec{\Delta}, \tau; \vec{x_0},t_0) = \sum_{\vec{x},\vec{y}}  e^{-i\vec{p}'\cdot(\vec{x}-\vec{x}_0)}e^{i\vec{\Delta}\cdot(\vec{y}-\vec{x}_0)} &\\
\times\text{Tr}[S_l(\vec{x_0},t_0;\vec{x},t_s+t_0)\gamma_5 S_l(\vec{x},t_s+t_0; \vec{x_0},t_0)\gamma_5] &\\
\;\;\times\text{Tr}\left[\hat{T}_{g\mathcal{R}\ell}(\vec{y},\tau+t_0)\right] &\;,
\end{split}
\end{equation}
where $\hat{T}_{g\mathcal{R}\ell}$ is the gluon EMT projected to the basis defined in Eqs.~\eqref{eq:irrep3basis} and~\eqref{eq:irrep6basis}. Quark smearing at the source and sink is left implicit. Gluon operators are computed on $2511$ configurations, with the gauge fields first subjected to $200$ steps of Wilson flow~\cite{Luscher:2010iy,Narayanan:2006rf,Lohmayer:2012hs} up to flow time $t_{\text{flow}}/a^2=2$ in order to reduce noise due UV fluctuations. The three-point functions are formed as with the disconnected quark contributions, averaging over $1024$ sources on each configuration, vacuum subtracting, and using all momenta with $|\vec{\Delta}|^2\leq25(2\pi/L)^2$ and $|\vec{p}'|^2\leq 10(2\pi/L)^2$, and all sink times $t_s$ and operator insertion times $\tau$. 

We fit the resulting summed averaged ratios $\bar{\Sigma}_{g\mathcal{R}ct}$, fitting the linear term only as in the case of the disconnected bare matrix elements. We perform fits to summed ratios constructed with $\tau_{\text{cut}}\geq 4$, avoiding contact terms due to the extension of the flowed operator definition, and to all ranges of $t_s$ of length $4$ or more in the range $[t_{s,\text{min}},t_{s,\text{max}}]$. 
As motivated in Appendix~\ref{app:bare-gluon}, we take $t_{s,\text{max}}=25$ and choose different values of $t_{s,\text{min}}$ for different $t$-bins due to differing extents of excited state contamination. Model averaging with AIC weights~\cite{Jay:2020jkz} yields the bare matrix elements used to produce the renormalized GFFs of Sec.~\ref{sec:renormGFFs}.

\section{RENORMALIZATION}
\label{sec:renorm}

In order to obtain renormalized GFFs $A_i^{\pi}(t)$, $D_i^{\pi}(t)$ for $i\in\{g,u,d,s\}$, we need to compute the renormalization coefficients of $\hat{T}_{g,\mu\nu}$, $\hat{T}_{q,\mu\nu}$, and $\hat{T}_{v,\mu\nu}$. We consider a renormalization scheme defined in the chiral limit where $\mathrm{SU}(3)$ isospin symmetry is exact. In this case, the nonsinglet operator $\hat{T}_{v,\mu\nu}$ is protected from mixing with the gluon operator, and therefore can be multiplicatively renormalized as
\begin{equation} \label{eq:isovectorR}
\hat{T}^{R}_{v,\mu\nu}(\mu^2) = Z^{R}_{v}(\mu^2,\mu_R^2) \hat{T}^{B}_{v,\mu\nu}(\mu_R^2) ,
\end{equation}
where $\hat{T}^{R}_{v,\mu\nu}$ and $\hat{T}^{B}_{v,\mu\nu}$ denote the renormalized and bare operators. $Z^{R}_{v}(\mu^2,\mu_R^2)$ denotes a factor that renormalizes the operator in a renormalization scheme $R$, and runs the result from scale $\mu_R$ to scale $\mu$. 

In contrast, the isosinglet quark and gluon EMTs mix under renormalization, and renormalized operators in scheme $R$ can be defined as
\begin{equation} \label{eq:mixingmatrix}
\begingroup
\renewcommand*{\arraystretch}{1.5}
 \begin{pmatrix}
    \hat{T}^{R}_{q,\mu\nu} \\ 
    \hat{T}^{R}_{g,\mu\nu} 
    \end{pmatrix}(\mu^2)
    = \begin{pmatrix}
    Z^R_{qq} & Z^R_{qg} \\
    Z^R_{gq} & Z^R_{gg}
    \end{pmatrix}(\mu^2,\mu_R^2)
    \begin{pmatrix}
    \hat{T}^{B}_{q,\mu\nu} \\
    \hat{T}^{B}_{g,\mu\nu} 
    \end{pmatrix}(\mu_R^2)\;.
    \endgroup
\end{equation}
To determine the renormalization factor in Eq.~\eqref{eq:isovectorR} and matrix in Eq.~\eqref{eq:mixingmatrix}, we first compute the renormalization coefficients of each irrep in the RI-MOM scheme~\cite{Martinelli:1994ty,Martinelli:1993dq}, match them to $\MS$, and run to the scale $\mu^2 = (2~\text{GeV})^2$. This is the scheme and scale in which we present the renormalized quark and gluon GFFs in Sec.~\ref{sec:renormGFFs}. The matching and running is equivalent to multiplication by the conversion factors $\mathcal{C}_{ij}^{\text{RI}/\overline{\text{MS}}}(\mu^2,\mu_R^2)$ computed to 2-loop order in $\alpha_s=g^2/4\pi$ in Ref.~\cite{Panagopoulos:2020qcn}. The inverse nonsinglet renormalization factor can be expressed in terms of the RI-MOM nonsinglet renormalization coefficient $R_{v\mathcal{R}}^{\text{RI}}$ with $\mathcal{R}\in\{\tau_1^{(3)},\tau_3^{(6)}\}$ as 
\begin{equation} \label{eq:RenormMSvec}
(Z^{\overline{\text{MS}}}_{v\mathcal{R}})^{-1}(\mu^2) = \mathcal{C}_{v}^{\text{RI}/\overline{\text{MS}}}(\mu^2,\mu_R^2)R_{v\mathcal{R}}^{\text{RI}}(\mu_R^2) \;,
\end{equation}
while the inverse of the isosinglet matrix can be written as\footnote{In our convention, the matrix $\mathcal{C}_{ij}^{\text{RI}/\overline{\text{MS}}}$ of Eq.~\eqref{eq:rimom_rcs} is the inverse of $\mathcal{C}_{ij}^{\text{RI}'}$  defined in Ref.~\cite{Panagopoulos:2020qcn}, setting to zero mixing with operators that vanish in matrix elements of physical states~\cite{Joglekar:1975nu}. In Eq.~\eqref{eq:RenormMSvec}, $\mathcal{C}_{v}^{\text{RI}/\overline{\text{MS}}} =|\mathcal{C}_{qq}^{\text{RI}/\overline{\text{MS}}}\mathcal{C}_{gg}^{\text{RI}/\overline{\text{MS}}}-\mathcal{C}_{qg}^{\text{RI}/\overline{\text{MS}}}\mathcal{C}_{gq}^{\text{RI}/\overline{\text{MS}}}|/\mathcal{C}_{gg}^{\text{RI}/\overline{\text{MS}}}$.}
\begin{equation} \label{eq:rimom_rcs}
\begin{split}
\begingroup
\renewcommand*{\arraystretch}{1.5}
\begin{pmatrix}
Z_{qq\mathcal{R}}^{\overline{\text{MS}}} & Z_{qg\mathcal{R}}^{\overline{\text{MS}}} \\
Z_{gq\mathcal{R}}^{\overline{\text{MS}}} & Z_{gg\mathcal{R}}^{\overline{\text{MS}}}
\end{pmatrix}\endgroup ^{-1}(\mu^2)
&=
\begingroup
\renewcommand*{\arraystretch}{1.5}
\begin{pmatrix}
R_{qq\mathcal{R}}^{\text{RI}} & R_{qg\mathcal{R}}^{\text{RI}} \\
R_{gq\mathcal{R}}^{\text{RI}} & R_{gg\mathcal{R}}^{\text{RI}}
\end{pmatrix}\endgroup (\mu_R^2) \\
&\times\begingroup
\renewcommand*{\arraystretch}{1.5}\begin{pmatrix}
\mathcal{C}_{qq}^{\text{RI}/\overline{\text{MS}}} & \mathcal{C}_{qg}^{\text{RI}/\overline{\text{MS}}} \\
\mathcal{C}_{gq}^{\text{RI}/\overline{\text{MS}}} &
\mathcal{C}_{gg}^{\text{RI}/\overline{\text{MS}}}
\end{pmatrix}\endgroup(\mu^2,\mu_R^2)  \;.
\end{split}
\end{equation} 
In the remainder of this section, we discuss the details of this calculation, which proceeds through the following steps:
\begin{itemize}
    \item For each irrep $\mathcal{R}$, we compute the five different RI-MOM renormalization coefficients that appear in Eqs.~\eqref{eq:RenormMSvec} and~\eqref{eq:rimom_rcs}, $R_{v}^{\text{RI}}$ and $R_{ij}^{\text{RI}}$ with $i,j\in\{q,g\}$, using renormalization conditions defined in Sec.~\ref{subsec:rimom}. 
    \item The resulting $\text{RI}$ renormalization coefficients are multiplied by the conversion factors to form the nonsinglet contribution $\tilde{R}^{\text{RI}}_v \mathcal{C}^{\text{RI}/\overline{\text{MS}}}_{v}$ and the eight isosinglet contributions arising from the matrix multiplication in the right hand side of Eq.~\eqref{eq:rimom_rcs},  $\tilde{R}_{ij}^{\text{RI}}\mathcal{C}^{\text{RI}/\overline{\text{MS}}}_{jk}$, where no summation over $j$ is implied. The tilde notation indicates that the terms computed are equal to $R^{\text{RI}}_v \mathcal{C}^{\text{RI}/\overline{\text{MS}}}_{v}$ and $R_{ij}^{\text{RI}}\mathcal{C}^{\text{RI}/\overline{\text{MS}}}_{jk}$ up to lattice artifacts. 
    \item We extract $R^{\text{RI}}_v \mathcal{C}^{\text{RI}/\overline{\text{MS}}}_{v}$ and $R_{ij}^{\text{RI}}\mathcal{C}^{\text{RI}/\overline{\text{MS}}}_{jk}$ from $\tilde{R}^{\text{RI}}_v \mathcal{C}^{\text{RI}/\overline{\text{MS}}}_{v}$ and $\tilde{R}_{ij}^{\text{RI}}\mathcal{C}^{\text{RI}/\overline{\text{MS}}}_{jk}$ by modelling the dependence of the latter on lattice artifacts, as discussed in Sec.~\ref{subsec:Rfacfit}. The results are combined to obtain the isosinglet $\overline{\text{MS}}$ renormalization matrix of Eq.~\eqref{eq:rimom_rcs} and the nonsinglet renormalization coefficent of Eq.~\eqref{eq:RenormMSvec} for each irrep $\mathcal{R}$.
\end{itemize}

The RI-MOM renormalization coefficients are computed on ensemble B with parameters as in Table~\ref{tab:ensemble}. While it would be preferable to compute these on ensemble A to match the bare matrix elements presented in Sec.~\ref{sec:barematel}, far greater statistics than can be practically obtained would be required to make their extraction possible on that ensemble.
However, comparison of other renormalization factors on these ensembles~\cite{Mondal:2020ela} suggests that the resulting systematic uncertainty is no more significant than the other unquantified uncertainties in this calculation using a single ensemble of gauge fields.

\subsection{RI-MOM}
\label{subsec:rimom}

The RI-MOM scheme~\cite{Martinelli:1994ty,Martinelli:1993dq} defines renormalization coefficients by imposing, in a fixed gauge, that amputated forward-limit three-point functions with incoming fields of four-momentum $p$ are equal to their tree-level values at some energy scale $\mu_R^2 = p^2$. For each irrep, four such renormalization conditions are needed to solve for the four coefficients of the mixing matrix in Eq.~\eqref{eq:mixingmatrix}, and an additional one is needed to fix the renormalization of the nonsinglet operator.

Bare three-point functions for the flavor-singlet quark and gluon EMT operators can be constructed with either light quarks or gluons as the external states as
\begin{equation} \label{eq:nonamputated}
\begin{split}
    C_{i,\mu\nu}^{q}(p^2) &= \sum_{x,y,z}e^{ip(x-z)}\braket{\psi_u(x)\hat{T}_{i,\mu\nu}(y)\overline{\psi}_u(z)}\;, \\
    C_{i,\mu\nu\alpha\beta}^{g}(p^2) &= \sum_{x,y,z}e^{ip(x-z)}\braket{\text{Tr}[A_{\alpha}(x) A_{\beta}(z)]\hat{T}_{i,\mu\nu}(y)} \;,
\end{split}
\end{equation}
where the subscript $i\in\{g,q\}$ denotes the type of operator and the superscript denotes the type of external fields. The variables in the equations above, and in the rest of this section, are written in Euclidean space. The gluon fields are computed from link fields $U_{\mu}(x)$ fixed to Landau gauge and can be expressed up to discretization effects as
\begin{equation} \label{eq:gluonfields}
\begin{split}
A_{\mu}(p)&=\sum_x e^{-ip(x+a\hat{\mu}/2)}A_{\mu}(x+a\hat{\mu}/2)\;, \\
A_{\mu}(x+a\hat{\mu}/2)&=\frac{1}{2ig_0}\bigg[U_{\mu}(x)-U^{\dagger}_{\mu}(x)\\&\ \quad\quad\quad-\frac{1}{N_c}\text{Tr}(U_{\mu}(x)-U^{\dagger}_{\mu}(x))\bigg] \;,
\end{split}
\end{equation} 
where $N_c=3$, and $\hat{\mu}$ is a unit vector in the $\mu$-direction. The bare light quark and gluon propagators needed to amputate the three-point functions are expressed as
\begin{equation}
S^q(p^2) = \sum_{x,y}e^{ip(x-y)}\braket{\psi_u(x)\bar{\psi}_{u}(y)}\;,
\end{equation}
\begin{equation} \label{eq:gluonprop}
S^{g}_{\alpha\beta}(p^2) = \sum_{x,y}e^{ip(x-y)}\braket{\text{Tr}[A_{\alpha}(x)A_{\beta}(y)]} \, . 
\end{equation}
These are renormalized by
\begin{equation}
\begin{split}
    S^{q,R}(\mu_R^2)&=Z_q(\mu_R^2) S^{q}(p^2=\mu_R^2) \;,\\ S^{g,R}_{\alpha\beta}(\mu_R^2)&=Z_g(\mu_R^2) S^{g}_{\alpha\beta}(p^2=\mu_R^2) \, .
\end{split}
\end{equation}
The gluon field renormalization is defined as
\begin{equation}
    Z_g(\mu_R^2)=\frac{N_c^2-1}{2}\frac{3/\tilde{p}^2}{\sum_{\alpha}\braket{\text{Tr}[A_{\alpha}(\tilde{p})A_{\alpha}(-\tilde{p})]}}\bigg|_{\tilde{p}^2=\mu_R^2}  \;
\end{equation}
with lattice momenta
\begin{equation}
\tilde{p}_{\mu} =  \frac{a}{\pi}\text{sin}\left(\frac{p_{\mu}}{2a}\right) \;,
\end{equation}
while the quark field renormalization is defined as
\begin{equation}
Z_q(\mu_R^2)=\frac{i}{4 N_c\tilde{p}^2}\text{Tr}[(S^q)^{-1}(\tilde{p}^2)\cancel{\tilde{p}}]\bigg|_{\tilde{p}^2=\mu_R^2} \;.
\end{equation}

We consider the amputated three-point functions,
\begin{equation}\label{eq:amputated}
\begin{split}
C_{i,\mu\nu}^{q,\text{amp}}(p^2) &= (S^q(p^2))^{-1} C_{i,\mu\nu}^{q}(p^2) (S^q(p^2))^{-1}\;, \\
C_{i,\mu\nu\alpha\beta}^{g,\text{amp}}(p^2) &= (S^{g}_{\alpha\alpha'}(p^2))^{-1}C_{i,\mu\nu\alpha'\beta'}^{g}(p^2)(S^{g}_{\beta'\beta}(p^2))^{-1} \;,
\end{split}
\end{equation}
where summation over repeated indices is implied. The tree-level value of the amputated three-point functions with a gluon operator and quark external states or a quark operator with gluon external states are equal to zero at $\tilde{p}^2=\mu_R^2$, while $C^{q,\text{amp}}_{q,\mu\nu}$ is set equal to 
\begin{equation} 
\Lambda_{\mu\nu}^{q}(\tilde{p}) = \frac{1}{2}(\tilde{p}_{\mu}\gamma_{\nu}+\tilde{p}_{\nu}\gamma_{\mu})-\frac{1}{4}\cancel{\tilde{p}}\delta_{\mu\nu}
\end{equation}
at $\tilde{p}^2=\mu_R^2$, and $C^{g,\text{amp}}_{g,\mu\nu\alpha\beta}$ is set equal to
\begin{equation} \label{eq:gluontree}
\begin{split}
\Lambda_{\mu\nu\alpha\beta}^{g}(\tilde{p}) &= \frac{N_c^2-1}{2}(2\tilde{p}_{\mu}\tilde{p}_{\nu}\delta_{\alpha\beta}-\tilde{p}_{\alpha}\tilde{p}_{\nu}\delta_{\mu\beta}-\tilde{p}_{\alpha}\tilde{p}_{\nu}\delta_{\mu\beta}\\
&\quad \quad \quad \quad-\tilde{p}_{\beta}\tilde{p}_{\nu}\delta_{\mu\alpha} 
-\tilde{p}_{\beta}\tilde{p}_{\mu}\delta_{\nu\alpha} +\tilde{p}_{\alpha}\tilde{p}_{\beta}\delta_{\mu\nu} \\
&\quad \quad\quad\quad -\tilde{p}^2(\delta_{\alpha\beta}\delta_{\mu\nu} -\delta_{\alpha\mu}\delta_{\beta\nu}-\delta_{\alpha\nu}\delta_{\beta\mu})) 
\end{split} 
\end{equation}
at $\tilde{p}^2=\mu_R^2$.
To isolate the gauge-invariant part of the gluon three-point function, which is proportional to the first term of Eq.~\eqref{eq:gluontree}~\cite{Caracciolo:1989pt}, the momenta and indices must be chosen such that $\tilde{p}_{\alpha}=0$, $\alpha=\beta$, $\alpha\neq\mu$, and $\alpha\neq\nu$. The Landau-gauge gluon propagator is not invertible, but for this choice of four-momenta $\tilde{p}$ it takes a block diagonal form of an invertible and a noninvertible piece; for this choice of indices, only the invertible part is needed to amputate the three-point function.

The RI-MOM renormalization coefficients can be obtained by solving the following system of renormalization conditions:
\begin{equation} \label{eq:RenormCond}
\begin{split}
R_{qq}^{\text{RI}}(\mu_R^2) &= \frac{C^{q,\text{amp}}_{q,\mu\nu}}{Z_q \Lambda_{\mu\nu}^{q}}\bigg|_{\tilde{p}^2=\mu_R^2} \;,\\ 
R_{gg}^{\text{RI}}(\mu_R^2) &= \frac{C^{g,\text{amp}}_{g,\mu\nu\alpha\beta}}{Z_g \Lambda_{\mu\nu\alpha\beta}^{g}}\bigg|_{\tilde{p}_{\alpha}=0,\tilde{p}^2=\mu_R^2}^{\alpha=\beta,\alpha\neq\mu,\alpha\neq\nu} \;,\\
R_{qg}^{\text{RI}}(\mu_R^2) &= \frac{C^{g,\text{amp}}_{q,\mu\nu\alpha\beta}}{Z_g \Lambda_{\mu\nu\alpha\beta}^{g}}\bigg|_{\tilde{p}_{\alpha}=0,\tilde{p}^2=\mu_R^2}^{\alpha=\beta,\alpha\neq\mu,\alpha\neq\nu} \;, \\
R_{gq}^{\text{RI}}(\mu_R^2) &= \frac{C^{q,\text{amp}}_{g,\mu\nu}}{Z_q \Lambda_{\mu\nu}^{q}}\bigg|_{\tilde{p}^2=\mu_R^2} \;.
\end{split}
\end{equation}

The leading-order $\mathcal{O}(a)$ correction to $\Lambda_{\mu\nu}^{q}(\tilde{p})$ calculated in lattice perturbation theory~\cite{Capitani:1994qn,Gracey:2003mr} is
\begin{equation} \label{eq:extratensor}
\delta\Lambda_{\mu\nu}^{q}(\tilde{p}) = \frac{\tilde{p}_{\mu}\tilde{p}_{\nu}}{\tilde{p}^2}\cancel{\tilde{p}}-\frac{1}{4}\cancel{\tilde{p}}\delta_{\mu\nu} \;.
\end{equation}
It can be removed by projecting the three-point functions $C^{q,\text{amp}}_i$ for $i\in\{q,g\}$ to the $\Lambda_{\mu\nu}^{q}(\tilde{p})$ space. To accomplish this, an inner product $\langle\cdot, \cdot\rangle$ is introduced on the space of tensors with two Lorentz and two Dirac indices~\cite{PhysRevLett.126.202001}, and the renormalization conditions for $R_{qq}^{\text{RI}}$ and $R_{gq}^{\text{RI}}$ are recast as the matrix equations
\begin{equation} \label{eq:patrick}
\begingroup
\renewcommand*{\arraystretch}{1.2}
    \begin{pmatrix} \langle C^{q, \mathrm{amp}}_i, \Lambda^q\rangle \\ \langle C^{q, \mathrm{amp}}_i, \delta\Lambda^q\rangle \end{pmatrix} = Z_q 
    \begin{pmatrix} \langle \Lambda^q, \Lambda^q\rangle & \langle \delta\Lambda^q, \Lambda^q\rangle \\
    \langle \Lambda^q, \delta\Lambda^q\rangle & \langle \delta\Lambda^q, \delta\Lambda^q\rangle \end{pmatrix}
    \begin{pmatrix} R_{iq}^{\mathrm{RI}} \\ \delta R_{iq}^{\mathrm{RI}}\end{pmatrix} \;, \endgroup
\end{equation}
which are solved for $R_{iq}^{\mathrm{RI}}$ with $i\in\{q,g\}.$

To reduce statistical fluctuation in the computation of the renormalization factors involving gluon external states, one can substitute~\cite{Yang:2018bft,Shanahan:2018pib}
\begin{equation} 
\braket{\text{Tr}[A_{\mu}(p)A_{\nu}(-p)]} = \frac{1}{Z_g(p^2)}\frac{N_c^2-1}{2p^2}\left(\delta_{\mu\nu}-\frac{p_{\mu}p_{\nu}}{p^2}\right)
\end{equation}
for one of the gluon propagators in the denominator of the amputated three-point functions. This cancels the dependence of Eq.~\eqref{eq:RenormCond} on $Z_g$.

The renormalization condition for the nonsinglet contribution is
\begin{equation} \label{eq:RenormCondVec}
R_{v}^{\text{RI}}(\mu_R^2) =  \frac{C^{q,\text{amp}}_{v,\mu\nu}}{Z_q \Lambda_{\mu\nu}^{q}}\bigg|_{\tilde{p}^2=\mu_R^2} \;,
\end{equation}
where $C^{q,\text{amp}}_{v,\mu\nu}$ is the amputated three-point function of the nonsinglet operator defined in Eq.~\eqref{eq:isovectorT}. We project out the leading order lattice artifact of Eq.~\eqref{eq:extratensor} for $R_{v}^{\text{RI}}$ using the same approach as in the case of $R_{qq}^{\text{RI}}$ and $R_{gq}^{\text{RI}}$ described above.

\subsection{Fitting of renormalization coefficients}
\label{subsec:Rfacfit}

The quantities $\tilde{R}^{\text{RI}}_v \mathcal{C}^{\text{RI}/\overline{\text{MS}}}_{v}$ and $\tilde{R}_{ij}^{\text{RI}}\mathcal{C}^{\text{RI}/\overline{\text{MS}}}_{jk}$, formed by multiplying the computed RI-MOM coefficients [Eqs.~\eqref{eq:RenormCond} and~\eqref{eq:RenormCondVec}] by the matching factors, have a residual dependence on $(a\tilde{p})^2$, as well as on invariants of the hypercubic group, due to nonperturbative effects, lattice artifacts, and discretization effects. The $\overline{\text{MS}}$ renormalization factor contributions $R^{\text{RI}}_v \mathcal{C}^{\text{RI}/\overline{\text{MS}}}_{v}$ and $R_{ij}^{\text{RI}}\mathcal{C}^{\text{RI}/\overline{\text{MS}}}_{jk}$ are obtained from the above quantities by fitting and subtracting this contamination. 

Rather than modeling and fitting hypercubic breaking effects, the data is restricted to the subset for which $\mathcal{O}(a^2)$ hypercubic artifacts are expected to be suppressed by cutting the momenta on ``democracy" $\text{dem}\equiv \tilde{p}^{[4]}/(\tilde{p}^2)^2$~\cite{Boucaud:2003dx}, where
\begin{equation}
\tilde{p}^{[4]} = \sum_{i=1}^4 \tilde{p}_i^4 \;.
\end{equation} 
For the three-point functions with external quarks, the cut $\text{dem} \leq 0.3$ both results in acceptable fit quality and retains sufficient data to allow several thousand fits to be performed to different subsets to assess the systematic uncertainties, as discussed in the following subsections. For the three-point functions with external gluons, there is the additional constraint that $\tilde{p}_{\alpha}=0$, where $\alpha$ is the Euclidean vector index of the external gluon fields. This reduces the number of available momenta with low democracy, so the cut for these varies between $0.4$ and $0.5$ instead, as is discussed case-by-case in the following subsections. 

The remaining $(a\tilde{p})^2$ dependence of the renormalization factors (i.e., due to lattice artifacts and nonperturbative effects) can be modeled as
\begin{equation}
\label{eq:laurent}
\begin{split}
\tilde{R}^{\text{RI}}_{ij}\mathcal{C}^{\text{RI}/\overline{\text{MS}}}_{jk} (a^2\tilde{p}^2) & =  
 P_{n_1}(a^2\tilde{p}^2,\tilde{R}_{10},\tilde{R}_{11},\tilde{R}_{12},...) \\
& +P^{\text{log}}_{n_2}(a^2\tilde{p}^2,\tilde{R}_{20},\tilde{R}_{21},\tilde{R}_{22},...) \\
& +P^{\text{inv}}_{n_3}(a^2\tilde{p}^2,\tilde{R}_{30},\tilde{R}_{31},\tilde{R}_{32},...)\;,
\end{split}
\end{equation}
with $P_n$ being the polynomial function
\begin{equation}
P_n(x,Y_1,Y_2,Y_3,...) = \sum_{l=0}^n Y_l x^l \;, 
\end{equation}
$P^{\text{log}}_n$ the polynomial of a logarithm
\begin{equation}
P_n^{\text{log}}(x,Y_1,Y_2,Y_3,...) = P_n(\text{log}(x),Y_1,Y_2,Y_3,...) \;,
\end{equation}
and $P^{\text{inv}}_n$ the inverse polynomial
\begin{equation}
P^{\text{inv}}_n(x,Y_1,Y_2,Y_3,...) = \frac{1}{P_n(x,Y_1,Y_2,Y_3,...)} \;.
\end{equation}
The renormalization component $R_{ij}^{\text{RI}}\mathcal{C}_{jk}^{\text{RI}/\overline{\text{MS}}}$ can be extracted from Eq.~\eqref{eq:laurent} as
\begin{equation}
R_{ij}^{\text{RI}}\mathcal{C}_{jk}^{\text{RI}/\overline{\text{MS}}} = \tilde{R}_{10}+\tilde{R}_{20}+\frac{1}{\tilde{R}_{30}} \;.
\end{equation}
Different $\tilde{R}$ parameters for each $(i,j,k)$ in Eq.~\eqref{eq:laurent} are implied, and an identical equation can be used to model $\tilde{R}^{\text{RI}}_v\mathcal{C}_v^{\text{RI}/\overline{\text{MS}}}$. In practice, we find that simultaneously including all $P_{n_1}$, $P^{\text{log}}_{n_2}$, and $P^{\text{inv}}_{n_3}$ terms leads to overfitting. The different renormalization terms are instead modeled by different combinations of $P_{n_1}$, $P^{\text{log}}_{n_2}$, and $P^{\text{inv}}_{n_3}$, according to which momentum dependence(s) is (are) dominant. Specifically: 
\begin{itemize}
    \item Logarithmic terms ($P^{\text{log}}_{n_2}$) are included only for the renormalization terms that are multiplied by the flavor off-diagonal $\mathcal{C}^{\text{RI}/\overline{\text{MS}}}_{ij}$ for $i\neq j$. In all such cases, we find that a polynomial with $n_2=2$ is sufficient to describe the data, and that including higher-order terms leads to overfitting. Logarithmic terms are not included for the renormalization coefficients multiplied by the flavor-diagonal $\mathcal{C}^{\text{RI}/\overline{\text{MS}}}_{jj}$, because the logarithmic dependence for those terms is a subleading correction to the matching coefficient.
    \item Polynomial terms ($P_{n_1}$) are included for renormalization coefficients multiplied by the flavor-diagonal $\mathcal{C}^{\text{RI}/\overline{\text{MS}}}_{jj}$ (including the single isovector contribution), excluding those with unflowed external gluon fields, as discussed below. Including polynomial terms for coefficients multiplied by the flavor off-diagonal $\mathcal{C}^{\text{RI}/\overline{\text{MS}}}_{ij}$ does not alter the final results, since the logarithmic dependence dominates for these terms. We find that setting $n_1=1$ in the polynomial is sufficient, and including higher-order terms yields orders-of-magnitude smaller AIC weights.
    \item The inverse polynomial term ($P^{\text{inv}}_{n_3}$) is included in the fit ansatz for all renormalization coefficients with unflowed external gluon fields. This is necessary because of the strong discretization artifacts of the gluon propagator in the denominator of these coefficients, when it is constructed from unflowed link fields. The parameters of the inverse polynomial term are always chosen such that the term is monotonic in the $(a\tilde{p})^2$ region in which the data is fit. The degree of the polynomial $n_3$ used is different between the different coefficients, and is set to be the integer that yields the highest $p$-value for each fit. Consistent results are obtained when including several different choices of $n_3$ in the model averaging.
\end{itemize}
Further details are provided in the following discussions for each renormalization coefficient.

\begin{figure*}
\centering
\subfloat[\centering  ]
{{\includegraphics[width=0.48\textwidth]{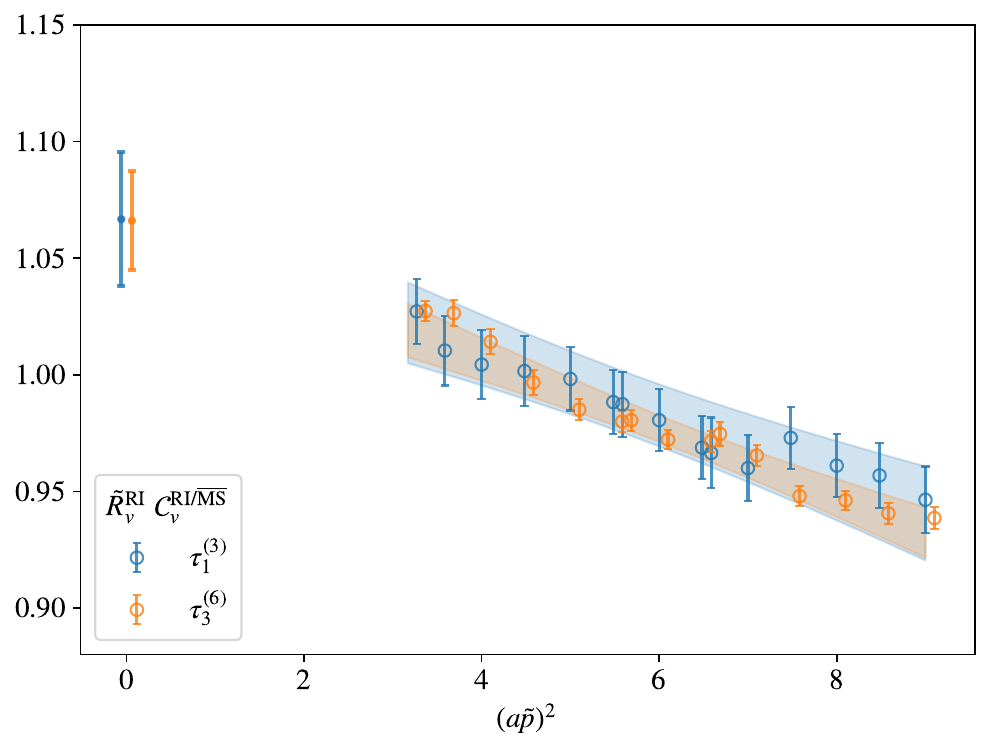}}\label{subfig:Rvfits}} \!
\subfloat[\centering  ]
{{\includegraphics[width=0.48\textwidth]{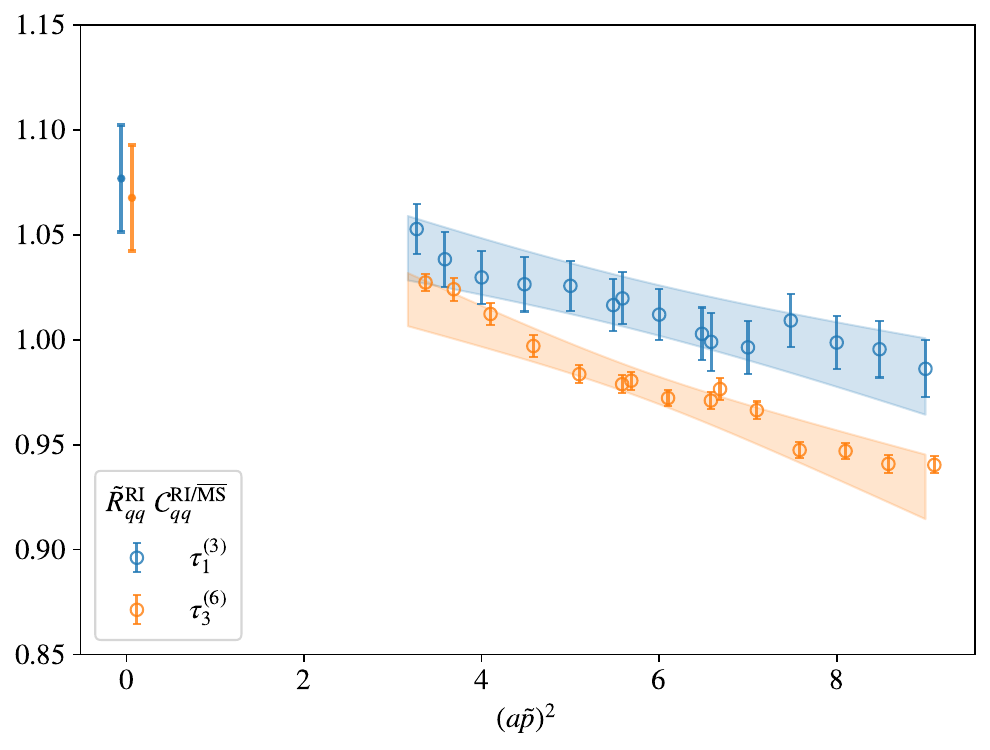}}\label{subfig:RqqZqqfits}} \\
\subfloat[\centering  ]
{{\includegraphics[width=0.48\textwidth]{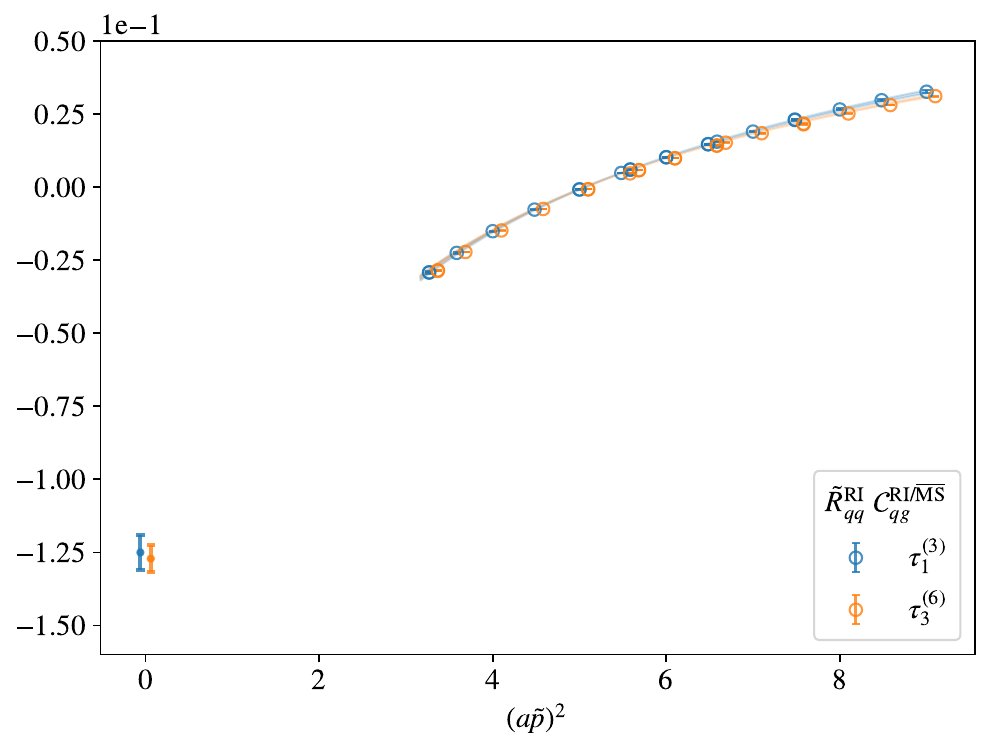}}\label{subfig:RqqZqgfits}}
\caption{The computed contributions $\tilde{R}_v^{\text{RI}}\mathcal{C}^{\text{RI}/\overline{\text{MS}}}_{v}$ (a), $\tilde{R}_{qq}^{\text{RI}}\mathcal{C}^{\text{RI}/\overline{\text{MS}}}_{qq}$ (b), and $\tilde{R}_{qq}^{\text{RI}}\mathcal{C}^{\text{RI}/\overline{\text{MS}}}_{qg}$ (c) are shown as empty markers for irreps $\tau_1^{(3)}$ (blue) and $\tau_3^{(6)}$ (orange). Fits to various sets of points are performed as described in Sec.~\ref{subsec:Rv}. The model-averaged fits to the data are shown as bands, and the resulting estimates of the renormalization contributions $R_v^{\text{RI}}\mathcal{C}^{\text{RI}/\overline{\text{MS}}}_{v}$, $R_{qq}^{\text{RI}}\mathcal{C}^{\text{RI}/\overline{\text{MS}}}_{qq}$, and $R_{qq}^{\text{RI}}\mathcal{C}^{\text{RI}/\overline{\text{MS}}}_{qg}$ are shown with filled markers at $(a\tilde{p})^2=0$. We note that $R_{qq}^{\text{RI}}\mathcal{C}^{\text{RI}/\overline{\text{MS}}}_{qg}$ (c) does not correspond to the fit band continued to $(a\tilde{p})^2=0$, due to the presence of logarithmic terms in the model ansatz, as described in Secs.~\ref{subsec:Rfacfit} and~\ref{subsec:Rv}. The markers for $\tau_3^{(6)}$ are shifted slightly on the horizontal axis for visibility.}
\label{fig:Rqqallfits}
\end{figure*}

\subsection{$R_v$ and $R_{qq}$}
\label{subsec:Rv}

\begin{figure*}[t]
\centering
\subfloat[\centering  ]
{{\includegraphics[width=0.48\textwidth]{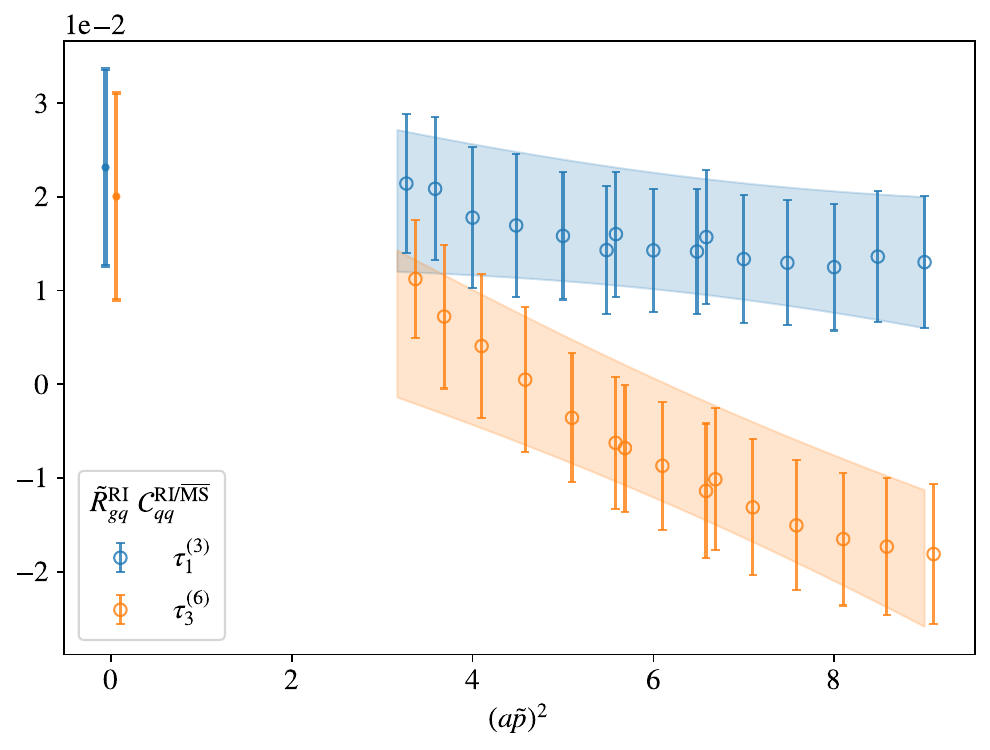}}} \!
\subfloat[\centering  ]
{{\includegraphics[width=0.48\textwidth]{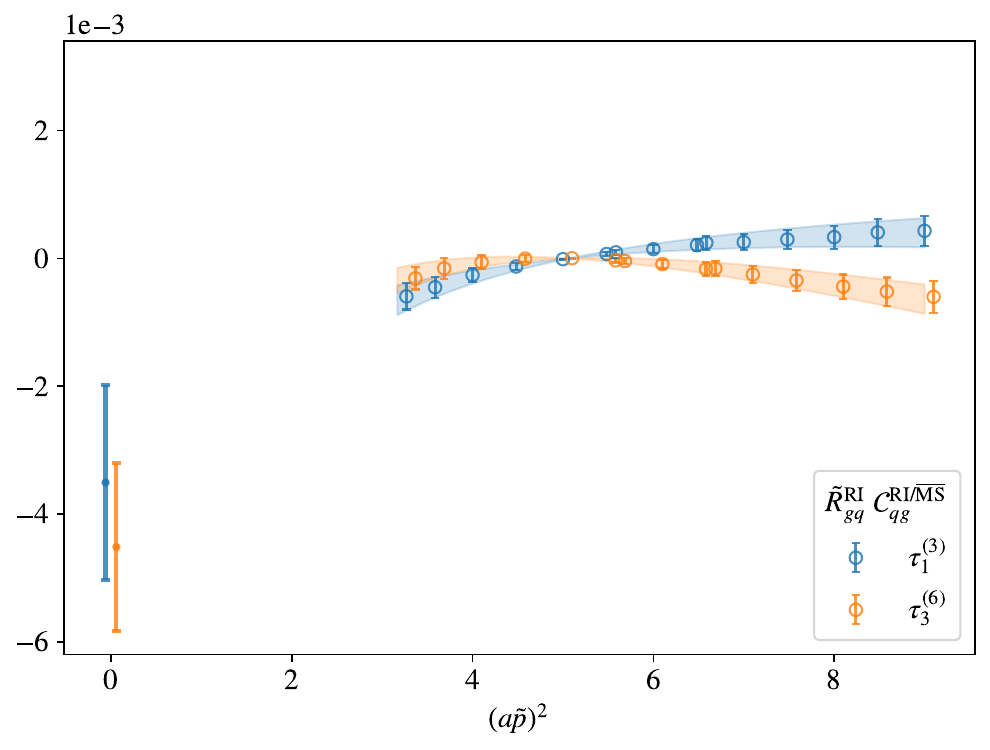}}}
\caption{The computed $\tilde{R}_{gq}^{\text{RI}}\mathcal{C}^{\text{RI}/\overline{\text{MS}}}_{qq}$ (a) and $\tilde{R}_{gq}^{\text{RI}}\mathcal{C}^{\text{RI}/\overline{\text{MS}}}_{qg}$ (b), and their corresponding fits, performed as discussed in Sec.~\ref{subsec:Rgq}. The notation is the same as in Fig.~\ref{fig:Rqqallfits}. We note that $R_{gq}^{\text{RI}}\mathcal{C}^{\text{RI}/\overline{\text{MS}}}_{qg}$ (b) does not correspond to the fit band continued to $(a\tilde{p})^2=0$, due to the presence of logarithmic terms in the model ansatz, as described in Secs.~\ref{subsec:Rfacfit} and~\ref{subsec:Rgq}.}
\label{fig:Rgqfits}
\end{figure*}

The renormalization condition for $R_{v\mathcal{R}}^{\text{RI}}$, Eq.~\eqref{eq:RenormCondVec}, depends on the three-point function of the nonsinglet quark EMT, projected to the Euclidean version of the corresponding irrep $\mathcal{R}$ basis of Eqs.~\eqref{eq:irrep3basis} and~\eqref{eq:irrep6basis}, with external $u$ quark fields. Before amputation, this can be expressed for operator $\ell$ as
\begin{equation}
C_{v\mathcal{R}\ell}^q(p^2) = C_{u\mathcal{R}\ell}^q(p^2) +C_{d\mathcal{R}\ell}^q(p^2)-2C_{s\mathcal{R}\ell}^q(p^2) \;,
\end{equation}
where
\begin{equation}
\begin{split}
C^{q}_{u\mathcal{R}\ell}(p^2)\!=\!\sum_{x,y,z}\!e^{ip(x\text{-}z)}\vev{\psi_u\!(x)\overline{\psi}_u\!(y)(\gamma\overleftrightarrow{D})_{\mathcal{R}\ell}(y)\psi_u\!(y)\overline{\psi}_u\!(z)},\\ 
C^{q}_{d\mathcal{R}\ell}(p^2)\!=\!\sum_{x,y,z}\!e^{ip(x\text{-}z)}\vev{\psi_u\!(x)\overline{\psi}_d\!(y)(\gamma\overleftrightarrow{D})_{\mathcal{R}\ell}(y)\psi_d\!(y)\overline{\psi}_u\!(z)},\\
C^{q}_{s\mathcal{R}\ell}(p^2)\!=\!\sum_{x,y,z}\!e^{ip(x\text{-}z)}\vev{\psi_u\!(x)\overline{\psi}_s\!(y)(\gamma\overleftrightarrow{D})_{\mathcal{R}\ell}(y)\psi_s\!(y)\overline{\psi}_u\!(z)},
\end{split}
\end{equation}
and $x,y,z,p$ are four-vectors. The $u$ three-point function results in a connected and a disconnected contribution
\begin{equation}
\begin{split}
C^{q}_{u\mathcal{R}\ell}(p^2) =& \sum_{x,y,z}e^{ip(x\text{-}z)}\text{Tr}[S_l(x,y)(\gamma\overleftrightarrow{D})_{\mathcal{R}\ell}(y)S_l(y,z)] \\
-&\sum_{x,y,z}e^{ip(x\text{-}z)}\text{Tr}[S_l(x,z)]\text{Tr}[(\gamma\overleftrightarrow{D})_{\mathcal{R}\ell}(y)S_l(y,y)] ,
\end{split}
\end{equation}
where $l=u/d$, while the $d$ and $s$ three-point functions are disconnected
\begin{equation}
\begin{split}
C^{q}_{d\mathcal{R}\ell}(p^2)\!=\!-\!\sum_{x,y,z}e^{ip(x\text{-}z)}\text{Tr}[S_l(x,z)]\text{Tr}[(\gamma\overleftrightarrow{D})_{\mathcal{R}\ell}(y)S_l(y,y)] ,\\
C^{q}_{s\mathcal{R}\ell}(p^2)\!=\!-\!\sum_{x,y,z}e^{ip(x\text{-}z)}\text{Tr}[S_l(x,z)]\text{Tr}[(\gamma\overleftrightarrow{D})_{\mathcal{R}\ell}(y)S_s(y,y)] .
\end{split}
\end{equation}
The connected contribution is computed using the sequential source method, inverting through the operator, on $240$ configurations of ensemble B. The light and strange disconnected contributions are computed on $20000$ configurations, using hierarchical probing~\cite{Stathopoulos:2013aci} with $16$ Hadamard vectors and $Z_4$ noise~\cite{doi:10.1080/03610919008812866}. Both connected and disconnected contributions are formed using quark propagators computed on Landau-gauge-fixed configurations, and projected to all momenta with 4-wave vector components $n_{\mu}$ such that $1\leq n_{\mu}\leq 5$. Uncertainties are propagated using bootstrap resampling with $200$ bootstrap ensembles, and are dominated by the disconnected contribution.

\begin{figure*}[t]
\centering
\subfloat[\centering  ]
{{\includegraphics[width=0.48\textwidth]{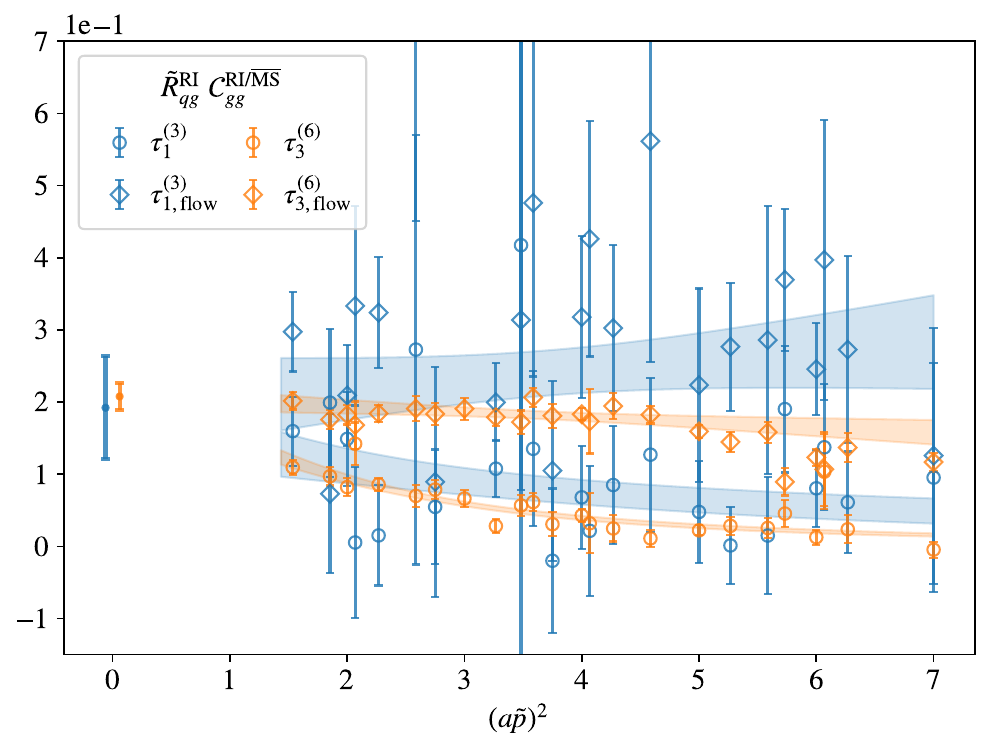}}} \!
\subfloat[\centering  ]
{{\includegraphics[width=0.48\textwidth]{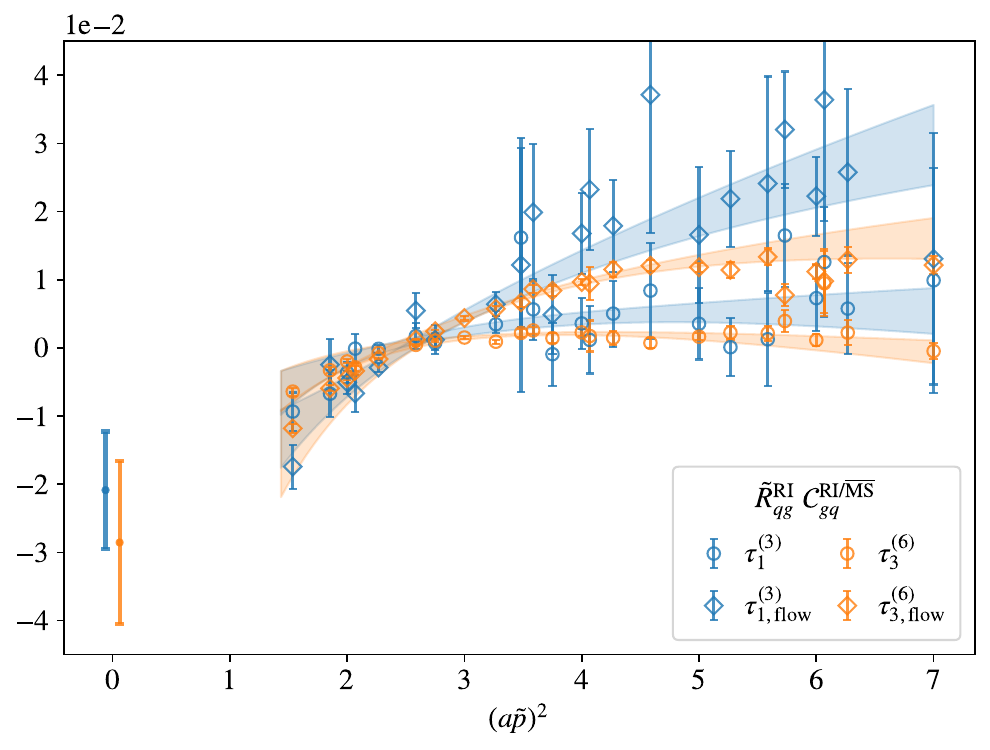}}}
\caption{Data for $\tilde{R}^{\text{RI}}_{qg}\mathcal{C}_{gg}^{\text{RI}/\overline{\text{MS}}}$ (a) and $\tilde{R}^{\text{RI}}_{qg}\mathcal{C}_{gq}^{\text{RI}/\overline{\text{MS}}}$ (b) with unflowed external gluon fields are shown with empty markers, while the flowed ones are shown as empty diamonds. The two sets of data are fit simultaneously as described in Sec.~\ref{subsec:Rqg}, and the extracted values for $R^{\text{RI}}_{qg}\mathcal{C}_{gg}^{\text{RI}/\overline{\text{MS}}}$ and $R^{\text{RI}}_{qg}\mathcal{C}_{gq}^{\text{RI}/\overline{\text{MS}}}$ for each of the two irreps are shown as filled markers at $a^2\tilde{p}^2=0$. We note that $R_{qg}^{\text{RI}}\mathcal{C}^{\text{RI}/\overline{\text{MS}}}_{gq}$ (b) does not correspond to the fit bands continued to $(a\tilde{p})^2=0$, due to the presence of logarithmic terms in the model ansatz, as described in Secs.~\ref{subsec:Rfacfit} and~\ref{subsec:Rqg}.}
\label{fig:Rqgfits}
\end{figure*}

$R_v^{\text{RI}}\mathcal{C}^{\text{RI}/\overline{\text{MS}}}_{v}$ for each hypercubic irrep is extracted by fitting a linear model [$P_1$ in Eq.~\eqref{eq:laurent}] to $\tilde{R}_v^{\text{RI}}\mathcal{C}^{\text{RI}/\overline{\text{MS}}}_{v}$, as shown in Fig.~\ref{subfig:Rvfits}. The results for each operator $\ell$ within each irrep are averaged to improve statistics. Fits are performed to all possible sets of $4$ or more points with  $3\leq a^2\tilde{p}^2\leq 9$, and all fits with $p$-value $>0.01$ are model-averaged based on their AIC weights.

$R^{\text{RI}}_{\mathcal{R}qq}$ is constrained by solving the corresponding renormalization condition of Eq.~\eqref{eq:RenormCond} using the three-point function
\begin{equation}
C_{q\mathcal{R}\ell}^q(p^2) = C_{u\mathcal{R}\ell}^q(p^2) +C_{d\mathcal{R}\ell}^q(p^2)+C_{s\mathcal{R}\ell}^q(p^2) \;,
\end{equation}
which also receives a connected and a disconnected contribution. Its computation proceeds as that of $R_{v}^{\text{RI}}$ discussed above. It contributes to the $\overline{\text{MS}}$ renormalization matrix of Eq.~\eqref{eq:rimom_rcs} in two ways: through $\tilde{R}^{\text{RI}}_{qq}\mathcal{C}^{\text{RI}/\overline{\text{MS}}}_{qq}$ and $\tilde{R}^{\text{RI}}_{qq}\mathcal{C}^{\text{RI}/\overline{\text{MS}}}_{qg}$. The fitting procedure applied to extract $R^{\text{RI}}_{qq}\mathcal{C}^{\text{RI}/\overline{\text{MS}}}_{qq}$ is identical to that of $R_{v}^{\text{RI}}\mathcal{C}^{\text{RI}/\overline{\text{MS}}}_{v}$, and the corresponding results are shown in Fig.~\ref{subfig:RqqZqqfits}. $\tilde{R}^{\text{RI}}_{qq}\mathcal{C}^{\text{RI}/\overline{\text{MS}}}_{qg}$ is modeled using a logarithmic ansatz [$P_2^{\text{log}}$ in Eq.~\eqref{eq:laurent}], as
as discussed in Sec.~\ref{subsec:Rfacfit}. The same fitting and averaging scheme as for $R_v^{\text{RI}}\mathcal{C}^{\text{RI}/\overline{\text{MS}}}_{v}$ and $R_{qq}^{\text{RI}}\mathcal{C}^{\text{RI}/\overline{\text{MS}}}_{qq}$ are used, with the data and fits presented in Fig.~\ref{subfig:RqqZqgfits}.

\subsection{$R_{gq}$}
\label{subsec:Rgq}

We calculate $R^{\text{RI}}_{gq\mathcal{R}}$ on $20,000$ configurations of ensemble B by forming vacuum-subtracted three-point functions of the gluon EMT projected to irrep $\mathcal{R}$ with external quark propagators. The operator is constructed using gluon link fields flowed to $t_{\text{flow}}/a^2 = 1.2$ to match the physical scale of the flow radius used for the bare gluon operator on the finer ensemble A of Table~\ref{tab:ensemble}, as detailed in Sec.~\ref{subsec:gluonbare}. The quark propagator calculations are as in Sec.~\ref{subsec:Rv}. When solving the renormalization conditions of Eq.~\eqref{eq:RenormCond} for each irrep, we average all operators $\ell$. $R^{\text{RI}}_{gq}\mathcal{C}_{qq}^{\text{RI}/\overline{\text{MS}}}$ is modeled by a linear fit [$P_1$ in Eq.~\eqref{eq:laurent}], as shown in the left panel of Fig.~\ref{fig:Rgqfits}.  $R^{\text{RI}}_{gq}\mathcal{C}_{qg}^{\text{RI}/\overline{\text{MS}}}$, shown in the right panel, is modeled by a logarithmic ansatz [$P^{\text{log}}_2$ in Eq.~\eqref{eq:laurent}], as discussed in Sec.~\ref{subsec:Rfacfit}. The fitting ranges and error propagation via bootstrapping and model averaging are the same as for the fits described in Sec.~\ref{subsec:Rv}.

\begin{figure*}[t]
\centering
\subfloat[\centering  ]
{{\includegraphics[width=0.48\textwidth]{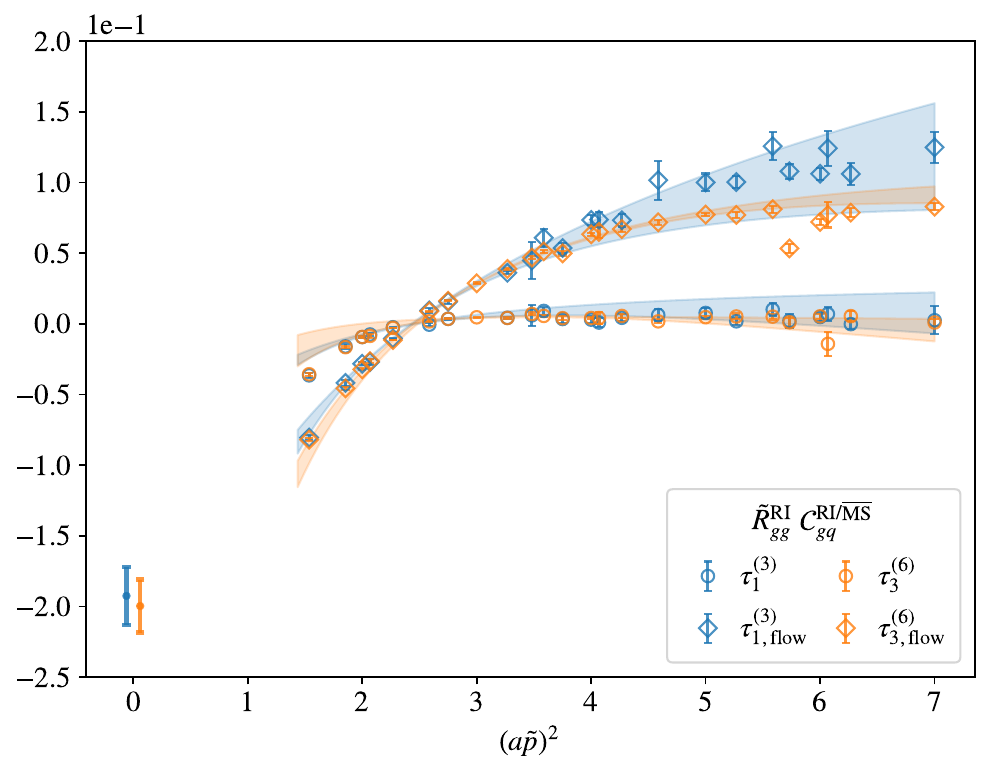}\label{fig:Rggfitsa}}} \!
\subfloat[\centering  ]
{\label{fig:Rggfitsb}{\includegraphics[width=0.48\textwidth]{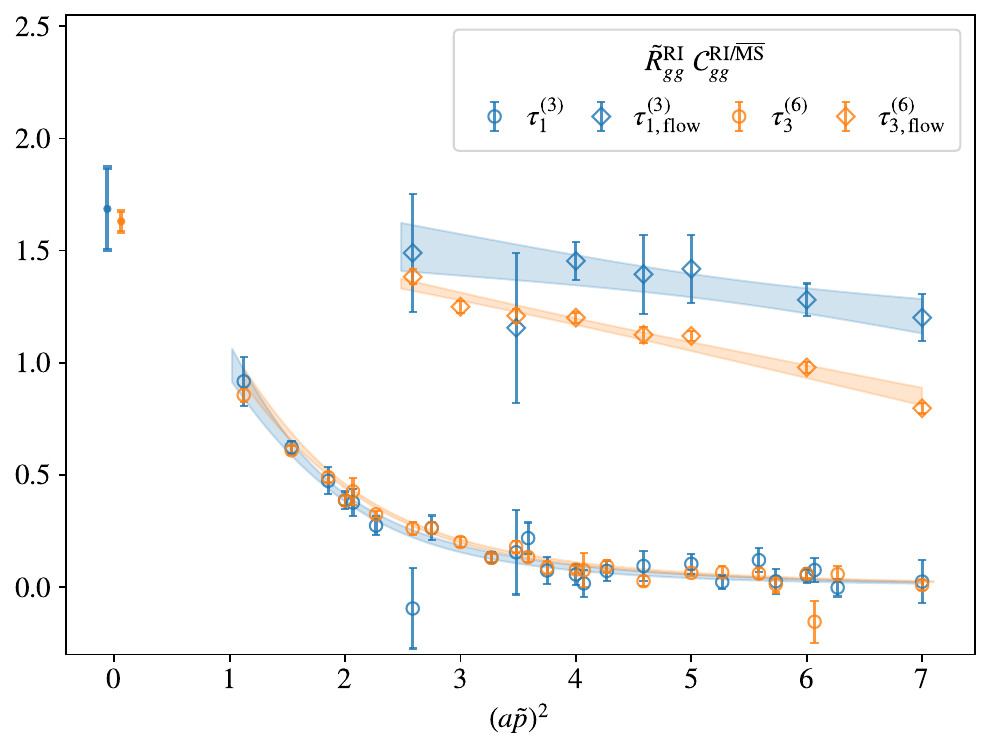}}} 
\caption{Data for $\tilde{R}_{gg}^{\text{RI}}\mathcal{C}_{gq}^{\text{RI}/\overline{\text{MS}}}$ (a) and $\tilde{R}_{gg}^{\text{RI}}\mathcal{C}_{gg}^{\text{RI}/\overline{\text{MS}}}$ (b), and corresponding fits performed as described in Sec.~\ref{subsec:Rgg}. The notation is the same as in Fig.~\ref{fig:Rqgfits}. We note that $R_{gg}^{\text{RI}}\mathcal{C}^{\text{RI}/\overline{\text{MS}}}_{gq}$ (a) does not correspond to the fit bands continued to $(a\tilde{p})^2=0$, due to the presence of logarithmic terms in the model ansatz, as described in Secs.~\ref{subsec:Rfacfit} and~\ref{subsec:Rgg}.}
\label{fig:Rggfits}
\end{figure*}

\subsection{$R_{qg}$}
\label{subsec:Rqg}

$R^{\text{RI}}_{qg}$ is calculated on $20,000$ configurations, using three-point functions formed by the disconnected singlet quark operator computed as described in Sec.~\ref{subsec:Rv}. The gluon propagators are formed as defined in Eq.~\eqref{eq:gluonprop} for the diagonal indices $\alpha=\beta$ using gluon fields computed from Eq.~\eqref{eq:gluonfields}, and are projected to all four-momenta $p=2\pi n/L$ that are subject to the constraints $\sum_{\mu} n_{\mu}^2 \leq 36$, $n_{\tau} = 0$, and dem $\leq 0.5$. We form $R^{\text{RI}}_{qg}$ in two ways, using gluon propagators computed on gluon link fields that are either flowed (to $t_{\text{flow}}/a^2 = 1.2$) or unflowed before gauge fixing. The two approaches are expected to have different discretization artifacts, but to give consistent results once the  $(a\tilde{p})^2$ dependence has been fit~\cite{Shanahan:2018pib,Alexandrou:2020sml}. In both cases, within each irrep we average all operators and gluon field polarizations $\alpha$ that yield the same $R^{\text{RI}}_{qg}(a^2\tilde{p}^2)$ as defined in the renormalization condition of Eq.~\eqref{eq:RenormCond}. 

$\tilde{R}^{\text{RI}}_{qg}\mathcal{C}_{gg}^{\text{RI}/\overline{\text{MS}}}$ formed with unflowed external gluon fields is modeled by $P^{\text{inv}}_2$ of Eq.~\eqref{eq:laurent}, as discussed in Sec.~\ref{subsec:Rfacfit}. For $\tilde{R}^{\text{RI}}_{qg}\mathcal{C}_{gg}^{\text{RI}/\overline{\text{MS}}}$ formed with flowed external gluon fields, we find that the discretization effects become milder with increasing flow time, and are well-described by a linear fit [$P_1$ in Eq.~\eqref{eq:laurent}] when $t_{\text{flow}}/a^2=1.2$. The fitting of both the flowed and unflowed versions yield consistent results for  $R^{\text{RI}}_{qg}\mathcal{C}_{gg}^{\text{RI}/\overline{\text{MS}}}$, and the final result is obtained by combined fits to both, sharing only the constant parameter. Fits are performed to all $\tilde{p}$ with $\tilde{p}_{\text{min}} \leq \tilde{p}\leq\tilde{p}_{\text{max}}$, varying the boundaries to all points between $1.5 \leq a^2\tilde{p}_{\text{min}}^2 \leq 3.5$ and $a^2\tilde{p}_{\text{max}}^2 \leq 7$. 

$\tilde{R}^{\text{RI}}_{qg}\mathcal{C}_{gq}^{\text{RI}/\overline{\text{MS}}}$ with unflowed external gluon fields similarly demonstrates beyond-linear discretization effects along with logarithmic dependence, and is found to admit high-quality fits when the ansatz $P_2^{\text{log}} + P_1^{\text{inv}}$ of Eq.~\eqref{eq:laurent} is used, as discussed in Sec.~\ref{subsec:Rfacfit}. The equivalent quantity formed with external gluon fields flowed to $t_{\text{flow}}/a^2=1.2$ demonstrates primarily logarithmic dependence, and only  $P_2^{\text{log}}$ is included in the model, as discussed in Sec.~\ref{subsec:Rfacfit}. We extract $R^{\text{RI}}_{qg}\mathcal{C}_{gq}^{\text{RI}/\overline{\text{MS}}}$ from model averages over combined fits to the flowed and unflowed $\tilde{R}^{\text{RI}}_{qg}\mathcal{C}_{gq}^{\text{RI}/\overline{\text{MS}}}$ using the same fit ranges as for $R^{\text{RI}}_{qg}\mathcal{C}_{gg}^{\text{RI}/\overline{\text{MS}}}$. The data and averaged fits are shown in Fig.~\ref{fig:Rqgfits}.

\subsection{$R_{gg}$}
\label{subsec:Rgg}

We calculate $R^{\text{RI}}_{gg}$ on $20,000$ configurations by forming vacuum subtracted three-point functions of the flowed gluon operator with external gluon propagators. The computation of the gluon operator is as described in Sec.~\ref{subsec:Rgq} and of the gluon propagator as in Sec.~\ref{subsec:Rgg}. As in Sec.~\ref{subsec:Rqg}, we form $R^{\text{RI}}_{gg}$ using external gluon fields that are either unflowed or flowed to $t_{\text{flow}}/a^2=1.2$. In both cases, we average within each irrep all operators and gluon field polarizations $\alpha$ that yield the same $R^{\text{RI}}_{gg}(a^2\tilde{p}^2)$ as defined in the renormalization condition of Eq.~\eqref{eq:RenormCond}.

Similarly to the case of $\tilde{R}_{qg}^{\text{RI}}\mathcal{C}_{gq}^{\text{RI}/\overline{\text{MS}}}$ described in Sec.~\ref{subsec:Rqg}, the flowed version of $\tilde{R}_{gg}^{\text{RI}}\mathcal{C}_{gq}^{\text{RI}/\overline{\text{MS}}}$ demonstrates primarily logarithmic dependence and is fit using $P_2^{\text{log}}$ of Eq.~\eqref{eq:laurent}, while the unflowed version has strong discretization effects and is modeled using $P_2^{\text{log}}+P_4^{\text{inv}}$, as discussed in Sec.~\ref{subsec:Rfacfit}. $R_{gg}^{\text{RI}}\mathcal{C}_{gq}^{\text{RI}/\overline{\text{MS}}}$ is obtained by combined fits to both versions, with the same momentum democracy cut and fit ranges as in Sec.~\ref{subsec:Rqg}. The results are shown in Fig.~\ref{fig:Rggfitsa}.

As discussed in Sec.~\ref{subsec:Rfacfit}, $\tilde{R}_{gg}^{\text{RI}}\mathcal{C}_{gg}^{\text{RI}/\overline{\text{MS}}}$ with unflowed gluon fields as the external states is fit using $P_3^{\text{inv}}$, using the same democracy cut as the results above, but with $1<a^2\tilde{p}^2_{\text{min}}<2$. Restricting the fit to start at small momenta is found to be necessary due to the strongly decaying behavior. The flowed version of $\tilde{R}_{gg}^{\text{RI}}\mathcal{C}_{gg}^{\text{RI}/\overline{\text{MS}}}$ is well-described by a linear model [$P_1$ of Eq.~\eqref{eq:laurent}] when restricted to the dataset with momenta of dem $\leq 0.5$ and $a^2\tilde{p}^2_{\text{min}}>2$. $R_{gg}^{\text{RI}}\mathcal{C}_{gg}^{\text{RI}/\overline{\text{MS}}}$ is extracted from a combined fit to both versions, with the results shown in Fig.~\ref{fig:Rggfitsb}.

\subsection{Renormalization factors}
\label{subsec:renormfactors}

The final results for the isosinglet and gluon inverse $\overline{\text{MS}}$ renormalization matrices of the two irreps are
\begin{equation} \label{eq:renormmatrixnumbers}
\begin{aligned}
\tau_1^{(3)}:\;\; && R_{qq}^{\overline{\text{MS}}} &= \phantom{-}1.056(27), \; & R_{qg}^{\overline{\text{MS}}} &=0.067(71) \;,\\
&& R_{gq}^{\overline{\text{MS}}} &= -0.169(22), \; & R_{gg}^{\overline{\text{MS}}} &=1.68(18) \;,\\
\tau_3^{(6)}:\;\; && R_{qq}^{\overline{\text{MS}}} &= \phantom{-}1.039(28), \; & R_{qg}^{\overline{\text{MS}}} &=0.081(19) \;,\\
&& R_{gq}^{\overline{\text{MS}}} &= -0.180(23), \; & R_{gg}^{\overline{\text{MS}}} &=1.625(48) \;.
\end{aligned}
\end{equation}
We note that the large values for $R_{gg}^{\overline{\text{MS}}}$, which contribute to small values around $0.6$ for the corresponding component of the inverse matrix $Z_{gg}^{\overline{\text{MS}}}$ , are due to the flowing of the gluon operator; the bare gluon GFFs also increase with increasing flow time, which is compensated by the decreasing renormalization coefficient.

The results for the nonsinglet $\overline{\text{MS}}$ renormalization coefficients are
\begin{equation} \label{eq:renormvectornumbers}
\begin{split}
\tau_1^{(3)}\;\;:\;\; &
R_v^{\overline{\text{MS}}} = 1.067(29) \;,\\
\tau_3^{(6)}\;\;:\;\; &
R_v^{\overline{\text{MS}}} = 1.066(21) \;.
\end{split}
\end{equation}

\section{RENORMALIZED RESULTS}
\label{sec:renormGFFs}

The procedure described in Sec.~\ref{sec:barematel} yields a set of measurements which constrain the bare 2-dimensional vector of GFFs $\vec{G}^{\pi,B}_{i\mathcal{R} t}$ for each irrep $\mathcal{R} \in \{ \tau_1^{(3)}, \tau_3^{(6)} \}$ and flavor $i\in\{q,g,v\}$ separately as
\begin{equation} \label{eq:bareGFFfit}
    \mathbb{K}_{\mathcal{R}t} \vec{G}^{\pi,B}_{i\mathcal{R}t} = \text{\bf{ME}}_{i\mathcal{R}t} \;,
\end{equation}
where $\text{\bf{ME}}_{i\mathcal{R}t}$ is the $c$-dimensional vector of matrix element fits extracted from the lattice data using the summation method, and $\mathbb{K}_t$ is the kinematic coefficient matrix written as a ($c\times 2$)-dimensional matrix in the space of $c$-bins and pion GFFs.

For $i\in\{q,g\}$, the bare GFFs are defined in terms of the renormalized GFFs, $\vec{G}^{\pi}_{it}$, as
\begin{equation} \label{eq:baretorenorm}
\vec{G}^{\pi,B}_{i\mathcal{R}t} = \sum_{j\in\{q,g\}} R^{\overline{\text{MS}}}_{ij\mathcal{R}}\vec{G}^{\pi}_{jt}  \;,
\end{equation}
where $R^{\overline{\text{MS}}}_{ij\mathcal{R}}$ are the components of the renormalization matrix with values listed in Eq.~\eqref{eq:renormmatrixnumbers}.
Since both irreps are expected to yield the same $\vec{G}^{\pi}_{it}$ after renormalization, they can be fit simultaneously, with the individual irrep renormalization coefficients used as inputs in the fit. We therefore recast Eq.~\eqref{eq:bareGFFfit} and~\eqref{eq:baretorenorm} into a single linear system for both irreps
\begin{equation} \label{eq:firstmatrixcombined}
\mathbb{RK}_t^{\overline{\text{MS}}} \vec{G}_t^{\pi} = \vec{ME}_t \;,
\end{equation}
where
\begin{equation} \label{eq:stackedME}
\vec{ME}_t = \begin{pmatrix}
\vec{ME}_{q\tau_1^{(3)}t} \\
\vec{ME}_{q\tau_3^{(6)}t} \\
\vec{ME}_{g\tau_1^{(3)}t} \\
\vec{ME}_{g\tau_3^{(6)}t}
\end{pmatrix} \;,
\end{equation}
\begin{equation}
\mathbb{RK}_t^{\overline{\text{MS}}} =
\begingroup
\renewcommand*{\arraystretch}{1.5}
\begin{pmatrix}
R_{qq\tau_1^{(3)}}^{\overline{\text{MS}}}\mathbb{K}_{\tau_1^{(3)}t} &
R_{qg\tau_1^{(3)}}^{\overline{\text{MS}}}\mathbb{K}_{\tau_1^{(3)}t} \\
R_{qq\tau_3^{(6)}}^{\overline{\text{MS}}}\mathbb{K}_{\tau_3^{(6)}t} &
R_{qg\tau_3^{(6)}}^{\overline{\text{MS}}}\mathbb{K}_{\tau_3^{(6)}t} \\
R_{gq\tau_1^{(3)}}^{\overline{\text{MS}}}\mathbb{K}_{\tau_1^{(3)}t} &
R_{gg\tau_1^{(3)}}^{\overline{\text{MS}}}\mathbb{K}_{\tau_1^{(3)}t} \\
R_{gq\tau_3^{(6)}}^{\overline{\text{MS}}}\mathbb{K}_{\tau_3^{(6)}t} &
R_{gg\tau_3^{(6)}}^{\overline{\text{MS}}}\mathbb{K}_{\tau_3^{(6)}t} \;
\end{pmatrix}.
\endgroup
\end{equation}
We fit Eq.~\eqref{eq:firstmatrixcombined} for $\vec{G}^{\pi}_t$ by minimizing 
\begin{equation}
\begin{split}
\chi^2_t &= (\mathbb{RK}_t^{\overline{\text{MS}}} \vec{G}_t^{\pi} - \vec{ME}_t)^T \\ &\quad\times\text{Cov}(\vec{ME}_t)^{-1} (\mathbb{RK}_t^{\overline{\text{MS}}} \vec{G}_t^{\pi} - \vec{ME}_t) \;,
\end{split}
\end{equation}
where $\text{Cov}(\vec{ME}_t)$ is the covariance matrix of the matrix element fits for momentum bin $t$, computed from the bootstrap samples, and $\mathbb{RK}_t^{\overline{\text{MS}}}$ is assumed to be Gaussian with the mean and covariance determined by the RI-MOM fitting procedure of Sec.~\ref{sec:renorm}. The solution takes the analytic form
\begin{equation} \label{eq:analyticrenorm}
\begin{split}
\vec{G}^{\pi}_t &= \left[\mathbb{RK}_t^{\overline{\text{MS}},T} \text{Cov}(\vec{ME}_t)^{-1}\mathbb{RK}_t^{\overline{\text{MS}}}\right]^{-1} \\
&\quad\times\mathbb{RK}_t^{\overline{\text{MS}},T}\text{Cov}(\vec{ME}_t)^{-1}\vec{ME}_t\;,
\end{split}
\end{equation} 
and the errors are determined using Gaussian error propagation. This circumvents the d'Agostini bias~\cite{DAgostini:1993arp} by neglecting additional correlations between renormalized constraints induced by common factors of $R^{\overline{\text{MS}}}_{ij\mathcal{R}}$.

For the renormalized nonsinglet GFF for which mixing is not present, we follow a combined irrep fitting procedure that is a Bayesian version of the ``penalty trick''~\cite{DAgostini:1993arp}, and was introduced in Ref.~\cite{Pefkou:2021fni} for the renormalization of gluon GFFs when mixing with quarks is assumed to be negligible. For a detailed discussion of this procedure, see Appendix~6 of Ref.~\cite{Pefkou:2021fni}.

\begin{figure*}
\centering
\subfloat
{{\includegraphics[width=0.48\textwidth]{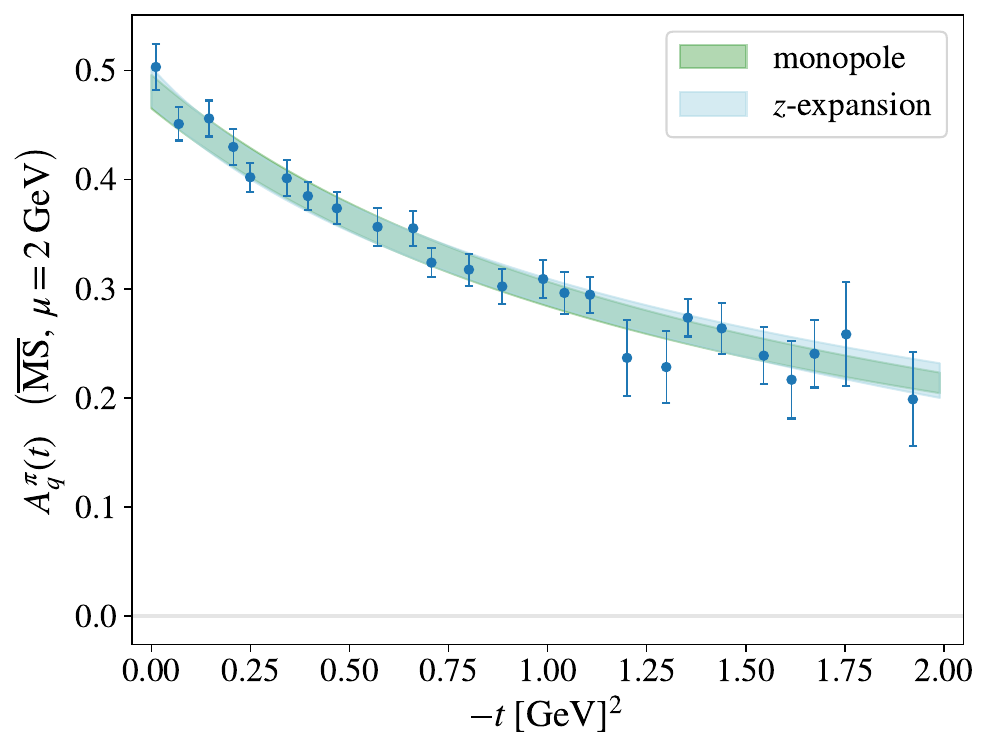} }}
\!
\subfloat
{{\includegraphics[width=0.48\textwidth]{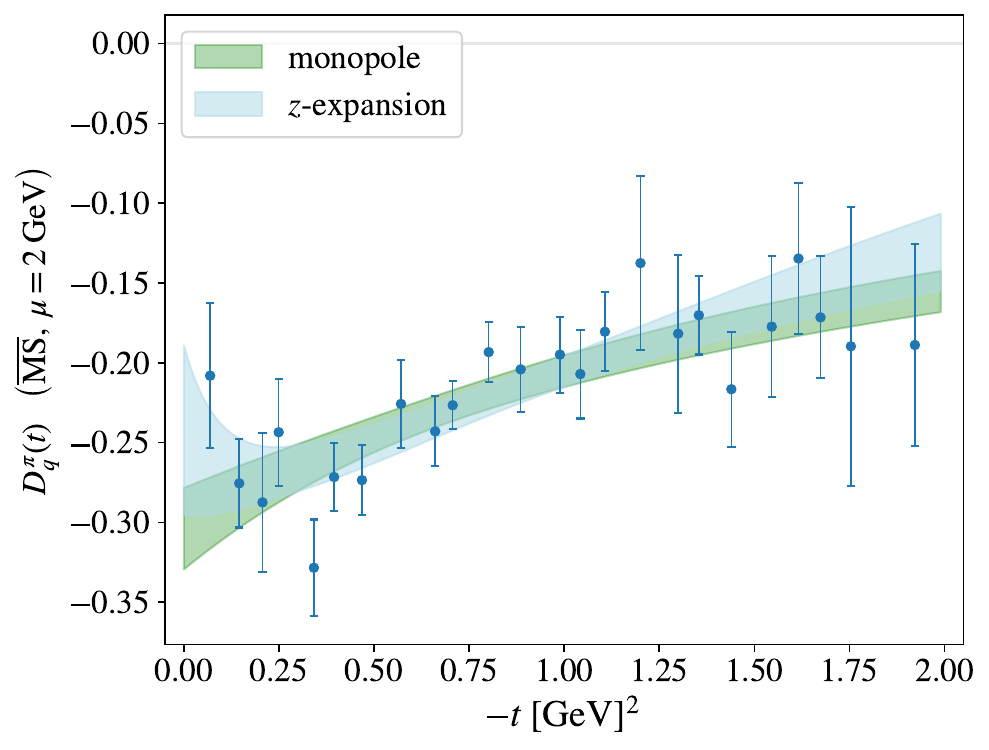} }} \\
\subfloat
{{\includegraphics[width=0.48\textwidth]{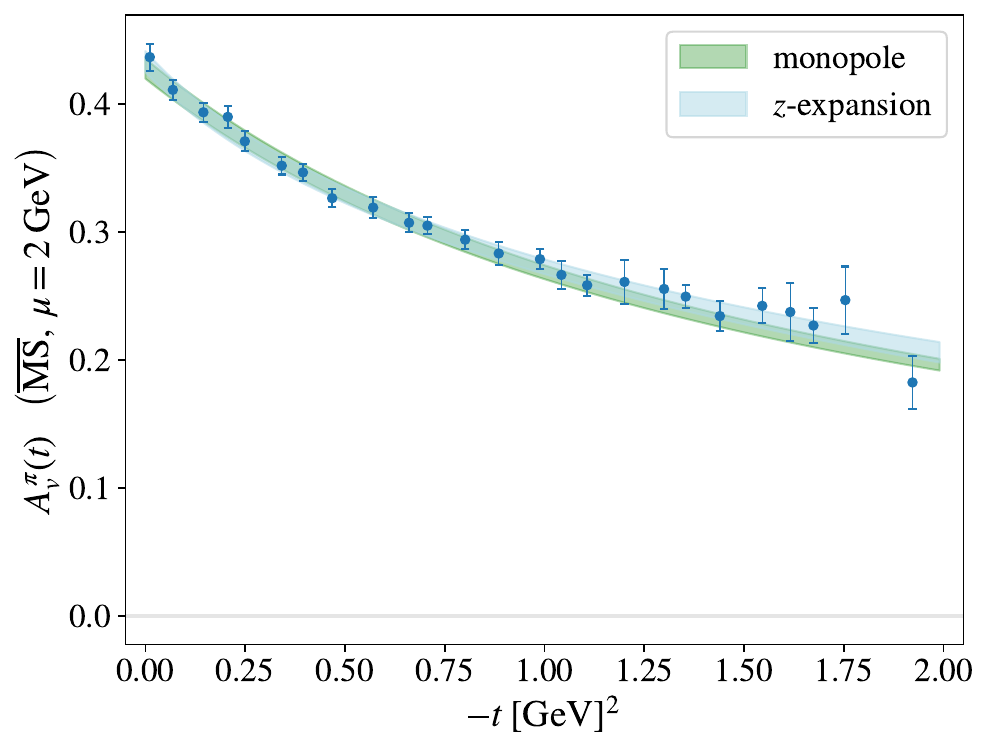} }}
\!
\subfloat
{{\includegraphics[width=0.48\textwidth]{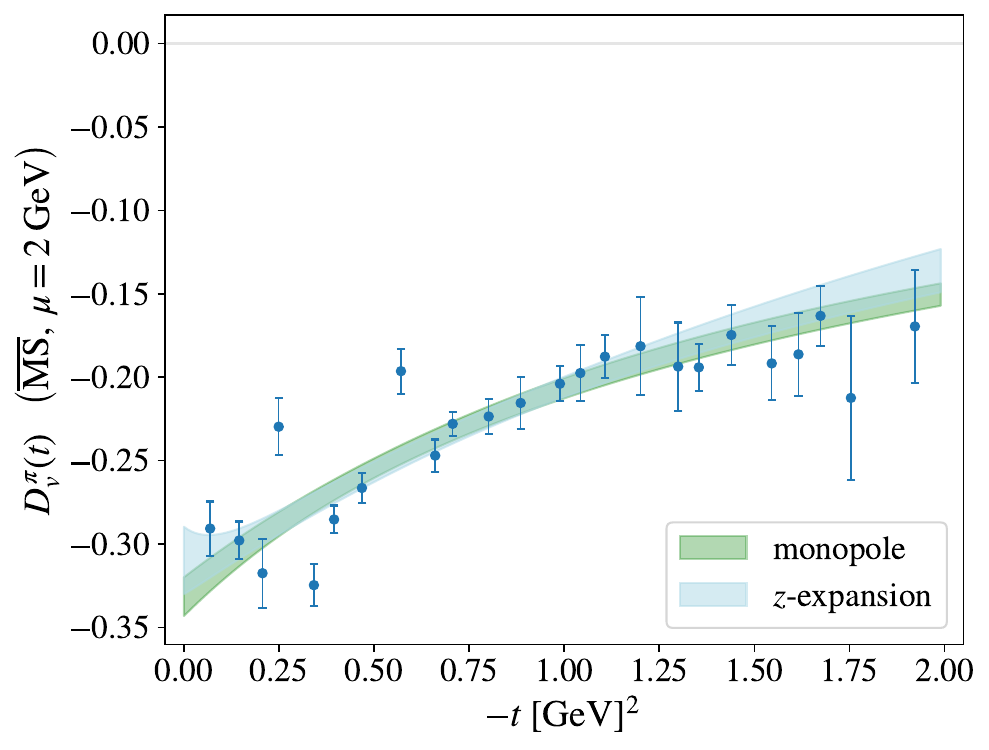} }} \\
{{\includegraphics[width=0.48\textwidth]{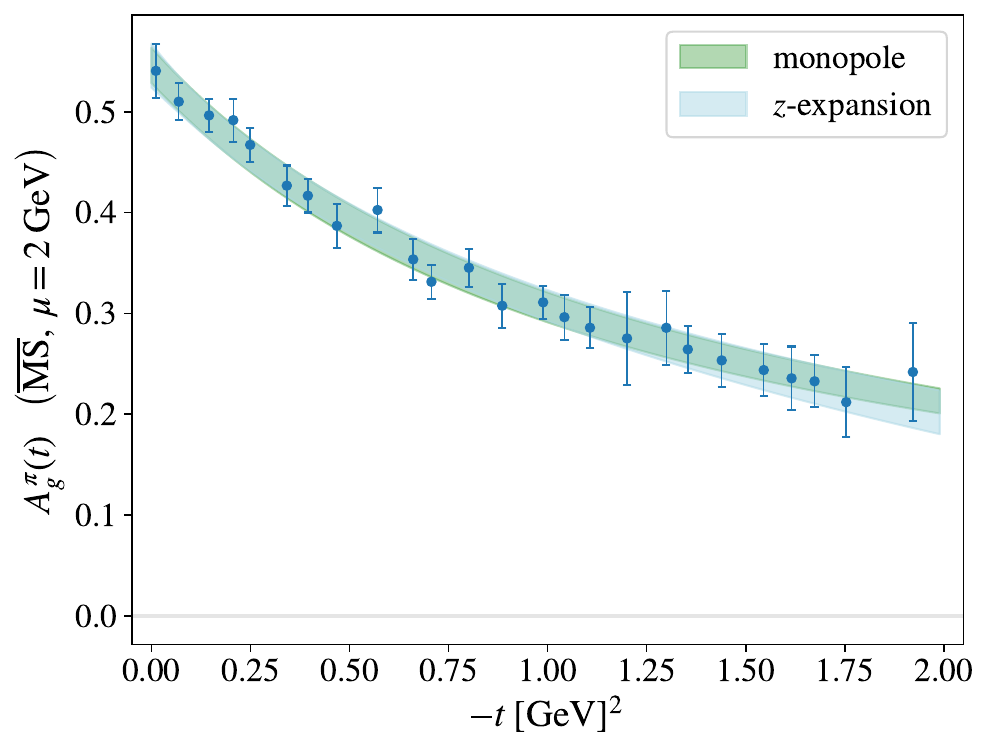} }}
\!
\subfloat
{{\includegraphics[width=0.48\textwidth]{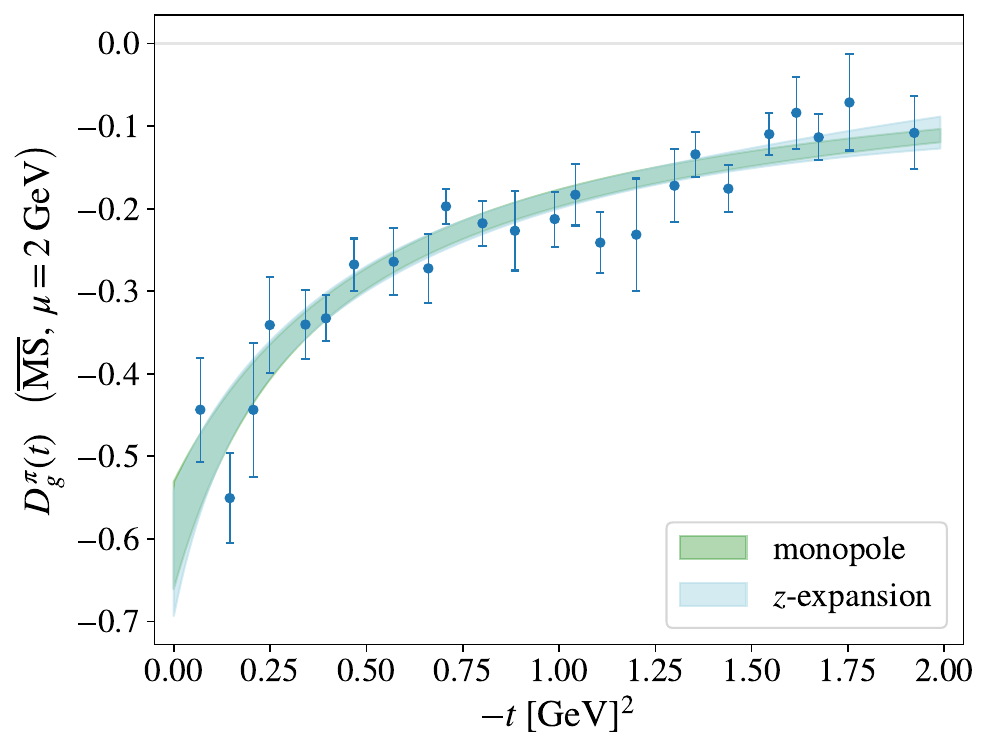} }}
\caption{The isosinglet (top), nonsinglet (center), and gluon (bottom) GFFs of the pion renormalized in the $\overline{\text{MS}}$ scheme at scale $\mu=2~\text{GeV}$. The $A^{\pi}_i(t)$ form factors are shown on the left and the $D^{\pi}_i(t)$ on the right. Fits using the monopole model of Eq.~\eqref{eq:multipole} and the $z$-expansion of Eq.~\eqref{eq:z-expansion} are shown, with fit parameters collected in Tables~\ref{tab:170pionparamsA} and~\ref{tab:170pionparamsD}.}
\label{fig:isoGFF}
\end{figure*}
\begin{figure*}[t]
\centering
\subfloat
{{\includegraphics[width=0.48\textwidth]{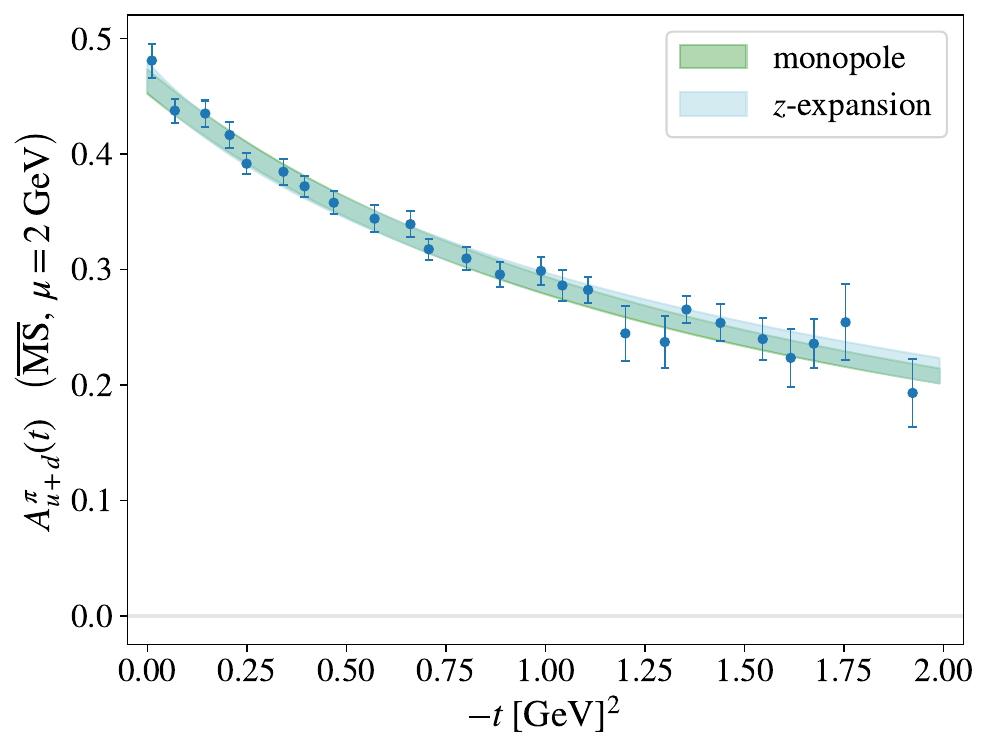} }}
\!
\subfloat
{{\includegraphics[width=0.48\textwidth]{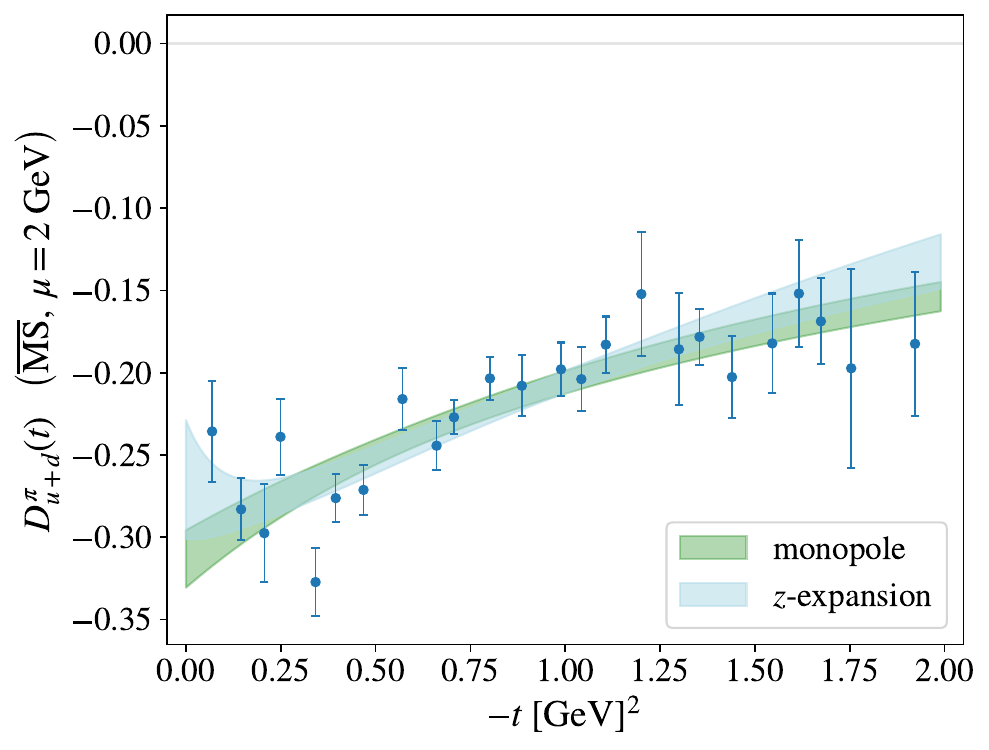} }} \\[-3ex]
\subfloat
{{\includegraphics[width=0.48\textwidth]{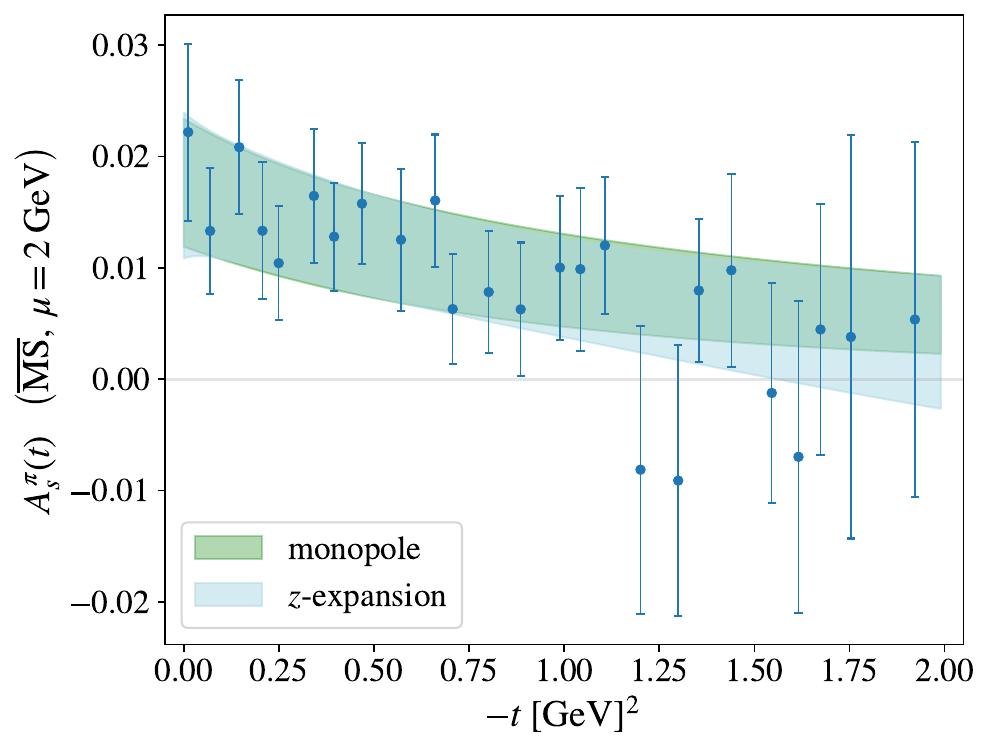} }}
\!
\subfloat
{{\includegraphics[width=0.48\textwidth]{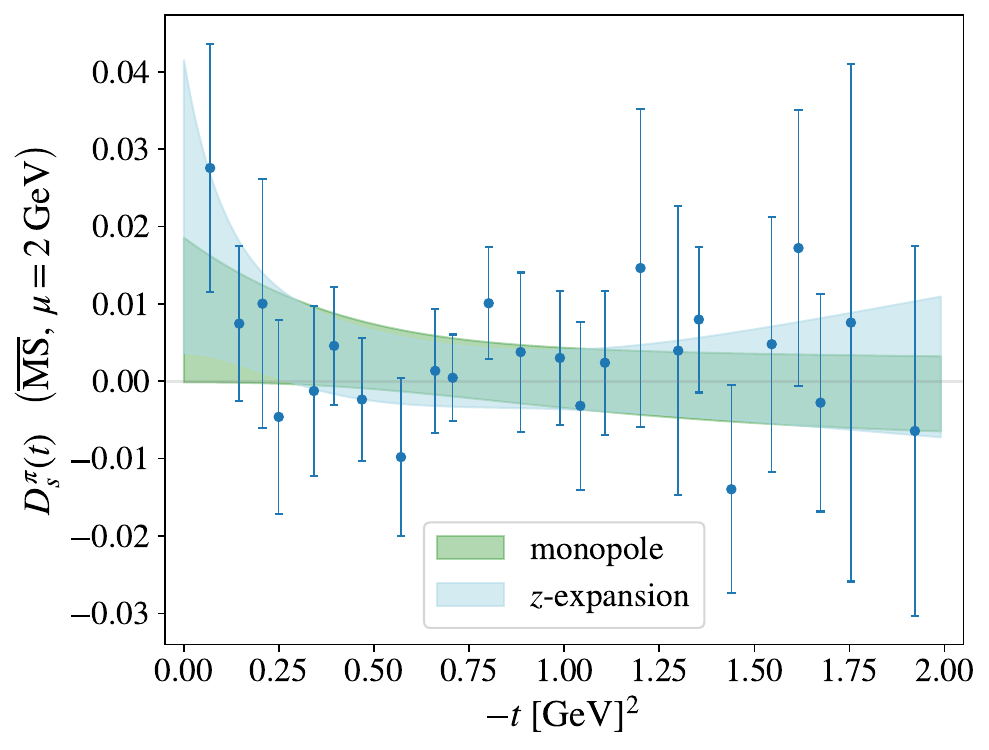} }}\\ [-0.5ex]
\caption{The light quark $u+d$ (top) and strange (bottom) GFFs of the pion, renormalized in the $\overline{\text{MS}}$ scheme at scale $\mu=2~\text{GeV}$. The notation is as in Fig.~\ref{fig:isoGFF}.}
\label{fig:flavorGFF}
\end{figure*}

\subsection{Pion gravitational form factors}
\label{subsec:renormGFFresults}

\begin{figure*}
\centering
\subfloat
{{\includegraphics[width=0.48\textwidth,keepaspectratio]{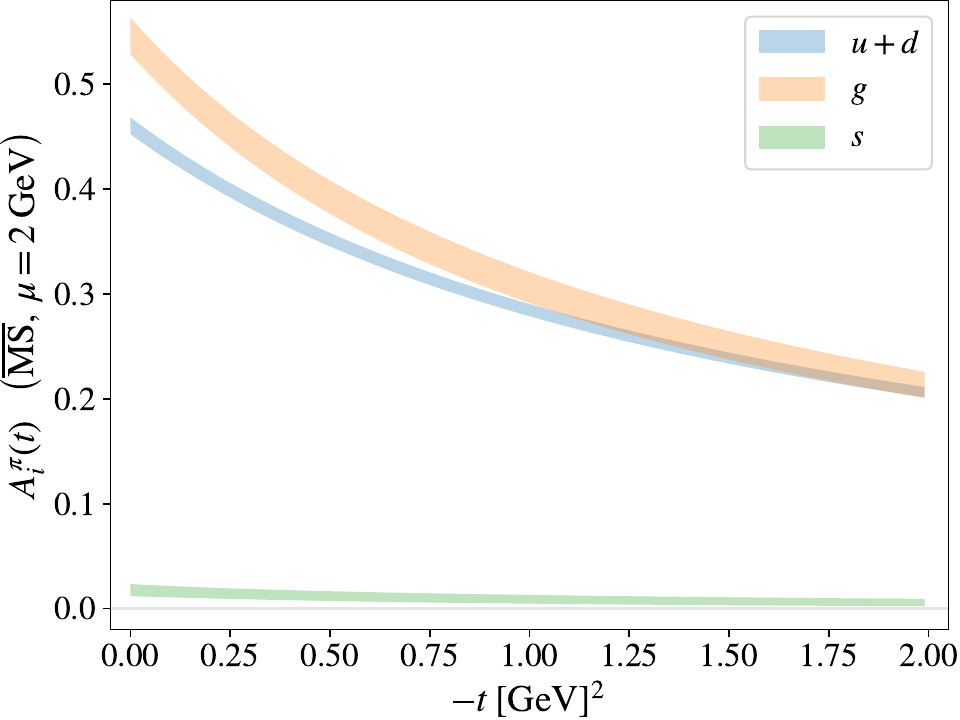} }}
\!
\subfloat
{{\includegraphics[width=0.48\textwidth,keepaspectratio]{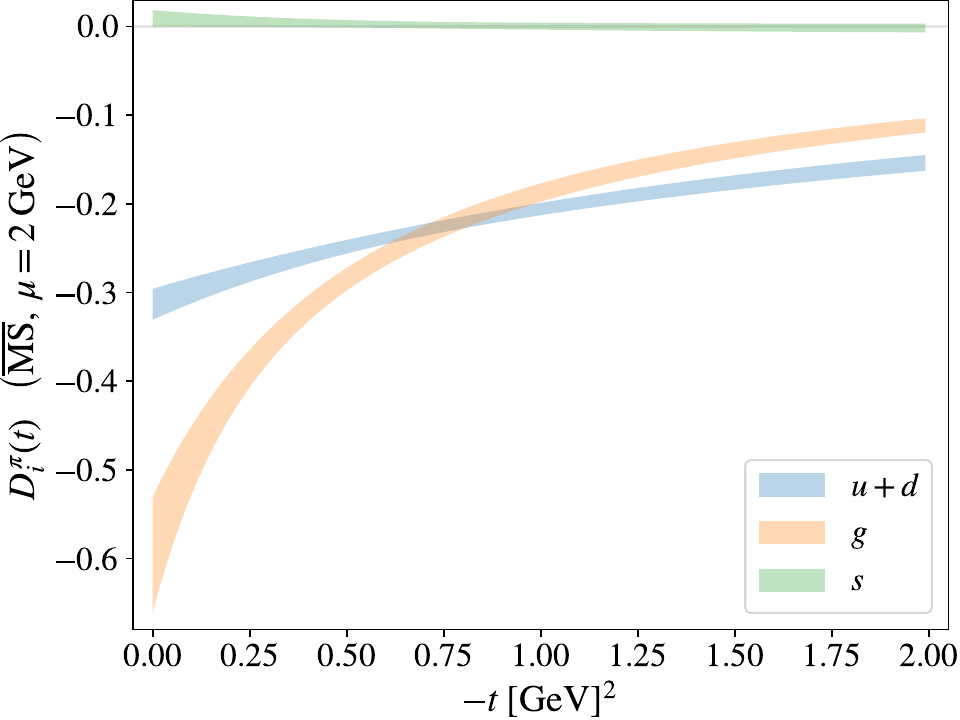} }} 
\caption{Pion GFF flavor decomposition using the monopole model of Eq.~\eqref{eq:multipole}, with fit parameters shown in Tables~\ref{tab:170pionparamsA} and~\ref{tab:170pionparamsD}.}
\label{fig:fitcompare}
\end{figure*}

\begin{figure*}[t]
\centering
\subfloat
{{\includegraphics[width=0.48\textwidth]{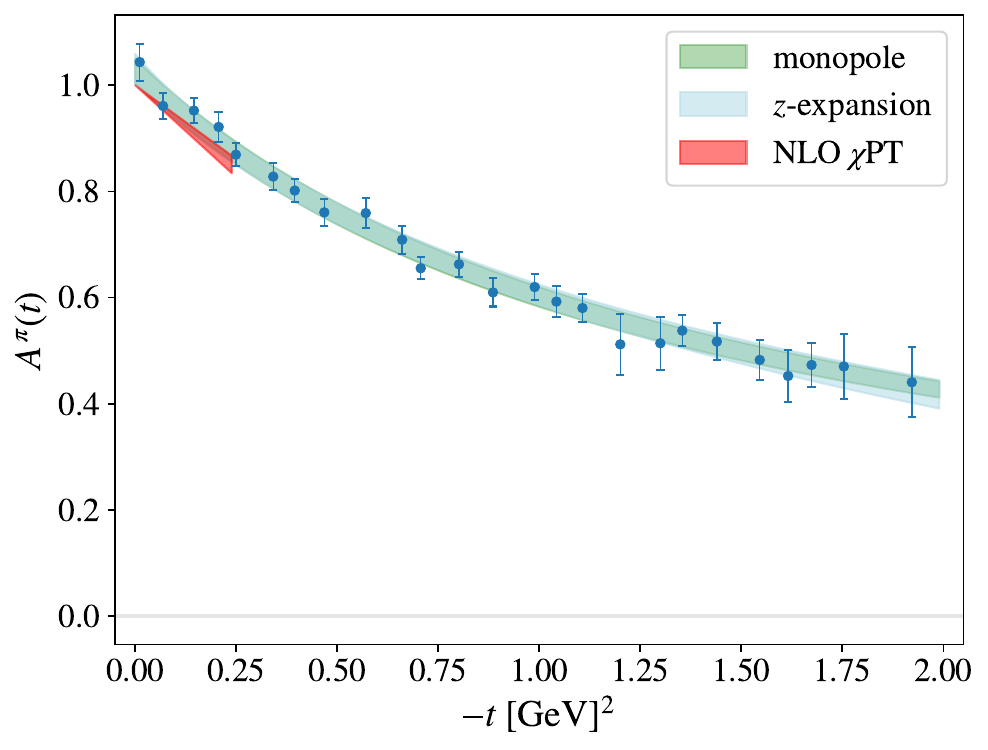} }}
\!
\subfloat
{{\includegraphics[width=0.48\textwidth]{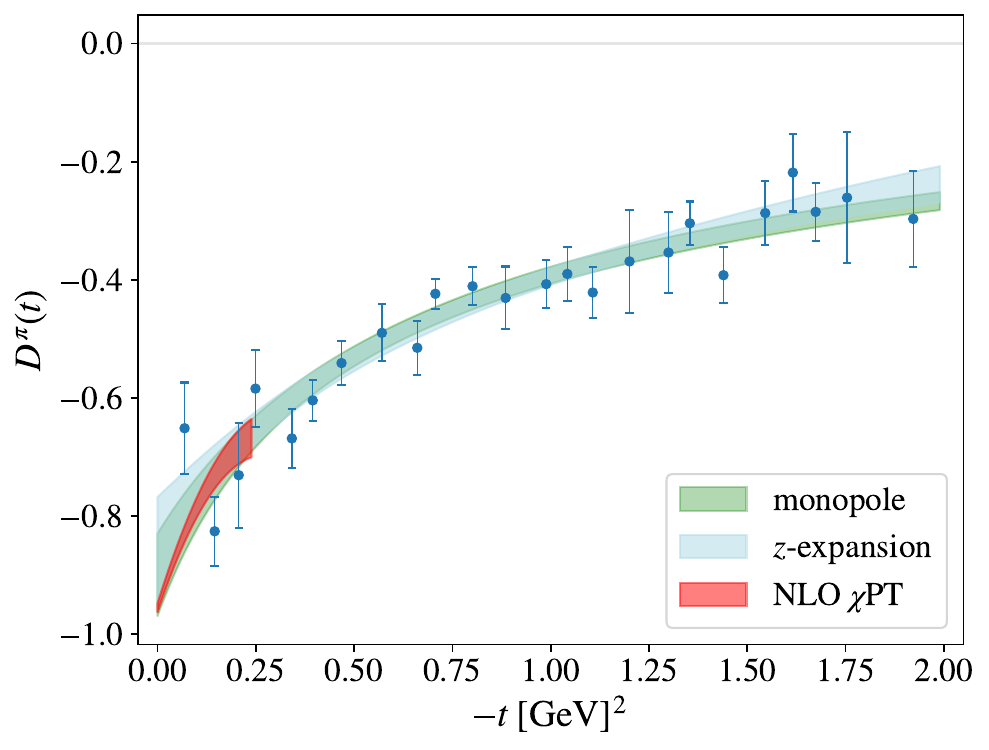} }} 
\caption{The scale- and scheme-independent total GFFs of the pion, obtained by summing  the gluon and quark contributions. The red bands show the next-to-leading order (NLO) $\chi$PT prediction for the low $|t|$ region, using a range of estimates for the low-energy constants, as presented in Ref.~\cite{Donoghue:1991qv}.}
\label{fig:totalGFF}
\end{figure*}

The results for the combined-irrep isosinglet, nonsinglet, and gluon pion GFFs renormalized in the $\overline{\text{MS}}$ scheme at $\mu=2~\text{GeV}$ are shown in Fig.~\ref{fig:isoGFF}, along with correlated fits to the data points using the $n$-pole model
\begin{equation}\label{eq:multipole}
\text{F}_n(t) = \frac{\alpha}{(1-t/\Lambda^2)^n}\;,
\end{equation}
where $\alpha$ and $\Lambda$ are free parameters. We set $n=1$ (monopole), which we find to be the choice that yields the highest $p$-value for all GFF fits. We test model dependence by also using the $z$-expansion~\cite{Hill:2010yb}
\begin{equation} \label{eq:z-expansion}
\text{F}_{z}(t) = \sum_{k=0}^{k_{\text{max}}}
\alpha_k[z(t)]^k \;,
\end{equation}
where
\begin{equation}
z(t) = \frac{\sqrt{t_{\text{cut}}-t}-\sqrt{t_{\text{cut}}-t_0}}{\sqrt{t_{\text{cut}}
-t}+\sqrt{t_{\text{cut}}-t_0}}\;,
\end{equation}
\begin{equation}
t_0 = t_{\text{cut}}\left(1-\sqrt{1+(2~\text{GeV})^2/t_{\text{cut}}}\right) \;,
\end{equation}
$t_{\text{cut}}=4m_{\pi}^2$, $\alpha_k$ are free parameters, and we set $k_{\text{max}}=2$. The nonsinglet GFFs are more precise than the isosinglet ones, due to the cancellation of correlated noise between the light-quark and strange disconnected contributions, and due to the fact that they do not mix with the less precise gluon contribution. The analogous results for each quark flavor, i.e., the light-quark ($u+d$) and strange ($s$) contributions, are presented in Fig.~\ref{fig:flavorGFF}, while Fig.~\ref{fig:fitcompare} includes a comparison of the gluon and different flavor quark monopole fits. The gluon GFFs are fit using the monopole and $z$-expansion models, while the light quark and strange GFFs and their fits are determined by taking linear combinations of the isosinglet and non-singlet components as
\begin{equation}
\begin{split}
\vec{G}^{\pi}_{u+d}(t) &= \frac{2}{3}\vec{G}^{\pi}_q(t)+\frac{1}{3}\vec{G}^{\pi}_v(t) \;,\\ 
\vec{G}^{\pi}_{s}(t) &= \frac{1}{3}\vec{G}^{\pi}_q(t)-\frac{1}{3}\vec{G}^{\pi}_v(t) \;.
\end{split}
\end{equation}
The parameters of the monopole and $z$-expansion fits are shown in Table~\ref{tab:170pionparamsA} for $A_i^{\pi}(t)$ and Table~\ref{tab:170pionparamsD} for $D_i^{\pi}(t)$. In Fig.~\ref{fig:totalGFF}, we present the renormalization scheme and scale independent total GFFs $A^{\pi}(t)$ and $D^{\pi}(t)$, obtained by sums of the corresponding isosinglet and gluon GFFs.  

While the joint fits discussed above have good fit quality, when alternatively solving for the flavor-singlet GFFs using the two irreps individually instead of performing a combined-irrep fit as shown in this section, we find some tension between the two irreps for the results of $D_q^{\pi}(t)$. The individual-irrep results, along with a discussion of this observation, are included in Appendix~\ref{app:renorm}.

\begin{table}
\begin{center}
\begin{tabular}{SD{:}{}{2.8}D{:}{}{2.8}D{:}{}{2.8}D{:}{}{2.8}}
\toprule
\multicolumn{1}{c}{monopole} & \multicolumn{1}{c}{$\alpha$} &\multicolumn{1}{c}{$\Lambda$~[GeV]} && \multicolumn{1}{c}{$\chi^2/\text{d.o.f.}$} \\ \midrule
{$A^{\pi}_g(t)$} & 0:.546(18) & 1:.129(41)  && 0:.9 \\[2pt]
{$A^{\pi}_q(t)$} & 0:.481(15) &  1:.262(37) && 1:.4 \\[2pt]
{$A^{\pi}_v(t)$} & 0:.4276(78) & 1:.300(22)  && 1:.2 \\[2pt]
\midrule\midrule
\multicolumn{1}{c}{$z$-expansion} & \multicolumn{1}{c}{$\alpha_0$} & \multicolumn{1}{c}{$\alpha_1$} & \multicolumn{1}{c}{$\alpha_2$} & \multicolumn{1}{c}{$\chi^2/$d.o.f.}\\ \midrule
{$A^{\pi}_g(t)$} & 0:.379(15) & -0:.551(31) & -0:.37(11) & 1:.0 \\[2pt]
{$A^{\pi}_q(t)$} & 0:.353(12) & -0:.430(25) & -0:.271(89) & 1:.5 \\[2pt]
{$A^{\pi}_v(t)$} & 0:.3187(60) &  -0:.359(15)  & -0:.205(45)& 1:.0 \\[2pt]
\bottomrule
\end{tabular}
\end{center}
\caption{\label{tab:170pionparamsA}
Fit parameters of the monopole and $z$-expansion parametrizations of the $t$-dependence of the pion GFFs $A_i^{\pi}(t)$ renormalized in the $\overline{\text{MS}}$ scheme at scale $\mu=2~\text{GeV}$.}
\end{table}

\begin{table}
\begin{center}
\begin{tabular}{SD{:}{}{2.8}D{:}{}{2.8}D{:}{}{2.8}D{:}{}{2.8}}
\toprule
\multicolumn{1}{c}{monopole} & \multicolumn{1}{c}{$\alpha$} &\multicolumn{1}{c}{$\Lambda$~[GeV]} && \multicolumn{1}{c}{$\chi^2/\text{d.o.f.}$} \\ \midrule
{$D^{\pi}_g(t)$} & -0:.596(65) & 0:.677(65)  && 1:.2 \\[2pt]
{$D^{\pi}_q(t)$} & -0:.304(26) &  1:.44(21) && 1:.0 \\[2pt]
{$D^{\pi}_v(t)$} & -0:.322(12) &  1:.286(76) && 3:.4 \\[2pt]
\midrule\midrule
\multicolumn{1}{c}{$z$-expansion} & \multicolumn{1}{c}{$\alpha_0$} & \multicolumn{1}{c}{$\alpha_1$} & \multicolumn{1}{c}{$\alpha_2$} & \multicolumn{1}{c}{$\chi^2/$d.o.f.}\\ \midrule
{$D^{\pi}_g(t)$} & -0:.265(14) & 0:.682(71) &-0:.38(36)& 1:.2 \\[2pt]
{$D^{\pi}_q(t)$} & -0:.244(11) & 0:.251(56)  &0:.61(30) & 0:.9 \\[2pt]
{$D^{\pi}_v(t)$} & -0:.2483(60) &  0:.308(27) & 0:.38(12) & 3:.5 \\[2pt]
\bottomrule
\end{tabular}
\end{center}
\caption{\label{tab:170pionparamsD}
Fit parameters of the monopole and $z$-expansion parametrizations of the $t$-dependence of the pion GFFs $D_i^{\pi}(t)$ renormalized in the $\overline{\text{MS}}$ scheme at scale $\mu=2~\text{GeV}$. We note the high $\chi^2$/d.o.f of the $D_v^{\pi}(t)$ fits, which are due to three data points fluctuating away from the trend of the curve, as seen in the center right panel of Fig.~\ref{fig:isoGFF}.}
\end{table}


\subsection{Forward limits}

Our results for the flavor decomposition of the momentum fraction and $D$-term at scale $\mu=2~\text{GeV}$ in the $\overline{\text{MS}}$ scheme using the monopole and $z$-expansion fit results of Tables~\ref{tab:170pionparamsA} and~\ref{tab:170pionparamsD} are shown in Table~\ref{tab:forwardlimit}. $A^{\pi}_g(0)$ is consistent with results obtained in a previous study of the gluon GFFs on an ensemble with $m_{\pi}=450~\text{MeV}$~\cite{Pefkou:2021fni,Shanahan:2018pib}, and with a  lattice extraction~\cite{ExtendedTwistedMass:2021rdx} in the forward limit with $N_f=2+1+1$ flavors at quark masses corresponding to the physical pion mass. Our result for $A^{\pi}_q(0)$ is however smaller than the result found in Ref.~\cite{ExtendedTwistedMass:2021rdx}. In contrast, we find a slightly larger contribution from gluons than from quarks to the momentum fraction of the pion at $\mu=2~\text{GeV}$. The separate quark-flavor contributions, $A^{\pi}_{u+d}(0)$ and $A^{\pi}_s(0)$,  are also smaller than those found in Ref.~\cite{ExtendedTwistedMass:2021rdx}. A possible explanation could be that the latter were computed on an $N_f=2+1+1$ ensemble at the physical quark mass, while our results were obtained on a single ensemble at $m_{\pi}\approx 170~\text{MeV}$ and could not be extrapolated to the physical point. Our results for $A^{\pi}_{q}(0)$ are also larger than what was found in Ref.~\cite{Loffler:2021afv} after extrapolation to the continuum limit. Our results for the total momentum fraction are slightly larger than the sum rule prediction, $A^{\pi}(0) = 1$.

\begin{table} 
\begin{center}
\begin{tabular}{S@{\hskip 0.2in}D{:}{}{2.7}@{\hskip 0.2in}D{:}{}{2.7}}
\toprule
& \multicolumn{1}{c}{monopole} & \multicolumn{1}{c}{$z$-expansion}  \\ \midrule
{$A_g^{\pi}(0)$} & 0:.546(18) & 0:.546(22)  \\[2pt]
{$A_q^{\pi}(0)$} & 0:.481(15) & 0:.485(18)  \\[2pt]
{$A_{u+d}^{\pi}(0)$} & 0:.463(11) & 0:.468(12)  \\[2pt]
{$A_s^{\pi}(0)$} & 0:.0176(57) & 0:.0174(66)  \\[2pt]
{$A^{\pi}(0)$} & 1:.026(23) & 1:.031(28) \\
\midrule\midrule
{$D_g^{\pi}(0)$} & -0:.596(65) & -0:.618(75) \\
{$D_q^{\pi}(0)$} & -0:.304(26) & -0:.242(53) \\
{$D_{u+d}^{\pi}(0)$} & -0:.313(17) & -0:.265(36) \\
{$D_s^{\pi}(0)$} & 0:.0092(94) & 0:.023(19) \\
{$D^{\pi}(0)$} & -0:.900(70) & -0:.860(92) \\
\bottomrule
\end{tabular}
\end{center}
\caption{\label{tab:forwardlimit}The flavor decomposition of the momentum fraction and the $D$ term of the pion, obtained from the monopole and $z$-expansion fits to the pion GFFs, renormalized at $\mu = 2\;\text{GeV}$ in the $\overline{\text{MS}}$ scheme, 
with parameters shown in Tables~\ref{tab:170pionparamsA} and~\ref{tab:170pionparamsD}.
}
\end{table}
\begin{figure}
    \centering
    \includegraphics[width=0.49\textwidth]{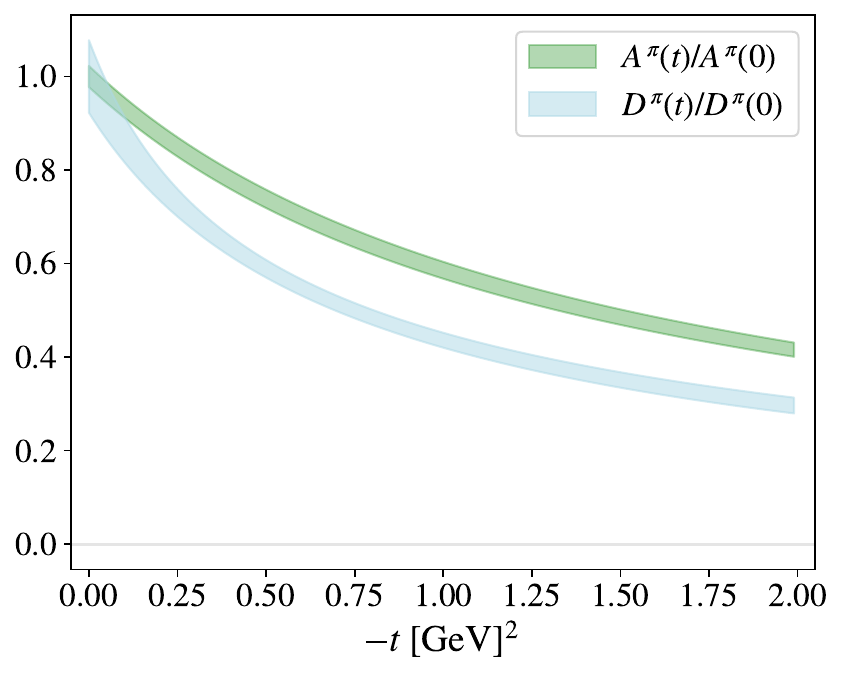}
    \caption{The pion GFFs rescaled by their central value at $t=0$, as obtained from the monopole model with fit parameters collected in Tables~\ref{tab:170pionparamsA} and~\ref{tab:170pionparamsD}.}
    \label{fig:ADpion}
\end{figure}

The total $D$ term obtained is consistent with $\chi$PT when the first chiral-symmetry breaking correction~\cite{Donoghue:1991qv}, which for the quark masses of this ensemble is estimated to result in $D^{\pi}(0)\approx -0.96$, is taken into consideration, and smaller in magnitude than the leading-order chiral limit prediction of $D^{\pi}(0)\approx -1$. Our result for $D^{\pi}_q(0)$ is statistically consistent with the result found in Refs.~\cite{Brommel:2005jC,Brommel:2007zz}, which was computed from several ensembles at heavier pion masses and extrapolated to the physical point, neglecting quark disconnected contributions and mixing with the gluon EMT. We find $D_g^{\pi}(0)$ to be smaller in magnitude than the result at $m_{\pi}\approx450~\text{MeV}$ neglecting mixing with quarks found in Ref.~\cite{Pefkou:2021fni}, [$-0.793(84)$].

Interesting physical comparisons can be made by considering the $t$-dependence of the GFFs, which we find to be different between $A^{\pi}$ and $D^{\pi}$ as seen in Fig.~\ref{fig:ADpion}, and consistent with the NLO $\chi$PT prediction~\cite{Donoghue:1991qv} in the small $|t|$ region, as seen in Fig.~\ref{fig:totalGFF}. Hadron radii associated with the spatial distributions of their physical properties have historically been defined in relation to the derivative of their form factors at $t=0$. For example, the magnitude of the derivative of the pion vector form factor $F_\text{v}^{\pi}$ is related to its charge radius, while that of $A^{\pi}(t)$ is related to its mass radius. We compare the relative sizes of the mass radius and the charge radius by taking the ratio
\begin{equation}
\frac{r^{\pi}_{\text{mass}}}{r^{\pi}_{\text{EM}}} \sim \sqrt{\frac{dA^{\pi}/dt|_{t=0}}{dF_\text{v}^{\pi}/dt|_{t=0}}} \;.
\end{equation}
Using the PDG averaged value for $dF_\text{v}^{\pi}/dt|_{t=0}$~\cite{ParticleDataGroup:2022pth} and the monopole fit of Sec.~\ref{subsec:renormGFFresults}, we obtain that the charge radius of the pion is approximately $1.6$ times larger than its mass radius. From equivalent relations for the mass radii of the individual constituents,
\begin{equation}
r^{\pi}_{i,\text{mass}} \propto \sqrt{\frac{1}{A^{\pi}_i(0)}\frac{dA_i^{\pi}}{dt}\bigg|_{t=0}} \;,
\end{equation}
we find $r^{\pi}_{g,\text{mass}}/r^{\pi}_{q,\text{mass}}\approx 1.1$, and $r_{q,\text{mass}}$ in agreement with a phenomenological extraction of the pion quark GFFs from experimental measurements~\cite{Kumano:2017lhr}. The equivalent quantity for $D^{\pi}$
\begin{equation}
\sqrt{\frac{1}{D^{\pi}_i(0)}\frac{dD_i^{\pi}}{dt}\bigg|_{t=0}} 
\end{equation}
is, however, approximately $2.5$ times smaller for $i=q$ than what was found in Ref.~\cite{Kumano:2017lhr}.

\section{SUMMARY AND CONCLUSION}
\label{sec:conc}

We present the first flavor decomposition of the $t$-dependence of the pion GFFs, calculated using a lattice QCD ensemble with quark masses yielding a close-to-physical pion mass. We constrain the gluon, isosinglet, and nonsinglet GFFs renormalized in the $\overline{\text{MS}}$ scheme at scale $\mu=2~\text{GeV}$, accounting for mixing between the gluon and isosinglet contributions. From our results, we perform the first extraction of renormalization scale- and scheme-independent hadron gravitational form factors from lattice QCD.

From fitting the $t$-dependence of $A^{\pi}$, we find indications of a smaller mass radius than charge radius for the pion. Our results are consistent with the momentum fraction sum rule, and with the NLO $\chi$PT prediction for the pion $D$ term. We observe some tension between $D_q^{\pi}(t)$ obtained by fitting each lattice operator irrep individually, as discussed in Appendix~\ref{app:renorm}. Even though this effect is not significant for this calculation, it is important to understand it for future extractions with higher precision. This result also highlights the value of considering different irreducible representations when extracting GFFs from lattice QCD. Another interesting feature in the analysis of the data is that excited state contamination affects the extraction of bare $D^{\pi}_g(t)$ significantly more than any of the other GFFs, as discussed in Appendix~\ref{app:bare}. No such effect was noted in a previous extraction of $A^{\pi}_g(t)$ and $D^{\pi}_g(t)$ at $m_{\pi}\approx 450~\text{MeV}$~\cite{Pefkou:2021fni}.

Future improvements, besides the repetition of the calculation on additional lattice ensembles in order to enable the continuum, physical quark mass, and infinite-volume limits to be taken, could include using a variational basis of hadron  interpolators~\cite{MICHAEL198558,LUSCHER1990222,PhysRevD.77.034501} to better control excited state contamination. Another improvement would be the use of gauge-invariant renormalization schemes~\cite{Costa:2021iyv,Spanoudes:2022gow} that allow the renormalization mixing matrix of the EMT to be computed nonperturbatively on large-volume ensembles.

As few experimental constraints on the pion GFFs exist to date, the results in this work, and future calculations with further improved systematics, provide particularly valuable information on hadron structure. The structure of the pion is of particular interest as the pseudo-Goldstone boson of dynamical chiral symmetry breaking. 
Taken together with the first constraints on proton GFFs from experimental measurements in recent years~\cite{Burkert:2018bqq,Duran:2022xag}, and the agreement with lattice QCD results for the gluon contribution~\cite{Pefkou:2021fni,Shanahan:2018nnv}, these developments mark a new milestone in our understanding of the gravitational properties of hadrons.

\begin{acknowledgements}
The authors thank Will Detmold, Fernando Romero-L\'opez, and Ross Young for useful feedback and suggestions. This work is supported in part by the U.S. Department of Energy, Office of Science, Office of Nuclear Physics, under grant Contract No. DE-SC0011090 and by Early Career Award No. DE-SC0021006, and has benefited from the QGT Topical Collaboration DE-SC0023646. P.E.S is supported in part by Simons Foundation grant 994314 (Simons Collaboration on Confinement and QCD Strings). This research used resources of the National Energy Research Scientific Computing Center (NERSC), a U.S. Department of Energy Office of Science User Facility operated under Contract No. DE-AC02-05CH11231, as well as resources of the Argonne Leadership Computing Facility, which is a DOE Office of Science User Facility supported under Contract DE-AC02-06CH11357, and the Extreme Science and Engineering Discovery Environment (XSEDE), which is supported by National Science Foundation grant number ACI-1548562. Computations were carried out in part on facilities of the USQCD Collaboration, which are funded by the Office of Science of the U.S. Department of Energy. The authors thank Robert Edwards, Rajan Gupta, Balint Jo{\'o}, Kostas Orginos, and the NPLQCD Collaboration for generating one of the ensembles used in this study.
The Chroma~\cite{Edwards:2004sx}, QLua~\cite{qlua},  QUDA~\cite{Clark:2009wm,Babich:2011np,Clark:2016rdz}, QDP-JIT~\cite{6877336}, and QPhiX~\cite{10.1007/978-3-319-46079-6_30} software libraries were used in this work. Code for disconnected diagrams was adapted from LALIBE~\cite{lalibe}, including the hierarchical probing implementation by Andreas Stathopoulos~\cite{Stathopoulos:2013aci}.
Data analysis used NumPy~\cite{harris2020array}, SciPy~\cite{2020SciPy-NMeth}, pandas~\cite{jeff_reback_2020_3715232,mckinney-proc-scipy-2010}, lsqfit~\cite{peter_lepage_2020_4037174}, and gvar~\cite{peter_lepage_2020_4290884}.
Figures were produced using matplotlib~\cite{Hunter:2007}.

\end{acknowledgements}

\onecolumngrid
\appendix

\section{FITS TO BARE RESULTS}
\label{app:bare}

This section presents additional details of our analysis of the bare matrix elements discussed in Sec.~\ref{sec:barematel}.
In the subsections below, we  discuss separately the connected quark, disconnected quark (both light and strange), and gluon GFFs. For each, we assess the degree of excited state contamination (ESC) and describe how the analysis hyperparameters were selected. Note that the physics of each contribution is different, and conclusions drawn from the analysis of one cannot be applied to another.

\begin{figure}
    \centering
    \subfloat[\centering  $\tau_1^{(3)}$]
    {{\includegraphics[width=0.9\textwidth]{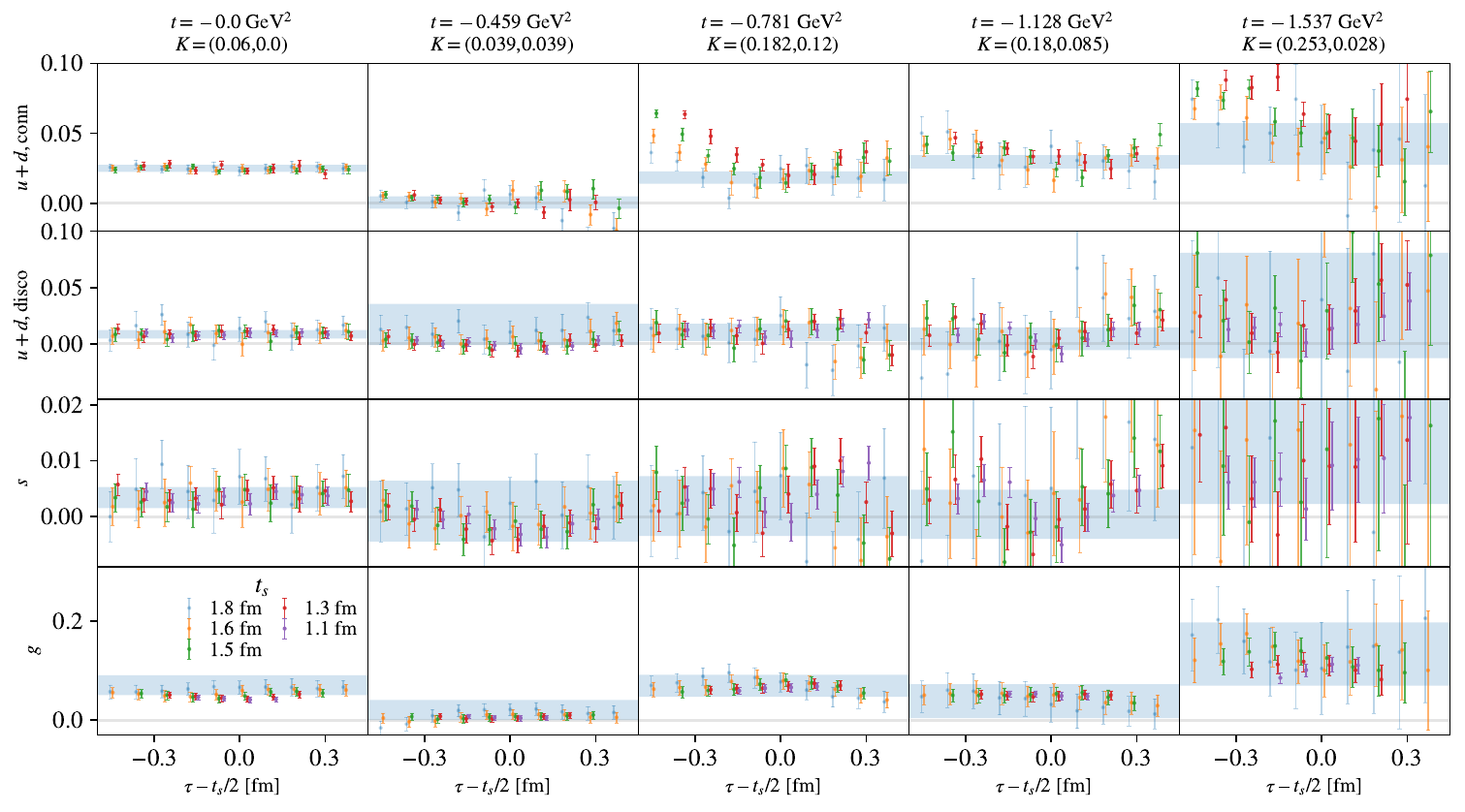}}} \\
    \subfloat[\centering  $\tau_3^{(6)}$]
    {{\includegraphics[width=0.9\textwidth]{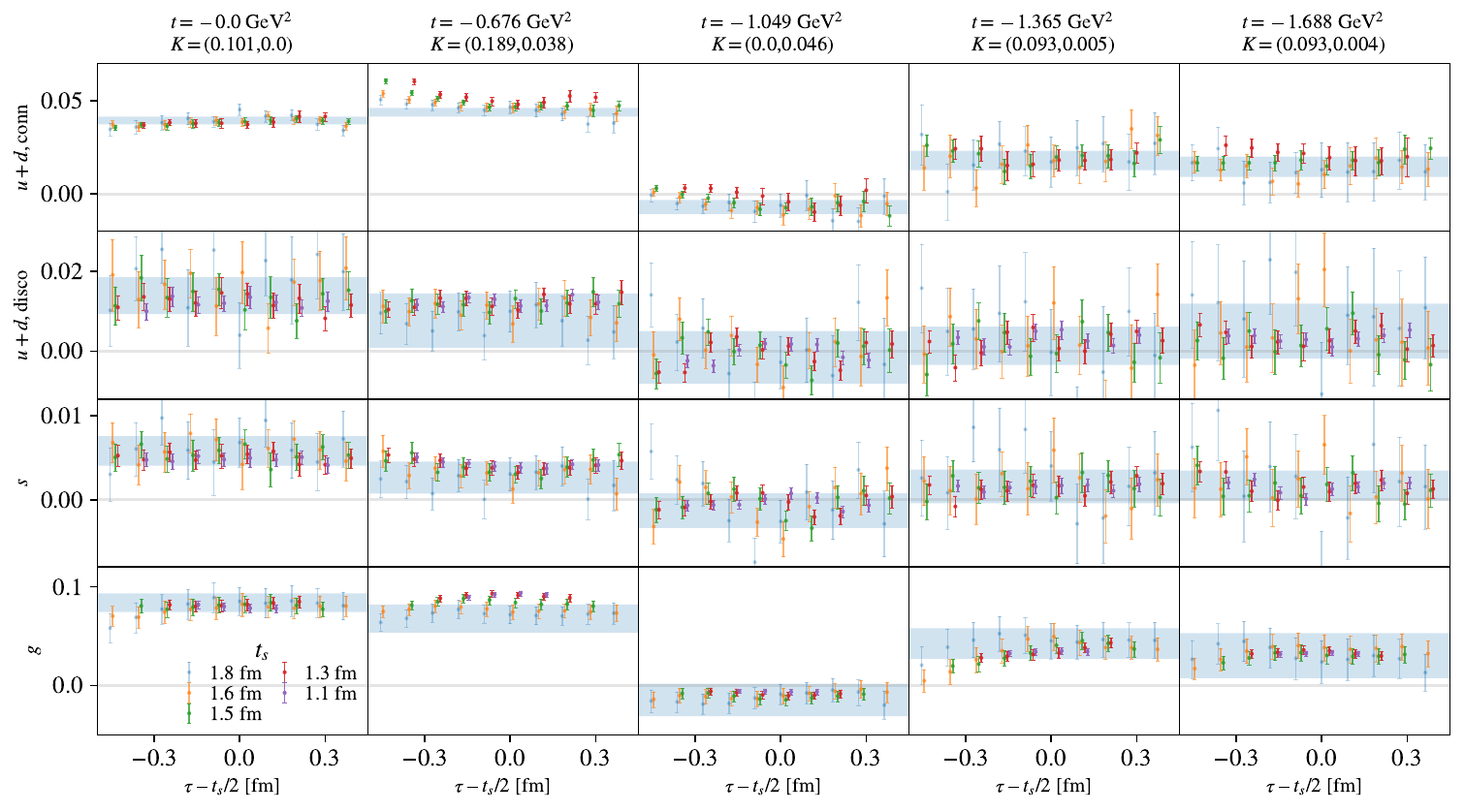} }}
    \caption{Examples of averaged ratios for $\tau_1^{(3)}$ (a) and $\tau_3^{(6)}$ (b). Each column represents a single ratio, with the corresponding $t$-value and $(K^A,K^D)$ coefficients indicated above. The rows represent the bare connected, light-quark disconnected, strange, and gluon contributions to the ratio. The overlaid bands are the results of fits to the ratios obtained via the summation method as described in Secs.~\ref{app:bare-conn},~\ref{app:bare-disco}, and~\ref{app:bare-gluon}.}
    \label{fig:bumps}
\end{figure}

In Fig.~\ref{fig:bumps}, we present examples of averaged ratios, defined in the text under Eq.~\eqref{eq:summation}, including the connected, disconnected, and gluon contributions and corresponding fits obtained from the summation method as described in the subsections below. These are selected to span the $t$-range for each of the two irreps, but due to the large number of ratios computed in this analysis, it is useful to consider ways of investigating ESC that do not treat the ratios individually. We define effective matrix elements from the summed, averaged ratios of matrix elements defined in Eq.~\eqref{eq:summation} and the surrounding text as 
\begin{equation} \label{eq:effectiveME}
\text{ME}_{i\mathcal{R}ct}^{\text{eff}}(t_s) = \partial_{t_s}\bar{\Sigma}_{i\mathcal{R}ct}(t_s)  \approx \frac{1}{\delta {t_s}} \left[ \bar{\Sigma}_{i\mathcal{R}ct}(t_s+\delta {t_s}) - \bar{\Sigma}_{i\mathcal{R}ct}(t_s) \right] \;.
\end{equation}
We use $\delta {t_s}=1$ for the gluon and quark disconnected data and $\delta {t_s}= 2$ for the quark connected data, since only sink times separated by two time slices are available for the latter.
The explicit $\tau_{\text{cut}}$ dependence of the summed ratios is dropped in this discussion.
From the effective matrix elements, which are formed directly from the data, one can obtain effective bare GFFs $A^{\pi,\text{eff}}_{i\mathcal{R}t}(t_s)$ and $D^{\pi,\text{eff}}_{i\mathcal{R}t}(t_s)$ for each flavor $r$, irrep $\mathcal{R}$, momentum bin $t$, and sink time $t_s$ by fitting the overconstrained system of linear equations,
\begin{equation} \label{eq:effectiveGFF}
\vec{K}^A_{\mathcal{R}t} A_{i\mathcal{R}t}^{\pi,\text{eff}}(t_s) + \vec{K}^D_{\mathcal{R}t} D_{i\mathcal{R}t}^{\pi,\text{eff}}(t_s) = \text{\bf{ME}}^{\text{eff}}_{i\mathcal{R}t}(t_s) \;,
\end{equation}
where $\vec{K}^A_{\mathcal{R}t}$ and $\vec{K}^D_{\mathcal{R}t}$, defined in Eq.~\eqref{eq:KAKDME}, are the kinematic coefficients of the GFFs in the decompositions of the effective bare matrix elements $\text{\bf{ME}}^{\text{eff}}_{i\mathcal{R}t}(t_s)$, written as vectors in the space of $c$-bins.

The effective GFFs must be interpreted with more caution than effective masses, due to the fit involved in their definition.
In the high-statistics limit, the effective GFFs as functions of sink time $t_s$ will plateau to the value of the bare GFFs at times when ESC becomes negligible. Note that effective GFFs are used only as a consistency check and as a tool to determine appropriate analysis hyperparameters for the analysis described in the main text.

\subsection{Quark connected}
\label{app:bare-conn}

\begin{figure*}
\centering
\includegraphics[width=0.98\textwidth]{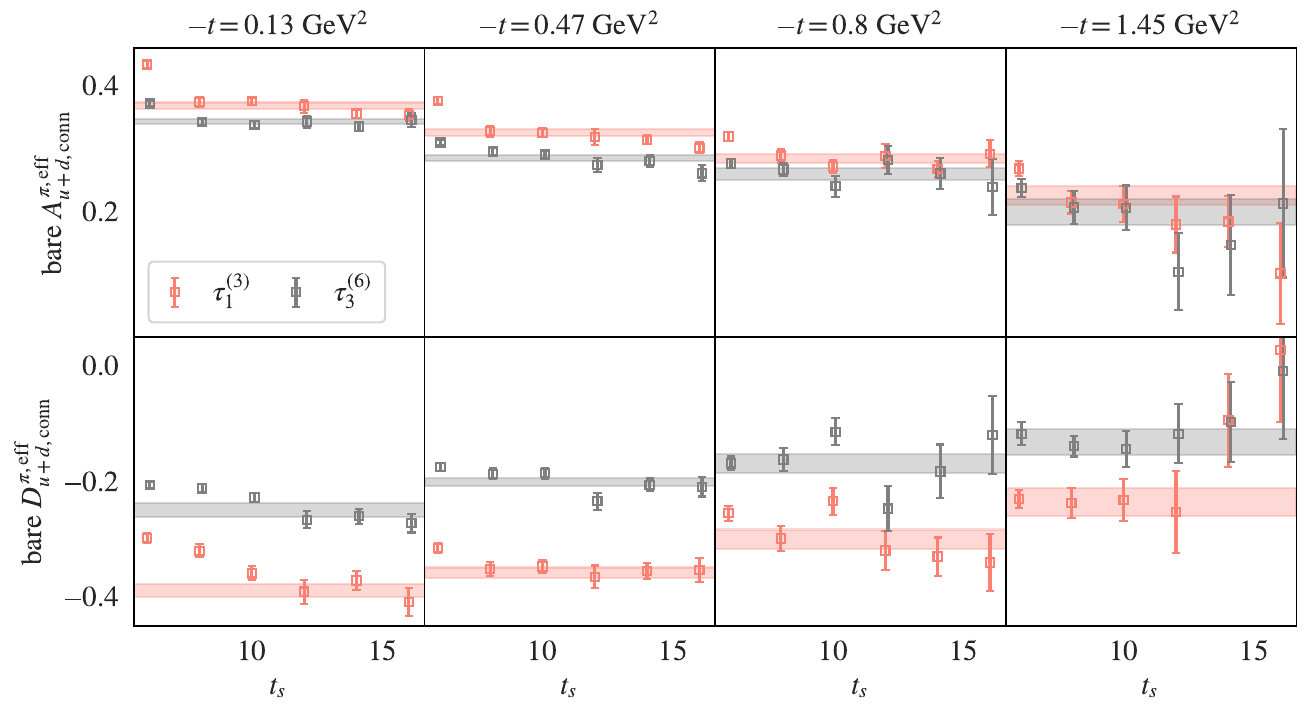}
\caption{
    Effective bare GFFs, defined in Eq.~\eqref{eq:effectiveGFF}, for the connected light-quark contribution,
    computed from summed ratios with $\tau_{\text{cut}} = 2$.
    Examples are chosen from the full set of 25 $t$-bins to span the momentum range and to represent different qualitative behaviors observed across all bins.
    Overlaid bands are not fit to the data shown, but are rather bare GFFs computed from the bare matrix elements used in the main analysis.
}
\label{fig:conn_GFFeff}
\end{figure*}

Inspection of the effective GFFs computed from the light-quark connected data suggests that ESC is a relatively small effect in this channel.
Figure~\ref{fig:conn_GFFeff} shows several examples.
Overall, most bins exhibit clear evidence of ESC decaying away at early $t_s$ followed by a plateau with fluctuations that may be attributed to signal-to-noise effects.
As described in Sec.~\ref{subsec:connbare}, for the main analysis we use the summation method to extract the bare matrix elements, including the lowest-order exponential term of Eq.~\eqref{eq:summation} in the fitting, averaging over fit ranges $\tau_\text{cut} \in\{2,3,4\}$, $t_{s,\text{max}}\in\{16,18\}$, and $t_{s,\text{min}} \in\{6,8,10\}$, restricting to include at least five distinct  $t_s$ in each fit.
To remove unstable outlier fits, fits not satisfying $(\delta \text{ME})_\text{med} / \sqrt{10} < \delta \text{ME} < \sqrt{10} (\delta \text{ME})_\text{med}$ are discarded, where $\delta \text{ME}$ is the error of $\text{ME}$ for the fit and $(\delta \text{ME})_\text{med}$ is the median error over the pool of fits to be averaged. 
Comparing with the fits to the summation method results (overlaid in Fig.~\ref{fig:conn_GFFeff}), we find that these are consistent with the effective GFFs. The bare GFF fits appear to be resilient against the signal breakdown at late $t_s$ apparent in some effective GFFs. Taken together, this provides a cross-check of the analysis employed in the main text.

\subsection{Quark disconnected}
\label{app:bare-disco}

\begin{figure}
    \centering
    \subfloat[\centering  ]
    {{\includegraphics[width=0.9\textwidth]{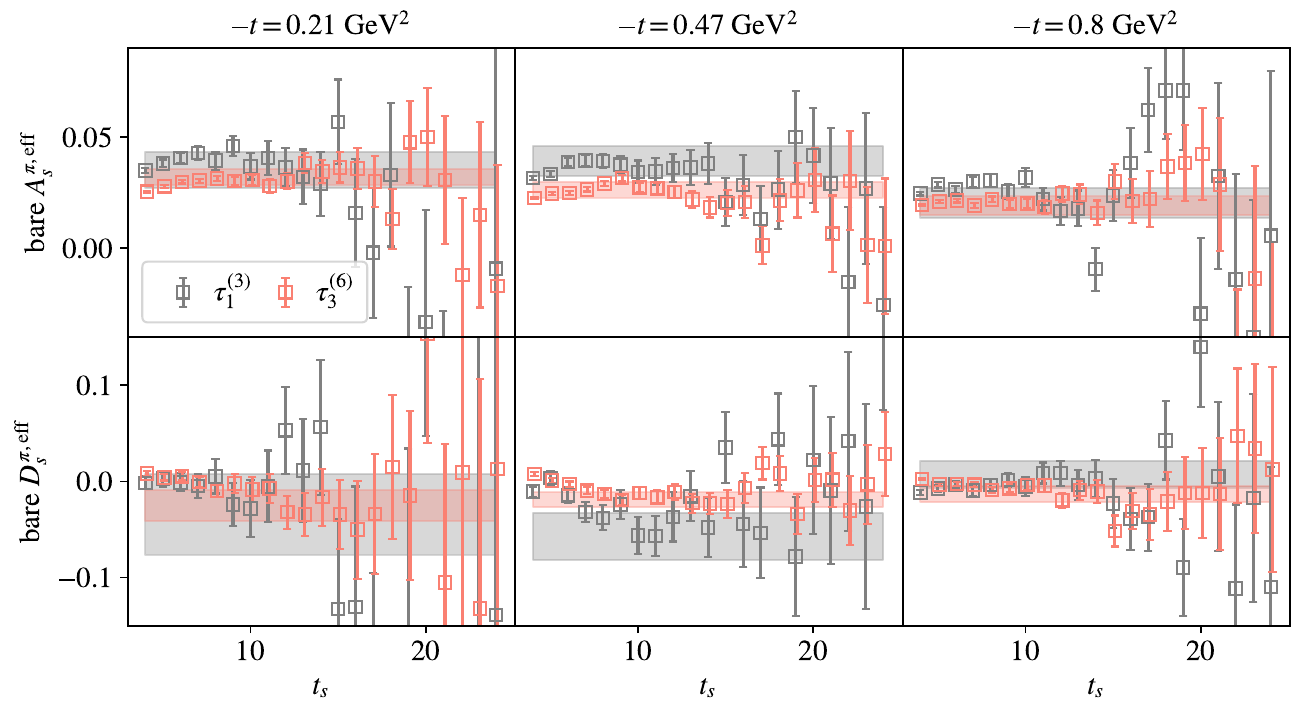} }\label{fig:strangejustify}} \\
    \subfloat[\centering  ]
    {{\includegraphics[width=0.9\textwidth]{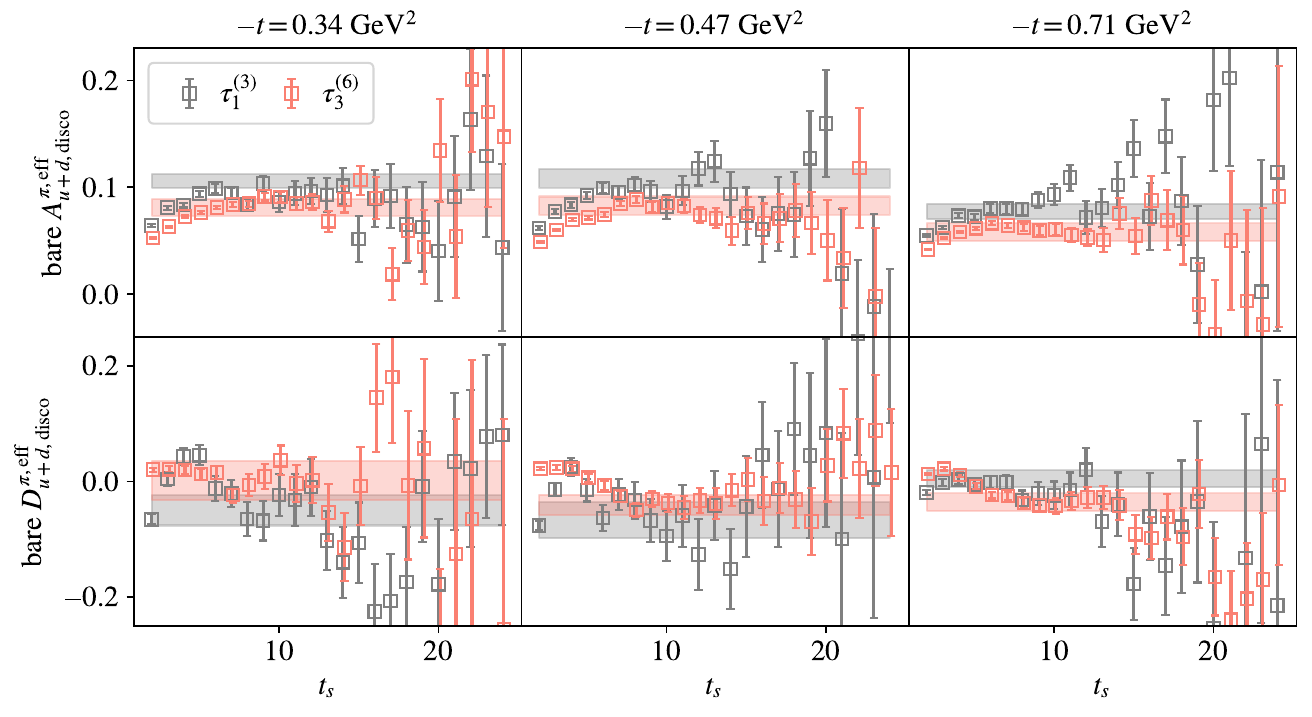} }\label{fig:discojustify}}
    \caption{Examples of bare strange (a) and light (b) quark disconnected effective GFFs, computed from Eq.~\eqref{eq:effectiveGFF} using summed ratios with $\tau_{\text{cut}} = 2$. The bands are not fit to the data shown but correspond to the bare GFFs used to produce the final results of the main text.}
    \label{fig:alldiscojustify}
\end{figure}

Similarly to the connected quark contribution, the disconnected light and strange quark contributions show mild ESC, but stronger signal-to-noise degradation. Effective GFF examples for different $t$-bins are shown in Fig.~\ref{fig:strangejustify} for the strange quark contribution and in Fig.~\ref{fig:discojustify} for the light quark piece. Consistent bare GFFs are obtained when varying the parameter $t_{s,\text{min}}$ of the summation fits. For the main analysis, we choose $t_{s,\text{min}}=7$ for $\tau_1^{(3)}$ and $t_{s,\text{min}}=11$ for $\tau_3^{(6)}$, which are the choices that yield the highest $p$-values for the majority of the $t$-bins in the bare GFF fits of both the light and the strange disconnected quark contributions. Summation fits are performed to all $t_s$-ranges extending over $4$ or more time slices in the window $[t_{s,\text{min}},t_{s,\text{max}}]$, and model-averaged based on their AIC weight. The upper bound is set to $t_{s,\text{max}}=24$, which is when approximately half of the $c$-bin ratios become consistent with non-Gaussian noise, as determined by their fourth cumulant being greater than $1$, while $2 \leq \tau_{\text{cut}}\leq 9$. The same condition as for the connected contribution, $(\delta \text{ME})_\text{med} / \sqrt{10} < \delta \text{ME} < \sqrt{10} (\delta \text{ME})_\text{med}$, is imposed for the fits that are model averaged together. The resulting bare GFFs for different $t$-bins, also presented in Fig.~\ref{fig:bareGFFdisco}, are shown as bands in Fig.~\ref{fig:alldiscojustify}.

\subsection{Gluon}
\label{app:bare-gluon}

\begin{table}[h]
    \begin{ruledtabular}
    \begin{tabular}{c|rrrrrrr}
    $t$-bin \# & 2-3 &  4-6 & 7-10 & 11-15 & 16-17 & 18-19 & 20-25 \\ 
    $\; t_{s,\text{min}}  \;$ & 18 & 17 & 16 & 15 & 14 & 13 & 10
    \end{tabular}
    \end{ruledtabular}
    \caption{Choices of $t_{s,\text{min}}$ for the constraints with $K^{D}\neq0$ within each gluon $t$-bin. These are taken to be the same for both irreps. No choice is shown for the first $t$-bin since it doesn't contain enough constraints with $K^{D}\neq0$ to provide an estimate for $D$.}
    \label{tab:tmin}
\end{table}

\begin{figure}
    \centering
    \subfloat[\centering  ]
    {{\includegraphics[width=0.9\textwidth]{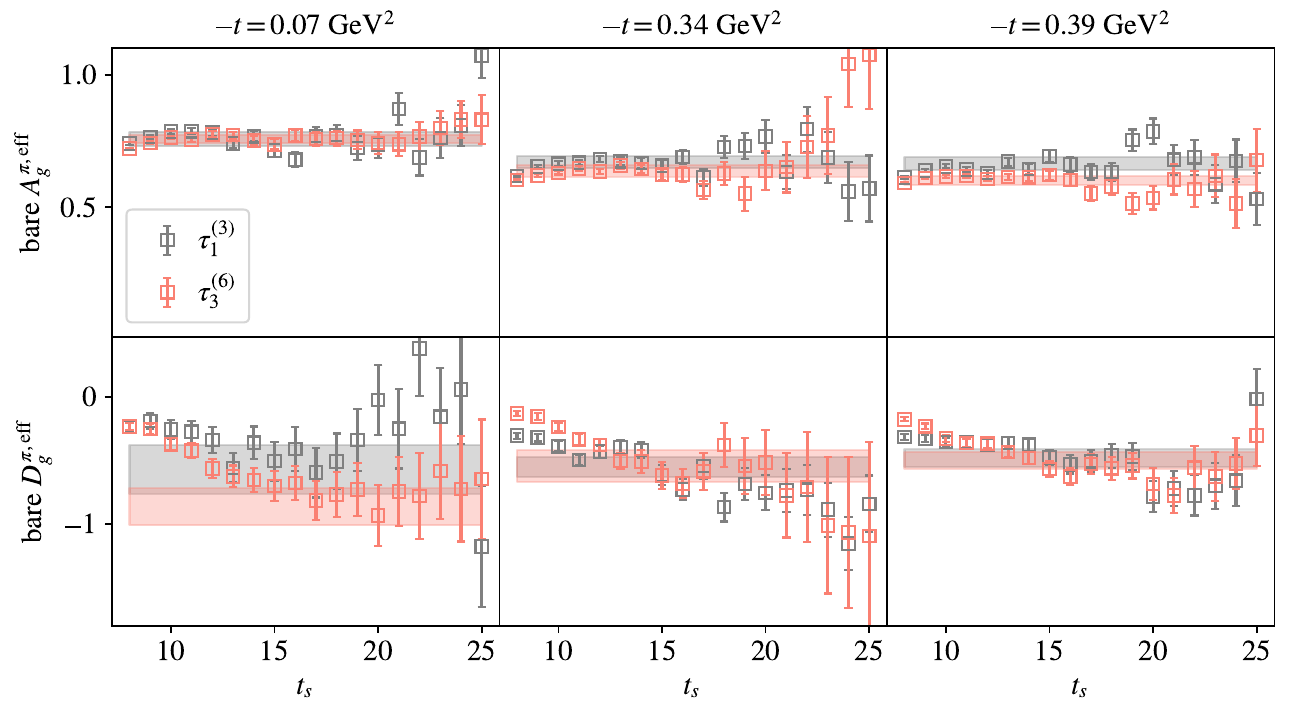}\label{fig:justifyglue1}}} \\
    \subfloat[\centering  ]
    {{\includegraphics[width=0.9\textwidth]{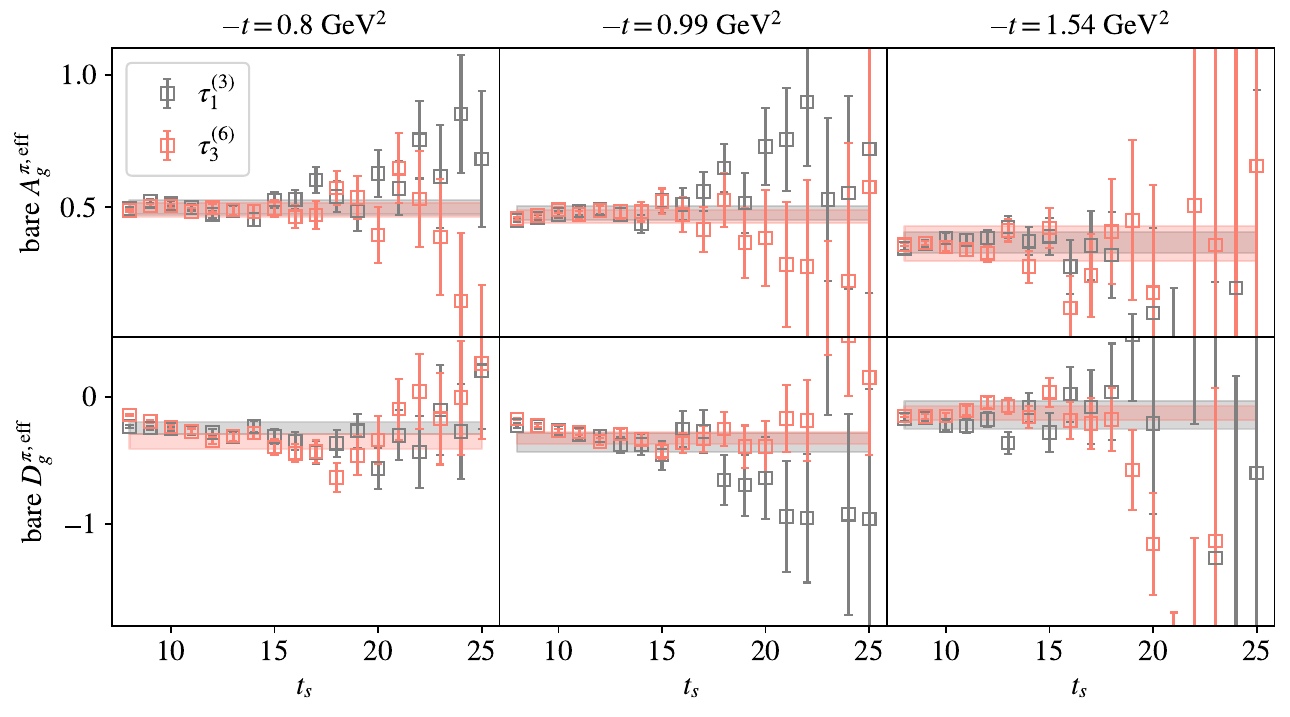}\label{fig:justifyglue2} }}
    \caption{Examples of gluon effective GFFs, computed from Eq.~\eqref{eq:effectiveGFF} using summed ratios with $\tau_{\text{cut}} = 4$. The effective $D_g^{\pi,\text{eff}}$, shown in the bottom panels of (a) and (b), does not reach a plateau until later $t_s$ values compared to $A$, a feature that is particularly prominent for smaller $t$-bins like those shown in (a). The bands are not fit to the data shown but correspond to bare GFFs obtained by fitting the bare matrix elements used to compute the results from the main text.}
    \label{fig:justifyglue}
\end{figure}

The bare gluon GFFs have a unique feature in comparison with the quark contributions: ESC affects $A$ and $D$ differently, as demonstrated in Fig.~\ref{fig:justifyglue1}. At the earliest $t_s$ values, the effective $D$ drifts downwards as a function of the sink time $t_s$, and does not reach a plateau region until later sink times compared with $A$. This behavior is present but more mild for larger momenta, as shown in Fig.~\ref{fig:justifyglue2}. To treat this effect in the main analysis, we use different hyperparameters for $c$-bins with and without contributions from $D$. For the $c$-bins with the coefficient of $D$, $K^D$, being zero, which only constrain $A$, we choose $t_{s,\text{min}}=10$ for both irreps and all $t$-bins, which is the approximate starting sink time of the $A^{\text{eff}}$ plateaus seen in Fig.~\ref{fig:justifyglue}. For the $c$-bins with $K^D\neq0$, we select $t_{s,\text{min}}$ as the point where $D^{\text{eff}}$ at different $t$-bins becomes consistent with a plateau. The chosen $t_{s,\text{min}}$ values, listed in Table~\ref{tab:tmin}, decrease as $|t|$ increases, as expected from the behavior of the corresponding $D^{\text{eff}}$ shown in Fig.~\ref{fig:justifyglue}. These $t_{s,\text{min}}$ values are used as the lowest possible bound in the summation fit ranges. The upper bound is set to $t_{s,\text{max}}=25$ using the same non-Gaussianity cut as for the disconnected results discussed in the above section. We set $4\leq \tau_{\text{cut}} \leq 9$ to avoid contact terms due to the flowed gauge links used in the computation of the gluon EMT. The bare gluon GFFs obtained using the results from the summation fits with these hyperparameter choices are shown as bands in Fig.~\ref{fig:justifyglue}.

\section{SINGLE-IRREP RESULTS}
\label{app:renorm}

\begin{table}[h]
\begin{center}
\begin{tabular}{SD{:}{}{2.7}D{:}{}{2.7}D{:}{}{2.7}D{:}{}{2.7}D{:}{}{2.7}D{:}{}{2.7}}
\toprule
\multicolumn{1}{c}{monopole} & \multicolumn{1}{c}{$\tau_1^{(3)}:\; \alpha$} &\multicolumn{1}{c}{$\Lambda$~[GeV]} & \multicolumn{1}{c}{$\chi^2/\text{d.o.f.}$}  & \multicolumn{1}{c}{$\tau_3^{(6)}:\; \alpha$} &\multicolumn{1}{c}{$\Lambda$~[GeV]} & \multicolumn{1}{c}{$\chi^2/\text{d.o.f.}$}\\ \midrule
{$A^{\pi}_g(t)$} & 0:.568(65) & 1:.107(47)  & 1:.2 & 0:.541(18) & 1:.152(43)  & 0:.63 \\[2pt]
{$D^{\pi}_g(t)$} & -0:.58(13) &  0:.72(13) & 1:.2 & -0:.72(12) &  0:.603(81) & 0:.52\\[2pt]
{$A^{\pi}_q(t)$} & 0:.525(36) & 1:.243(48)  & 0:.81 & 0:.459(14) & 1:.206(50)  & 0:.86 \\[2pt]
{$D^{\pi}_q(t)$} & -0:.442(59) & 1:.20(21)  & 1:.1 & -0:.271(33) & 1:.26(25)  & 0:.48 \\[2pt]
\bottomrule
\end{tabular}
\end{center}
\caption{\label{tab:monopolesingle}
Fit parameters of monopole fits to the pion GFFs obtained individually from each one of the two irreps, $\tau_1^{(3)}$ and $\tau_3^{(6)}$, renormalized in the $\overline{\text{MS}}$ scheme at scale $\mu=2~\text{GeV}$.
}
\end{table}

\begin{figure}
    \centering
    \includegraphics[width=\textwidth]{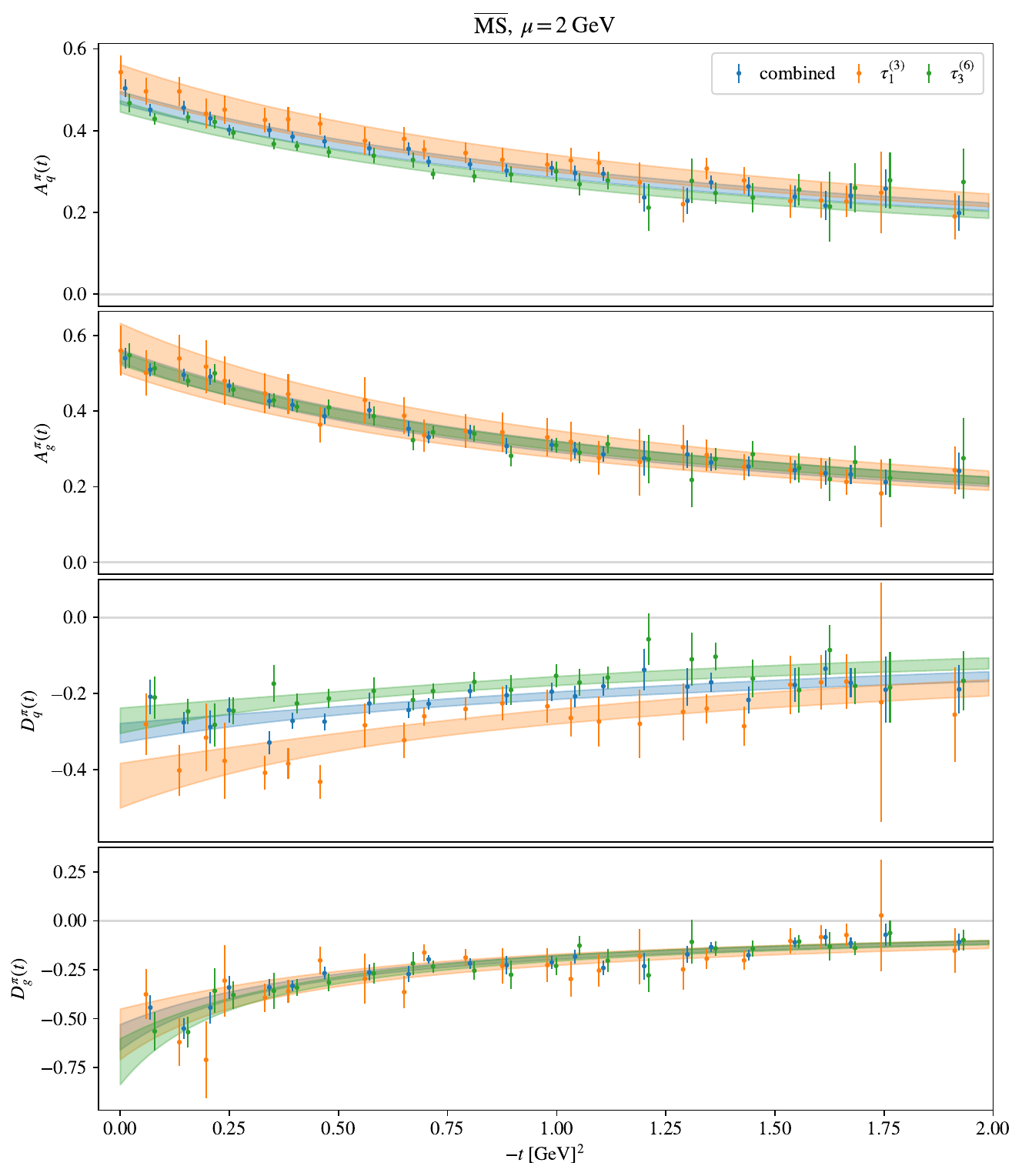}
    \caption{The renormalized isosinglet quark and gluon combined-irrep GFFs of Sec.~\ref{subsec:renormGFFresults} (blue points) shown alongside the GFFs obtained by fitting just the result for irrep $\tau_1^{(3)}$ (orange points) or $\tau_3^{(6)}$ (green points). The overlaid bands are monopole fits to the corresponding matching color data, with fit parameters tabulated in Table~\ref{tab:monopolesingle}.}
    \label{fig:Drenormtestl}
\end{figure}

\begin{figure}
    \centering
    \includegraphics[width=\textwidth]{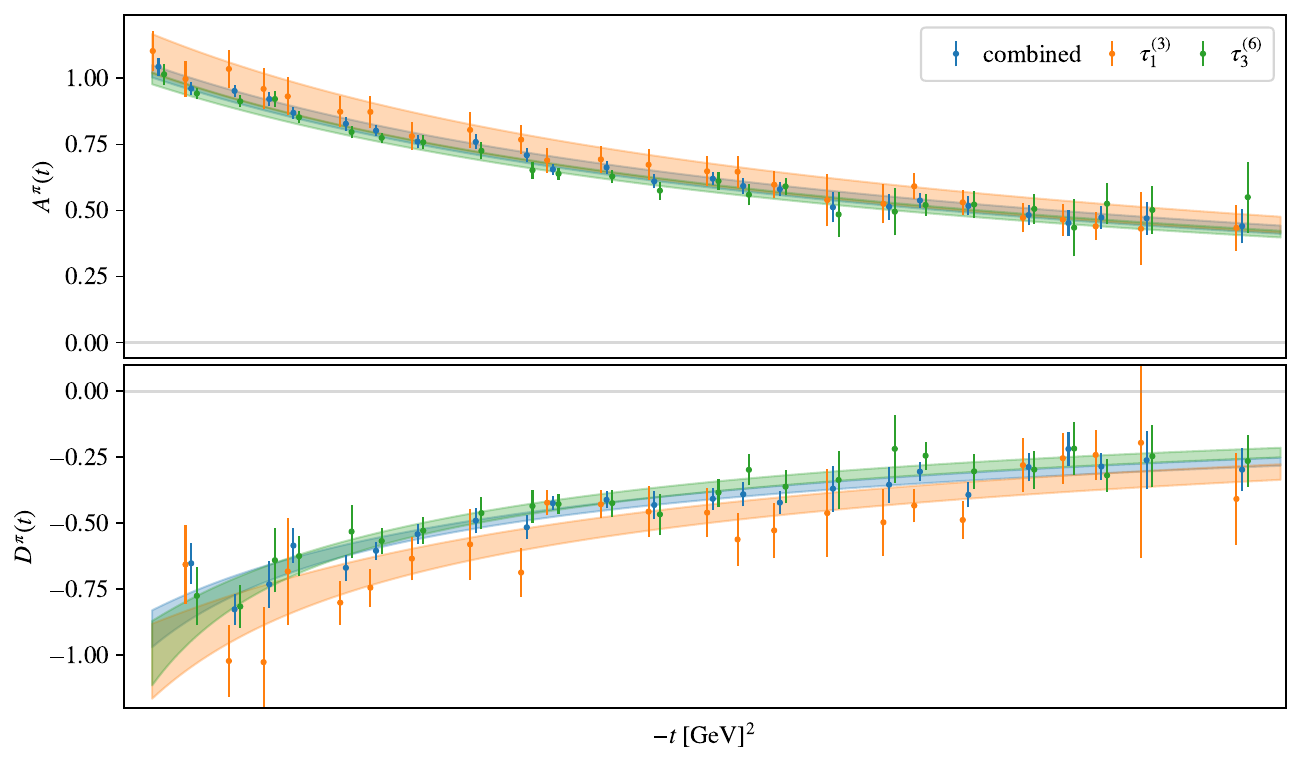}
    \caption{Same conventions as in Fig.~\ref{fig:Drenormtestl}, but for the renormalization scale- and scheme-independent total GFFs.}
    \label{fig:testotal}
\end{figure}

In this Appendix, we present the renormalized isosinglet and gluon GFFs obtained by considering the two irreps individually, and compare them against the combined-irrep results of Sec.~\ref{subsec:renormGFFresults}. The single-irrep results are obtained by solving Eq.~\eqref{eq:firstmatrixcombined} with
\begin{equation}
\vec{ME}_{\mathcal{R}t} = \begin{pmatrix}
\vec{ME}_{q\mathcal{R}t} \\
\vec{ME}_{g\mathcal{R}t} \\
\end{pmatrix}, \; \text{and} \;\;
\mathbb{RK}_{\mathcal{R}t}^{\overline{\text{MS}}} =
\begin{pmatrix}
R_{qq\mathcal{R}}^{\overline{\text{MS}}}\mathbb{K}_{\mathcal{R}t} &
R_{qg\mathcal{R}}^{\overline{\text{MS}}}\mathbb{K}_{\mathcal{R}t} \\
R_{gq\mathcal{R}}^{\overline{\text{MS}}}\mathbb{K}_{\mathcal{R}t} &
R_{gg\mathcal{R}}^{\overline{\text{MS}}}\mathbb{K}_{\mathcal{R}t} \\
\end{pmatrix} \;,
\end{equation}
for $\mathcal{R}\in\{\tau_1^{(3)},\tau_3^{(6)}\}$. The gluon and isosinglet results are shown in Fig.~\ref{fig:Drenormtestl}, and the total GFFs in Fig.~\ref{fig:testotal}.

\begin{figure}
    \centering
    \includegraphics[width=\textwidth]{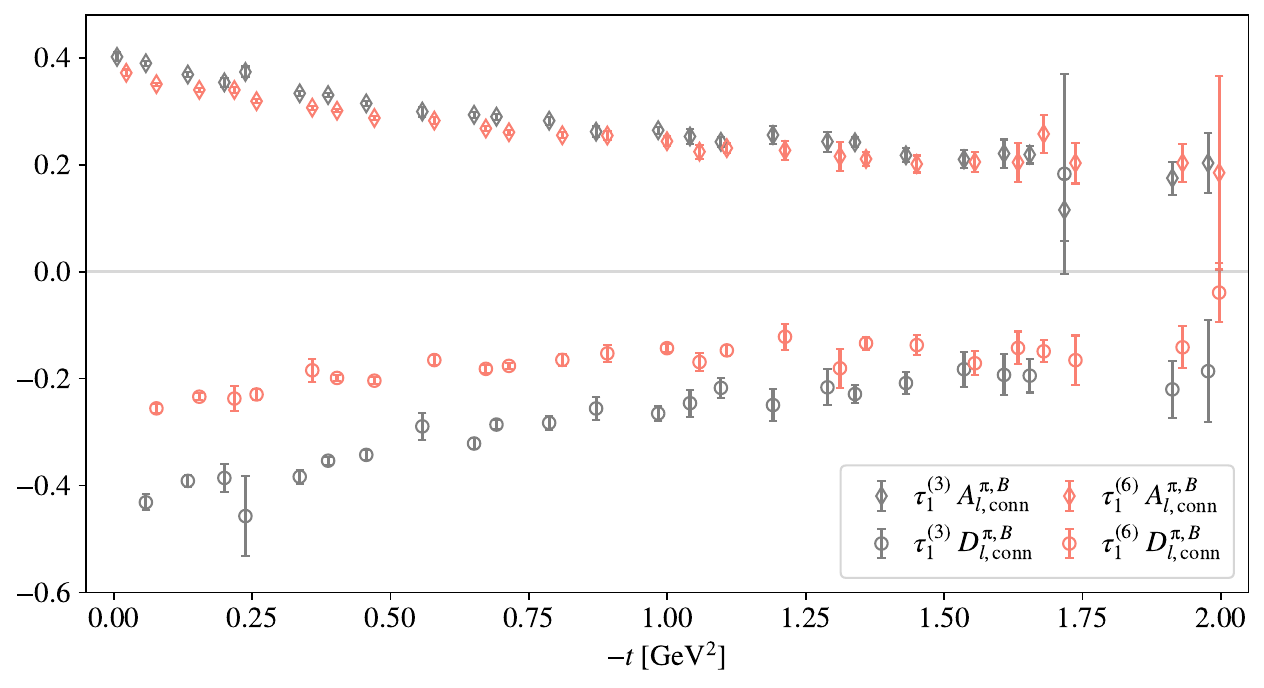}
    \caption{The connected contribution to the bare quark GFFs for the separate irreps. The tension between the $D^{\pi,B}_{l,\text{conn}}$ results for the two irreps is discussed in Appendix~\ref{app:renorm}.}
    \label{fig:conn_bare}
\end{figure}

Tension is observed between the two individual irrep results for $D^{\pi}_q(t)$. However, the combined fit procedure yields fits with better quality on average than the individual irrep ones, with $60\%$ of the $t$-bins having higher $p$-value than that for at least one of the single-irrep fits. 
Despite the fact that no simple expectation for the relationship between the bare GFFs of the two irreps for a single flavor contribution exists when mixing is present, we expect that the renormalized $D^{\pi}_q(t)$ inconsistency is due to the tension between the bare connected $D^{\pi,\text{conn},B}_{\mathcal{R}l}$ results of the two irreps, which are shown in Fig.~\ref{fig:conn_bare}. This could be due to different discretization artifacts, including hypercubic symmetry breaking effects arising from not further reducing the EMT to the ``little groups'' that it formally transforms under when it is boosted in different directions. Further investigations are needed to understand the origin of this tension.


\twocolumngrid
\bibliography{main}

\end{document}